\documentclass[a4paper,11pt]{article}
\usepackage{jcappub}
\usepackage{amsmath}
\usepackage{subcaption}
\usepackage{mathtools}
\usepackage{bm}
\usepackage[normalem]{ulem}
\usepackage{multirow}
\usepackage{comment}
\usepackage{xfrac}
\usepackage{enumitem}
\usepackage[pscoord]{eso-pic}
\usepackage{url}
\usepackage{tikz}
\usetikzlibrary{shapes.geometric, arrows, positioning}
\usepackage{ragged2e}
\usepackage{blindtext}
\usepackage{circuitikz}
\tikzstyle{process} = [rectangle, rounded corners, minimum width=3cm, minimum height=1.5cm, text centered, draw=black, fill=blue!20]
\tikzstyle{arrow} = [thick,->,>=stealth]
\tikzstyle{title} = [rectangle, text centered, minimum height=1cm]

\DeclareMathAlphabet{\mathpazoletters}{OT1}{ppl}{m}{it}

\newcommand{\placetextbox}[3]{
	\setbox0=\hbox{#3}
	\AddToShipoutPictureFG*{
		\put(\LenToUnit{#1\paperwidth},\LenToUnit{#2\paperheight}){\vtop{{\null}\makebox[0pt][c]{#3}}}
	}
}

\newcommand{\Neff}{N_\text{eff}}

\newcommand{\eqsp}{\;\,}

\title{Improved Big Bang Nucleosynthesis constraints on decaying massive relics

}

\author[a]{S.~Bianco,}
\author[b]{J.~Frerick,}
\author[c]{M.~Hufnagel,}
\author[a,d]{and K.~Schmidt-Hoberg}

\affiliation[a]{Deutsches Elektronen-Synchrotron DESY, Notkestr.~85, 22607 Hamburg, Germany}
\affiliation[b]{Dipartimento di Fisica, Sapienza Università di Roma \& Sezione INFN Roma1, Piazzale Aldo Moro 5, 00185, Roma, Italy}
\affiliation[c]{Service de Physique Th\'{e}orique, Universit\'{e} Libre de Bruxelles, C.P. 225, B-1050 Brussels, Belgium}
\affiliation[d]{CP3-Origins, University of Southern Denmark,
Campusvej 55, DK-5230 Odense M, Denmark}

\emailAdd{sara.bianco@desy.de}
\emailAdd{jonas.frerick@uniroma1.it}
\emailAdd{marco.hufnagel@ulb.be}
\emailAdd{kai.schmidt-hoberg@desy.de}

\abstract{
We present updated and improved Big Bang Nucleosynthesis~(BBN) constraints on heavy, long-lived beyond the Standard Model (BSM) relics $\phi$ decaying into pairs of Standard Model particles, covering a comprehensive set of two-body decay channels. We treat the leading effects of these injections in detail, discussing the modification of the neutron-to-proton ratio from hadronic interconversions, as well as hadro- and photodisintegration of the light elements. Our analysis incorporates several important refinements with respect to earlier work. We adopt up-to-date primordial abundance measurements, including the new $^4$He determination and the latest nuclear reaction rates. The hadronic and electromagnetic injection spectra are computed using \textsc{Pythia\,8}, providing a proper treatment of final-state radiation and hadronisation. We further implement an improved treatment of $p\leftrightarrow n$ interconversions, accounting for dynamical equilibrium, kaon-induced processes, and updated rates. Additionally, we make use of a refined hadrodisintegration formalism which allows us to also consider disintegration processes while BBN is still active. Together, these improvements yield updated exclusion contours on lifetime, mass, and abundance of the relic for each decay channel considered. Furthermore, we discuss the irreducible freeze-in contribution from inverse decays.
}

\begin{document}

\vspace*{0.1 cm}
\placetextbox{0.85}{0.97}{\small ULB-TH/26-21, DESY-26-066}

\maketitle

\flushbottom

\tableofcontents

\section{Introduction}

Big Bang Nucleosynthesis (BBN) represents one of the earliest and most precisely
testable predictions of the standard cosmological model. The successful explanation
of the primordial abundances of light elements as relics of nuclear reactions occurring in the first minutes of cosmic history, provides a stringent probe of both the Standard Model~(SM) of particle physics and the thermal history of the early
Universe~\cite{Wagoner:1966pv, Schramm:1977ne, Olive:1999ij, Iocco:2008va,Pitrou:2018cgg}. Any new physics active during or after the BBN epoch, corresponding
to temperatures below a few MeV, is therefore subject to tight observational constraints~\cite{Kawasaki:2018, Jedamzik:2006}.

A particularly well-motivated class of beyond the SM (BSM) scenarios features
massive, long-lived relics $\phi$ that are produced in the early Universe
and decay to SM particles on timescales comparable to or longer than the BBN epoch \cite{Ellis:1983ew,Weinberg:1982zq,Ellis:1990nb,Cyburt:2002uv,Boyarsky:2009ix,Gross:2018zha}. Even for small abundances of these relics at the time around and after BBN, their decay products can profoundly alter
the light-element abundances through several distinct mechanisms. At early times, the presence of the relic modifies the Hubble rate, thereby shifting the neutron-to-proton ratio
$n/p$ and thus the $^4$He yield. The decays after neutrino decoupling generally also change the effective number of neutrinos $N_\text{eff}$ as well as the time-temperature relation. At the same time, injected hadrons can additionally interconvert protons and neutrons via scatterings and annihilations, further changing the neutron-to-proton ratio. 
Once baryons, and in particular neutrons, are no longer immediately thermalised after their injection, their kinetic energy is sufficient to hadrodisintegrate light nuclei, especially $^4$He, while at even later epochs energetic photons and $e^\pm$ pairs
initiate electromagnetic cascades that photodisintegrate previously synthesised
nuclei. The interplay of these processes leads to a modification of predicted abundances that can be confronted with observations to place bounds on the relic's mass $m_\phi$, lifetime $\tau_\phi$, and primordial abundance $\mathpazoletters{Y}_\phi=n_\phi/s_\text{SM}$ with $n_\phi$ the relic's number density and $s_\text{SM}$ the SM entropy density. We are generally agnostic about the production mechanism of the relic $\phi$ and treat $\mathpazoletters{Y}_\phi$ and $\tau_\phi$ as independent in our study. We note however that not all combinations are physically realisable due to an irreducible freeze-in production via inverse decays, and we comment on the impact of this correlation on our results. Furthermore, we are able to constrain relic abundances that, if stable, would be significantly below the DM abundance.\footnote{Also gravitational particle production could be probed in our formalism \cite{Kolb:2023ydq}.}

Constraints on decaying relics from BBN have a long
history~\cite{Reno:1988, Dimopoulos:1987fz, Kawasaki:1994af, Jedamzik:1999di,
Kawasaki:2005, Jedamzik:2006}, with comprehensive analyses of both hadronic and electromagnetic energy injection spanning a wide range of masses and lifetimes~\cite{Kawasaki:2018, Forestell:2018txr}.
Note that while we concentrate on heavy relics in this work, a complementary line of investigation has focused on light, MeV-scale relics, for which the assumption that the decaying particle is non-relativistic at the time of its decay no longer holds fully and also the often utilised `universal photon spectrum' does not apply. In this regime it is therefore mandatory to evaluate the actual phase space distributions of the various species. Recent studies have made important progress in this direction, deriving refined BBN constraints on MeV-scale dark-sector states and mediators~\cite{Nollett:2013pwa, Poulin:2015opa,Hufnagel:2017dgo,Hufnagel:2018bjp,Depta:2019lbe, Sabti:2020yrt, Depta:2020wmr,Kawasaki:2020qxm, Depta:2020zbh, Giovanetti:2021izc}. The present work is complementary to these efforts: we focus on the regime of \emph{heavy} relics, $m_\phi \gtrsim \mathcal{O}(\mathrm{few}\;\mathrm{GeV})$, which are thoroughly non-relativistic at decay, and for which the dominant non-trivial effects are hadronic and electromagnetic energy injection at intermediate and late times.

In this paper, we present updated and improved BBN bounds on heavy decaying relics across a comprehensive set of SM two-body decay channels, except for neutrinos, whose treatment requires additional care and is left for future work.
Our analysis builds on and substantially refines the approach of refs.~\cite{Kawasaki:2005, Kawasaki:2018} through several important improvements. First, we adopt up-to-date abundances of the light elements including the updated measurement of the primordial $^4$He abundance from ref.~\cite{Yeh:2026pil}, which significantly tightens constraints from the helium yield. Second, we incorporate updated nuclear reaction rates relevant for standard BBN. Third, we implement a refined treatment of $p\leftrightarrow n$ interconversions, including a dynamical equilibrium approach, the contribution of kaon-induced interconversions, and updated rate computations. Fourth, we perform a proper treatment of final-state radiation~(FSR) and hadronisation using \textsc{Pythia\,8}~\cite{Sjostrand:2014zea}, providing a self-consistent and accurate modelling of the hadronic and electromagnetic injection spectra.

To calculate the light-element abundances we adapted \texttt{AlterBBN} \cite{Arbey:2011nf,Arbey:2018zfh} into a substantially extended version which we denote as \texttt{AlterAlterBBN}, a comprehensive code that can handle arbitrary background cosmologies \cite{Hufnagel:2017dgo,Hufnagel:2018bjp}, and takes into account the
injection of hadronic material, accounting for both $p\leftrightarrow n$
interconversions and hadrodisintegration during the BBN epoch. At late times, hadrodisintegration and photodisintegration are instead computed
with \texttt{ACROPOLIS}~\cite{Depta:2020mhj}. We additionally correct an error in the electromagnetic energy-loss curve, first identified in \cite{Bianco:2025boy}, which had led to artificially strong
hadrodisintegration rates by protons in earlier analyses. Finally, we refine
the treatment of elastic nucleon scattering in the hadronic
cascade~\cite{Cugnon:1996kh, Falter:2004uc, Larionov:2025wce}.

This paper is organised as follows: In section~\ref{sec:abundances}, we introduce the observational data available for BBN with emphasis on the latest results. We then introduce the three main sources of constraints in section~\ref{sec:injections}, namely neutron-proton interconversions \ref{sec:interconversion}, hadrodisintegration \ref{sec:hadrodis}, and photodisintegration \ref{sec:photodis}. In particular, we present a detailed account of the highly important dynamical equilibrium approximation when discussing the interconversions. In section~\ref{sec:pythia}, we discuss the use of $\texttt{PYTHIA}$ to obtain the composition of the injection after FSR and hadronisation. Combining all these ingredients, we show the resulting constraints for a plethora of decay channels in section~\ref{sec:results}, including a comparison to previous literature before wrapping up with our conclusions in section~\ref{sec:conclusion}. We provide several appendices with detailed explanations on technical details. Throughout this article, we work in natural units $\hbar=c=k_\text{B}=1$.

\section{The light-element abundances}\label{sec:abundances}

To describe the observational and theoretical values of the light-element
abundances, we use the conventional notation $N/N' \equiv Y_N/Y_{N'}$ with
$N,\,N' \in \{{}^1\text{H},\,\text{D},\,^3\text{He},\,^7\text{Li}\}$,
$Y_N \equiv n_N/n_b$, and $n_b$ the baryon number density. For $^4\text{He}$,
we follow the common notation $\mathcal{Y}_p \equiv 4Y_{^4\text{He}}$.

The primordial light-element abundances are inferred from several independent
astrophysical observations. The helium-4 mass fraction $\mathcal{Y}_p$ is
determined from emission lines of metal-poor H\,\textsc{ii}
regions~\cite{Aver:2015iza, Hsyu:2020uqb, Yeh:2026pil}, while the deuterium
abundance is measured from quasar absorption
systems~\cite{Cooke:2017cwo, Zavarygin:2017cov}. The helium-3 abundance is
constrained by measurements in the local interstellar
medium~\cite{Geiss2003}, and lithium-7 is inferred from the Spite plateau of
metal-poor halo stars~\cite{Cyburt:2008kw}. Combined with the
baryon density determined from the
CMB~\cite{Planck:2018vyg}, these observations place standard BBN in an
increasingly well-constrained setting. Throughout this work, we adopt the following
observed values, taking the latest recommended values for D$/^1$H and
$^3$He/D from ref.~\cite{ParticleDataGroup:2024cfk} and
ref.~\cite{Geiss2003}, respectively:
\begin{align}
    \text{D}/^1\text{H} &=(25.08 \pm 0.029)\times 10^{-5}\eqsp,
    \label{eq:DH_obs} \\
    ^3\text{He}/\text{D} &= (8.3 \pm 1.5) \times 10^{-1}\eqsp,
\end{align}
where the latter is used only as an upper limit.\footnote{The rationale behind argument is that it is expected that $^3$H/D only decreases after BBN has concluded \cite{Geiss2003}.} Note that we have used the D/$^1$H ratio suggested in the \emph{updated} 2025 version of \cite{ParticleDataGroup:2024cfk}. For $^4$He we use the recent
determination of ref.~\cite{Yeh:2026pil},
\begin{align}
    \mathcal{Y}_p &= (2.458 \pm 0.013) \times 10^{-1}\eqsp,
\end{align}
whose improved precision with respect to earlier measurements has a notable
impact on our constraints, as we discuss in Sec.~\ref{sec:results}. Finally,
the primordial lithium abundance inferred from the Spite plateau
reads~\cite{ParticleDataGroup:2024cfk}
\begin{align}
    ^7\text{Li}/^1\text{H} = (1.45 \pm 0.25) \times 10^{-10}\eqsp.
\end{align}
This value is known to be in tension with the standard BBN prediction -- the
so-called cosmological lithium problem -- for which an astrophysical resolution
currently appears most likely~\cite{Korn:2024gel, Gao:2020}. Since we do not
address this discrepancy in the present work, we conservatively exclude the
lithium abundance from our analysis pipeline.

On the theoretical side, we need a tool to track thermal BBN (potentially with a modified background cosmology) to compare to the observations. To this end, we have modified  \texttt{AlterBBN} \cite{Arbey:2011nf,Arbey:2018zfh} into a version we denote as \texttt{AlterAlterBBN} in the following.
This code can handle arbitrary background cosmologies \cite{Hufnagel:2017dgo,Hufnagel:2018bjp} and further incorporates the most recent set of nuclear reaction rates \cite{Gariazzo:2021iiu,Pisanti:2020efz,Tisma:2019acf,2014ApJ...785...96T,Mossa:2020gjc}, as well as the injection of hadronic material which we have added for this work. As further input for BBN, we take the value of the baryon-to-photon ratio $\eta$ obtained by the latest Planck measurements \cite{Planck:2018vyg}. This determination of $\eta$ has a strong correlation with $\Neff$, which we incorporate by following the prescription presented in \cite{Depta:2020wmr}. 

\section{Injections of SM particles during BBN}\label{sec:injections}

The injection of high-energy SM particles into the primordial plasma at temperatures below a few MeV is well known to significantly affect cosmological observables, in particular the abundances of light elements produced during BBN, through a variety of mechanisms.

In this work, we study in detail the impact of injecting different SM particles, excluding neutrinos, whose treatment requires special care. To accurately assess these effects, it is necessary to account for rapid decays as well as the hadronisation of coloured particles. These coloured particles do not have to originate from the primary decay but could also be induced by final-state radiation. As hadronisation proceeds on very short timescales, the cosmologically relevant species at the time of BBN are nucleons (protons and neutrons), mesons (in particular pions and kaons), photons, and electrons.

These particles can affect the formation of light elements via a number of different effects. The direct effects can be categorised as follows, ordered by the earliest times when they become relevant:
\begin{enumerate}
    \item \textbf{Interconversions} For $0.1\, \text{s} \lesssim t_\text{inj} \lesssim 10^2\,\text{s}$, the most stringent constraints on particle injections typically arise from changes to the neutron-to-proton ratio ($n/p$), which can be due to changes in the Hubble rate as well as hadronic interconversion reactions, where the latter typically dominate if present.
    Precisely tracking $n/p$ is essential whenever processes beyond the standard weak rates affect this ratio prior to BBN.
    \item \textbf{Hadrodisintegration} For $t_\text{inj} \gtrsim 10^2\,\text{s}$, injected nucleons are no longer thermalised immediately compared to scatterings and may disintegrate the previously formed light nuclei via hadronic interactions. 
    \item \textbf{Photodisintegration} For $t_\text{inj} \gtrsim 10^4\,\text{s}$, the universal spectrum formed in electromagnetic cascades contains sufficiently energetic photons to disintegrate the light elements produced during BBN.
\end{enumerate}
In addition, the injections generally lead to changes in the \textbf{background cosmology}, which need to be taken into account.
In the following, we will discuss these effects and their implementation in detail, pointing out improvements made compared to what has been done in the literature.

\subsection{Interconversions} \label{sec:interconversion}

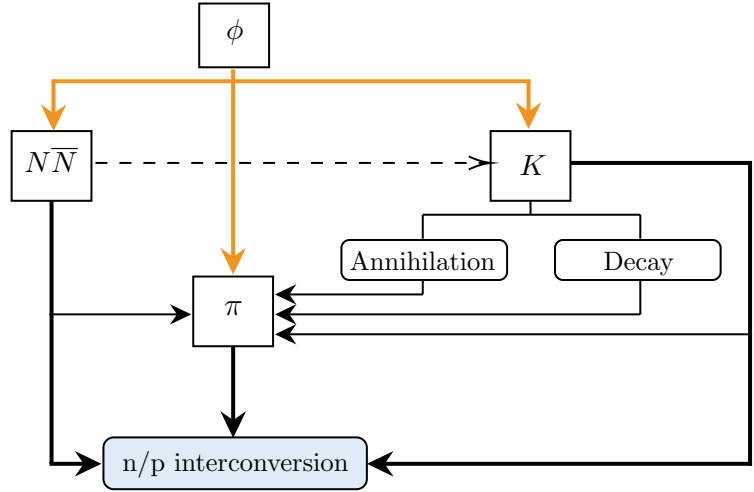
\begin{figure}

\tikzset{every picture/.style={line width=0.75pt}} 
\centering
\begin{tikzpicture}[x=0.75pt,y=0.75pt,yscale=-1,xscale=1]

\draw   (251,18) -- (286,18) -- (286,50) -- (251,50) -- cycle ;
\draw   (157,82.02) -- (197,82.02) -- (197,117) -- (157,117) -- cycle ;
\draw [color={rgb, 255:red, 239; green, 148; blue, 30 }  ,draw opacity=1 ][line width=1.5]    (177,57) -- (417,57) ;
\draw   (397,82.02) -- (437,82.02) -- (437,117) -- (397,117) -- cycle ;
\draw [color={rgb, 255:red, 239; green, 148; blue, 30 }  ,draw opacity=1 ][line width=1.5]    (178,57) -- (178,78) ;
\draw [shift={(178,82)}, rotate = 270] [fill={rgb, 255:red, 239; green, 148; blue, 30 }  ,fill opacity=1 ][line width=0.08]  [draw opacity=0] (13.4,-6.43) -- (0,0) -- (13.4,6.44) -- (8.9,0) -- cycle    ;
\draw [color={rgb, 255:red, 239; green, 148; blue, 30 }  ,draw opacity=1 ][line width=1.5]    (416,57) -- (416,77) ;
\draw [shift={(416,81)}, rotate = 270] [fill={rgb, 255:red, 239; green, 148; blue, 30 }  ,fill opacity=1 ][line width=0.08]  [draw opacity=0] (13.4,-6.43) -- (0,0) -- (13.4,6.44) -- (8.9,0) -- cycle    ;
\draw [color={rgb, 255:red, 239; green, 148; blue, 30 }  ,draw opacity=1 ][line width=1.5]    (268,51) -- (268,150) ;
\draw [shift={(268,154)}, rotate = 270] [fill={rgb, 255:red, 239; green, 148; blue, 30 }  ,fill opacity=1 ][line width=0.08]  [draw opacity=0] (13.4,-6.43) -- (0,0) -- (13.4,6.44) -- (8.9,0) -- cycle    ;
\draw   (248,155.02) -- (288,155.02) -- (288,190) -- (248,190) -- cycle ;
\draw  [dash pattern={on 4.5pt off 4.5pt}]  (199,98) -- (395,98) ;
\draw [shift={(397,98)}, rotate = 180] [color={rgb, 255:red, 0; green, 0; blue, 0 }  ][line width=0.75]    (10.93,-3.29) .. controls (6.95,-1.4) and (3.31,-0.3) .. (0,0) .. controls (3.31,0.3) and (6.95,1.4) .. (10.93,3.29)   ;
\draw [line width=0.75]    (363,124) -- (472,124) ;
\draw [line width=0.75]    (363,124) -- (363,136) ;
\draw [line width=0.75]    (472,124) -- (472,136) ;
\draw [line width=0.75]    (417,117) -- (417,124) ;
\draw [line width=0.75]    (363,157) -- (363,164) ;
\draw [line width=0.75]    (363,164) -- (292,164) ;
\draw [shift={(289,164)}, rotate = 360] [fill={rgb, 255:red, 0; green, 0; blue, 0 }  ][line width=0.08]  [draw opacity=0] (10.72,-5.15) -- (0,0) -- (10.72,5.15) -- (7.12,0) -- cycle    ;
\draw [line width=0.75]    (472,157) -- (472,174) ;
\draw [line width=0.75]    (472,174) -- (292,174) ;
\draw [shift={(289,174)}, rotate = 360] [fill={rgb, 255:red, 0; green, 0; blue, 0 }  ][line width=0.08]  [draw opacity=0] (10.72,-5.15) -- (0,0) -- (10.72,5.15) -- (7.12,0) -- cycle    ;
\draw [line width=1.5]    (177,117) -- (177,249) ;
\draw [line width=1.5]    (527,97) -- (527,250) ;
\draw [line width=1.5]    (437,98) -- (527,98) ;
\draw [line width=1.5]    (268,190) -- (268,232) ;
\draw [shift={(268,236)}, rotate = 270] [fill={rgb, 255:red, 0; green, 0; blue, 0 }  ][line width=0.08]  [draw opacity=0] (13.4,-6.43) -- (0,0) -- (13.4,6.44) -- (8.9,0) -- cycle    ;
\draw [line width=1.5]    (528,249) -- (350,249) -- (339,249) ;
\draw [shift={(335,249)}, rotate = 360] [fill={rgb, 255:red, 0; green, 0; blue, 0 }  ][line width=0.08]  [draw opacity=0] (13.4,-6.43) -- (0,0) -- (13.4,6.44) -- (8.9,0) -- cycle    ;
\draw [line width=0.75]    (244,174) -- (176,174) ;
\draw [shift={(247,174)}, rotate = 180] [fill={rgb, 255:red, 0; green, 0; blue, 0 }  ][line width=0.08]  [draw opacity=0] (10.72,-5.15) -- (0,0) -- (10.72,5.15) -- (7.12,0) -- cycle    ;
\draw [line width=1.5]    (198,249) -- (190,249) -- (176,249) ;
\draw [shift={(202,249)}, rotate = 180] [fill={rgb, 255:red, 0; green, 0; blue, 0 }  ][line width=0.08]  [draw opacity=0] (13.4,-6.43) -- (0,0) -- (13.4,6.44) -- (8.9,0) -- cycle    ;
\draw  [fill={rgb, 255:red, 221; green, 236; blue, 246 }  ,fill opacity=1 ] (203,240.95) .. controls (203,238.08) and (205.33,235.75) .. (208.2,235.75) -- (329.8,235.75) .. controls (332.67,235.75) and (335,238.08) .. (335,240.95) -- (335,256.55) .. controls (335,259.42) and (332.67,261.75) .. (329.8,261.75) -- (208.2,261.75) .. controls (205.33,261.75) and (203,259.42) .. (203,256.55) -- cycle ;
\draw   (322.25,140.68) .. controls (322.25,138.47) and (324.05,136.67) .. (326.27,136.67) -- (401.23,136.67) .. controls (403.45,136.67) and (405.25,138.47) .. (405.25,140.68) -- (405.25,152.73) .. controls (405.25,154.95) and (403.45,156.75) .. (401.23,156.75) -- (326.27,156.75) .. controls (324.05,156.75) and (322.25,154.95) .. (322.25,152.73) -- cycle ;
\draw   (429.25,140.68) .. controls (429.25,138.47) and (431.05,136.67) .. (433.27,136.67) -- (508.23,136.67) .. controls (510.45,136.67) and (512.25,138.47) .. (512.25,140.68) -- (512.25,152.73) .. controls (512.25,154.95) and (510.45,156.75) .. (508.23,156.75) -- (433.27,156.75) .. controls (431.05,156.75) and (429.25,154.95) .. (429.25,152.73) -- cycle ;
\draw [line width=0.75]    (528,184) -- (292,184) ;
\draw [shift={(289,184)}, rotate = 360] [fill={rgb, 255:red, 0; green, 0; blue, 0 }  ][line width=0.08]  [draw opacity=0] (10.72,-5.15) -- (0,0) -- (10.72,5.15) -- (7.12,0) -- cycle    ;
\draw (263,24.4) node [anchor=north west][inner sep=0.75pt]    {$\phi $};
\draw (162,88.4) node [anchor=north west][inner sep=0.75pt]    {$N\overline{N}$};
\draw (410,92.4) node [anchor=north west][inner sep=0.75pt]    {$K$};
\draw (262,166) node [anchor=north west][inner sep=0.75pt]  [font=\large]  {$\pi $};
\draw (325,141.02) node [anchor=north west][inner sep=0.75pt]  [font=\small] [align=left] {Annihilation};
\draw (452,141.02) node [anchor=north west][inner sep=0.75pt]  [font=\small] [align=left] {Decay};
\draw (211,241.02) node [anchor=north west][inner sep=0.75pt]  [font=\small] [align=left] {n/p interconversion};

\end{tikzpicture}
  \caption{Flowchart of the hadronic interconversion due to the relic decay, including secondary pion production from heavier hadrons.}
  \label{fig:flowchart}
\end{figure}

Mesons and baryons can interconvert neutrons 
and protons via charge-exchange and absorption reactions \cite{Reno:1988,Pospelov:2010cw,Kawasaki:2005,Kawasaki:2018,Boyarsky:2020dzc,Jung:2025dyo,Omar:2025jue}. Charged pions are one prominent example, with reactions such as $\pi^- p \;\to\; n  X$ or $\pi^+ n \;\to\; p  X$,
but kaons, antinucleons, and other hadrons contribute analogous processes.
The cross sections for the two pion reactions above are very similar, so which direction of the conversion dominates strongly depends on the available target nuclei and hence the pre-existing ratio of $n/p$. Once the primordial plasma approaches the temperatures relevant to BBN, it is strongly proton-dominated within the SM. The $\pi^- + p \to n$ channel is therefore generally more efficient at these temperatures due to more target nuclei, so hadronic injections generically \emph{increase} $n/p$ above its SM value.\footnote{Note that this assumes \emph{symmetric} hadron injections. While this is true for primary injections, secondary pion injections generally have a small asymmetry. This, however, is negligible and does not impact this conclusion.}

Interconversion reactions impact $n/p$ for all the temperatures at which exotic decays occur, potentially well above the BBN epoch. However, what directly controls light-element abundances is the $n/p$ ratio at $T \lesssim 100\;\text{keV}$, when nucleosynthesis occurs. Any increase in $n/p$ 
at that point translates almost linearly into an enhanced ${}^4\text{He}$ mass fraction $\mathcal{Y}_p$, since virtually every neutron present is eventually incorporated into helium.

\subsubsection{Relevant timescales}
To see which particles are relevant, it is instructive to estimate the timescales for different processes, in particular the scattering, decay and thermalisation rates. Interconversions are only possible for sufficiently long-lived particles where a significant amount of hadronic interactions occur before decay. 
Assuming a typical hadron-nucleon (interconversion) cross section of $\sim 10\,$mb \cite{Reno:1988,Jung:2025dyo} and taking the value of the baryon-to-photon ratio~$\eta\sim 6\cdot 10^{-10}$ at the time of CMB, we can estimate the typical scattering rate for a given hadron~as
\begin{align}
    \Gamma_\text{had} = \langle \sigma v \rangle_{n/p}^\text{had}\, n_{n/p}\sim \langle \sigma v \rangle_{n/p}^\text{had}\, \eta n_\gamma \sim 6\cdot 10^6\,\text{s}^{-1}\left(\frac{T}{\text{MeV}}\right)^3\left(\frac{\sigma_{n,p}^\text{had}}{10\,\text{mb}}\right)\;.
\end{align}
The only hadrons that do not decay much more rapidly than this scattering timescale are from the set $[ \pi^\pm,K^\pm, K_L, n, \bar{n}, p, \bar{p}]$ with\footnote{In most decay channels, the limit on the proton lifetime is even longer.}~\cite{ParticleDataGroup:2024cfk}
\begin{align}
    \tau_{\pi^\pm}&=(2.6033\pm0.0005)\times 10^{-8}\,\text{s}\\
    \tau_{K^\pm}&=(1.2380\pm0.0020)\times 10^{-8}\,\text{s}\\
    \tau_{K_L}&=(5.116\pm0.021)\times 10^{-8}\,\text{s}\\
    \tau_{n/\bar{n}}&=(878.4\pm0.5)\,\text{s}\\
    \tau_{p/\bar{p}}&>9\cdot10^{29}\,\text{yr}\;.
\end{align}
As interconversions are only relevant early on, neutrons (and of course protons) can be considered stable in this discussion, whereas the charged pions and kaons, as well as the long-lived neutral kaon, may dominantly scatter or decay depending on the temperature considered. All other mesons and baryons decay instantaneously w.r.t.~the relevant timescales and therefore are not considered here.\footnote{In fact, our procedure lets these short-lived particles decay immediately as discussed in section~\ref{sec:pythia}.}

Due to the electromagnetic interactions
the injected charged hadrons are rapidly scattered down to thermal energies.\footnote{This simplifies the analysis as otherwise we would have to account for the (energy-dependent) time dilation of relativistic unstable particles as well as the energy dependence of the scattering cross sections.} 
More specifically, taking the expression for the Coulomb scattering energy loss rate in the case of $T\lesssim m_e$ for relativistic particles \cite{Kawasaki:2005,Bianco:2025boy}, we can estimate the thermalisation timescale to be $\sim 10^{-12}\,\text{s}$ even for temperatures as low as 0.1~MeV.

\begin{figure}
    \centering
    \includegraphics{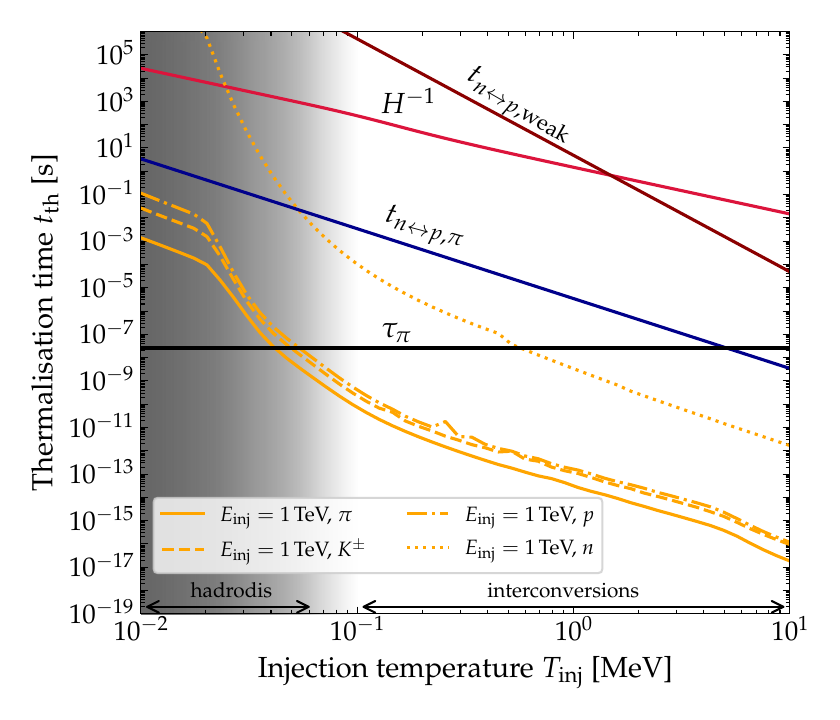}
    \caption{Display of relevant timescales for hadrons during the temperature frame of interest. Apart from the thermalisation time, we also show the decay time $\tau_\pi$ (black), the approximate pion-driven interconversion rates (dark blue), the Hubble rate (red), and the weak interconversion rate (dark red).}
    \label{fig:therm_check}
\end{figure}

We have made this statement more quantitative for different injections in figure~\ref{fig:therm_check}, where we have checked the time it takes for a particle to scatter down to energies similar to the temperature for different initial energies.\footnote{To be specific, we define
\begin{equation}
    t_\text{th}(E_\text{inj})=\int_{E_\text{inj}}^{3T}\text{d}E\frac{\text{d}t}{\text{d}E}\;.
\end{equation}}
We compare these times to other typical timescales at play. For simplicity, we only focus on pionic rates, in particular the interconversion rates of thermal pions in dark blue, assuming (conservatively) that pions scatter on all baryons instead of just protons \emph{or} neutrons and taking a cross section of $\langle \sigma v\rangle =1\,$mb.\footnote{Note that pion interconversion cross sections are typically 1-2 orders of magnitude below the other hadronic cross sections we will encounter later.} As we take the pion cross section to be temperature independent, the interconversion rate just inherits the approximate $T^3-$scaling of the baryon abundance. In reality, we will use temperature-dependent thermal cross sections and, in particular, the $\pi^--p$ rate will be enhanced by Coulomb interactions.

We also show the pion decay rate at rest in black. The grey gradient shows where interconversions lose their importance w.r.t.~hadrodisintegration (cf.~\ref{sec:hadrodis}). The exact transition depends on the mass, lifetime, and abundance of the relic as well as on the type of injection. 
In practice, we have found that ignoring interconversions below temperatures of $100\,$keV is a good approximation and balances the run time of the code and the accuracy of the light-element determination.

Additionally, since we typically consider relic masses which allow for nucleons in the final state, we ignore hadrodisintegration via mesons, as they typically decay before interacting, which gives a huge suppression compared to injected nucleons. For a more quantitative statement, we refer to the note at the end of section~\ref{sec:hadrodis}.

Finally, we have also plotted the Hubble timescale ($\sim T^2$) and the timescale of weak interconversions ($\sim T^5$) for comparison. It is fairly clear that all processes discussed before are happening essentially instantaneously w.r.t.~the cosmological background evolution. 
In conclusion, we see that thermalisation (or decay) is always happening before interconversions for all charged particles and even for neutrons in the temperature window of BBN.

These considerations do not hold for the long-lived component of the neutral kaon, which is known to be stopped less efficiently \cite{Reno:1988,Jung:2025dyo}.\footnote{Note that the neutron is still stopped efficiently due to magnetic moments interactions for temperatures as low as $100\,$keV \cite{Kawasaki:2005, Bianco:2025boy}. This is earlier than what our figure naively seems to imply; the reason is that disintegration processes will become more relevant than thermalisation well before thermalisation becomes less efficient than the Hubble rate.
} 
However, since we find this particle to be a subdominant component of the overall injection, we resort to the conservative approximation that the neutral kaons also thermalise rapidly (see also refs.~\cite{Pospelov:2010cw,Boyarsky:2020dzc,Akita:2024ork}).
To see that this is conservative, note that as discussed above, the interconversions are preferentially converting protons to neutrons for similar cross sections due to the larger proton target density at the time when $n/p$ actually affects the light-element abundances and for regions in parameter space that are close to the allowed region.
In addition, a fully relativistic treatment would lead to two partially compensating effects. On the one hand, the relativistic time dilation will effectively increase the lifetime of the kaon in the cosmic frame. This effective decrease in decay rate makes all scattering rates more relevant in turn, as long as only a small fraction of kaons scatters before decay. On the other hand, the scattering cross section decreases with energy~\cite{Jung:2025dyo}, which has the opposite effect. Overall, the time dilation will be the dominant effect, such that we somewhat \textit{underestimate} the effect of neutral kaons by treating them as thermalised, making this approximation conservative.

An essential ingredient in the calculation of the interconversion rates are the actual abundances of the participating hadrons. As we will discuss below, we will start out with a certain injection rate of the different hadrons, but we need to take into account scatterings, annihilations and decays to obtain the actual abundances.
As we have seen above, many of the relevant timescales which enter this calculation are rather short and certainly significantly shorter than the fusion rates during BBN. It turns out that tracking the hadrons and BBN in the same code without approximations requires very short time steps compared to the duration of BBN, which in turn leads to prohibitively long run times. 
We therefore briefly investigate an important concept that facilitates the treatment of the interconversions: the approximation of dynamical equilibrium. Let us begin with a generic Boltzmann equation for the number density of a species in an FLRW Universe
\begin{align}
    \frac{\text{d}n}{\text{d}t}+3Hn&= S - \Gamma n\;,
\end{align}
where $S$ is an arbitrary source term and $\Gamma$ is a (fast) rate which depletes the number density.\footnote{See also figure~\ref{fig:therm_check} where it is clear that the interconversion rates are always much faster than the Hubble rate.}  If all interactions happen much faster than the Hubble rate, these processes are driven to equilibrium instantaneously w.r.t.~the cosmological evolution. This equilibrium only changes with the background expansion and the change in the source terms, which are slow compared to the interactions. In our case, this line of reasoning is possible since the relic and the background evolution of the Universe are essentially unaffected by the presence of the additional hadrons, while the hadrons themselves react rapidly due to the strong nuclear reactions. Thus, we can solve the time dependence of the background quantities separately and feed them into the source term, which carries all the time dependence for the hadronic number densities. We can therefore set 
\begin{align}
    \frac{\text{d}n}{\text{d}t}+3Hn\simeq 0\\
    \Rightarrow n^{\text{de}}\simeq \frac{S}{\Gamma}\;,
\end{align}
where we have introduced the superscript $^{\text{de}}$ to denote a quantity in dynamical equilibrium.
We can make intuitive sense of this expression if $S$ is described by a decay, $S=\Gamma_\text{dec} n_\text{source}$. In this case, the particle of interest tracks the evolution of the decaying particle rescaled by the fraction of injection rate over the depletion rate, $\Gamma_\text{dec}/\Gamma$. 

Having established the properties of the injected particles as well as how to infer their abundance via dynamical equilibrium, we will next discuss the individual injections in more detail. Since all particles share the temperature of the plasma, we know that the available kinetic energy is insufficient to produce heavier hadrons from lighter ones, i.e.\ in all processes involving only pions in the initial state, neither kaons nor baryons can be produced. Conversely, for nucleons and kaons, we will find that decays, annihilations, and nuclear scattering can produce secondary pions, see also figure~\ref{fig:flowchart} for a schematic representation.\footnote{We neglect the tiny but non-zero branching ratio into kaonic final states in nucleonic annihilations.} With this in mind, we will first focus our discussion on the heavier particles, starting with (anti-)nucleons.

\subsubsection{Nucleon-antinucleon pairs}

Let us start this section by pointing out that, depending on the injection time, nucleons can lead to either hadrodisintegration or baryonic interconversions. 
At early times, the nucleons quickly scatter down to thermal energies and subsequently participate in interconversion processes. At later times, if their energy before the first scattering is still above the $^4$He binding energy, they will instead mainly contribute to disintegration. As we will see later, this transition occurs roughly at $T=0.1\,$MeV.

As baryon number is conserved in the decays, we find that baryons and anti-baryons are always injected in pairs.
As the anti-nucleons will eventually annihilate on the background neutrons and protons, we have four different constellations:
\begin{enumerate}
    \item $\bar{p}p+p_\text{BG}\to p + X$ (no interconversion)
    \item $\bar{p}p+n_\text{BG}\to p + X$ (interconversion)
    \item $\bar{n}n+n_\text{BG}\to n + X$ (no interconversion)
    \item $\bar{n}n+p_\text{BG}\to n + X$ (interconversion)\;.
\end{enumerate}
In other words, a background nucleon is annihilated away and replaced by the partner of the annihilated antinucleon, which, depending on the process, leads to an effective interconversion. As an example, let us discuss the case of an injected proton-antiproton pair. If the antiproton annihilates on a (background) proton, the neutron-to-proton ratio is not affected. However, if it annihilates on a neutron instead, we effectively replace a neutron by a proton, and thus, we have an effective interconversion process.

In the following, we will discuss step by step how to arrive at the required interconversion rates. More details on what we have implemented in our numerical scheme can be found in appendix~\ref{app:new_xsec}.

The contribution of the injected nucleon-antinucleon pairs to the total conversion rate of the background neutron and protons can be written as 
\begin{align}\label{eq:nuc_inter}
    \frac{\text{d}n_{p}}{\text{d}t}\Bigg|_{n\bar{n},p\bar{p}}= -\frac{\text{d}n_{n}}{\text{d}t}\Bigg|_{n\bar{n},p\bar{p}}=\bigg(\langle \sigma v  \rangle^{\bar{p}}_{n\to p}n_{\bar{p}}  n_n - \langle \sigma v \rangle^{\bar{n}}_{p\to n} n_{\bar{n}} n_p\bigg)\;,
\end{align}
where the superscript of the cross section denotes the hadron going into the reaction and the subscript indicates the induced process (in this case corresponding to the processes 2.\ and 4.\ listed above). 
To solve this equation, the anti-baryon number densities are required.

The Boltzmann equation for the anti-baryon number densities, in turn, can be written as
\begin{align}
    \frac{\text{d}n_{\bar{n}}}{\text{d}t}+3Hn_{\bar{n}}&=S_{\bar{n}}-\langle \sigma v\rangle^{\bar{n}}_{n\to n}n_{n}-\langle \sigma v\rangle^{\bar{n}}_{p\to n}n_p
\end{align}
There are analogous expressions for the anti-protons. Using the fact that dynamical equilibrium is a very good approximation, as described above, this implies for the number density
\begin{align}
   n^{\text{de}}_{\bar{n}}&=\frac{S_{\bar{n}}}{\langle \sigma v\rangle^{\bar{n}}_{n\to n}n_{n}+\langle \sigma v\rangle^{\bar{n}}_{p\to n}n_p}\;,\label{eq:dyn_eq_bar}
\end{align}
Note that actually solving this equation requires the knowledge of $n_n$ and $n_p$, so that we have a coupled system of equations. Given that the annihilation cross sections are very similar, a good first estimate can be obtained by setting the cross sections equal, in which case the expression only depends on the known baryon number density $n_b = n_n + n_p$. We show the cross sections in appendix~\ref{app:new_xsec} figure~\ref{fig:km_comp} (left bottom).
We will discuss how to fully solve the system of equations later.\\[2mm]

\noindent {\bf{Secondary pions:}}
These annihilation processes will, in general, also inject pions into the plasma.\footnote{We neglect the small non-zero branching ratio into kaons  \cite{CrystalBarrel:1993uxb}.} For proton-antiproton annihilations, we can find an approximate value of 1.55 $\pi^+\pi^-$ pairs per process in the literature \cite{Klempt:2005pp,Golubeva:2018mrz}. From the latter reference, we can also infer that $\bar{n}n$ annihilations are approximately identical. Furthermore, we can extend that symmetry to $\bar{p}n$ and $\bar{n}p$ annihilations, respectively (modulo charge conjugation). From table~II of the same paper, we can extract an approximate number of $2.18\pi^-$ ($2.18\pi^+$) and $1.18\pi^+$ ($1.18\pi^-$) for $\bar{p}n$ ($\bar{n}p$) annihilations.\footnote{We have neglected branching ratios of less than $0.01\%$ which should change the average number of pions per annihilation at most at the percent level.} From these numbers, we can already predict that typically, these secondary pions will be subleading to the main pion source coming from hadronisation, but for completeness, we will still add them in our full calculation. Adding all the different contributions, the source terms for pions originating from anti-baryon annihilations can be expressed as
\begin{align}
    S_{\pi^+,\text{ann}}^{\bar{n}+\bar{p}}&=1.55\langle \sigma v\rangle^{\bar{n}}_{n\to n}  n_{\bar{n}}n_n+1.55\langle \sigma v\rangle^{\bar{p}}_{p\to p} n_{\bar{p}}n_p+2.18\langle \sigma v\rangle^{\bar{n}}_{p\to n} n_{\bar{n}}n_p + 1.18\langle\sigma v\rangle^{\bar{p}}_{n\to p} n_{\bar{p}}n_n\\
    S_{\pi^-,\text{ann}}^{\bar{n}+\bar{p}}&=1.55\langle \sigma v\rangle^{\bar{n}}_{n\to n} n_{\bar{n}}n_n+1.55\langle \sigma v\rangle^{\bar{p}}_{p\to p} n_{\bar{p}}n_p+1.18 \langle \sigma v\rangle^{\bar{n}}_{p\to n} n_{\bar{n}}n_p + 2.18\langle \sigma v\rangle^{\bar{p}}_{n\to p} n_{\bar{p}}n_n\;.
\end{align}
We will come back to this (potentially asymmetric) source term of pions below.

\subsubsection{Kaons}

In this section, we will consider the effect of injected neutral and charged kaons. Let us start the discussion with the charged ones. 
\subsubsection*{$\mathbf{K^\pm}$}
While both types of charges participate equally in decays and annihilation, there is a major difference in the nuclear scattering: the positively charged kaons do \emph{not} interact with the nucleons (as long as they are stopped efficiently). Firstly, note that the ``direct'' conversions $K^\pm + n/p\to K^0/\Bar{K}^0+p/n$ have an energy threshold of $5.2\,$MeV, which forbids the process at early times where energy loss is efficient.\footnote{While we will start our simulation at sufficiently high temperatures such that the threshold could be overcome, we discard this possibility since we are only interested in the later stages of the $n/p$ evolution. Furthermore, \cite{Jung:2025dyo} demonstrated that the inverse case with $K_L$ in the initial state is allowed but typically has a smaller cross section than the processes including hyperons.} Secondly, the way to actually have thresholdless interconversions is via higher baryonic resonances, which then quickly decay to nucleons (and typically, accompanying pions). Since there are no baryons with positive strangeness, these processes are only available to negatively charged kaons.\footnote{Of course, anti-baryons can have positive strangeness, but production of those would require even higher threshold energies.} 
We will follow the discussion of \cite{Pospelov:2010cw} to accurately track the injection of kaons.

Let us now investigate the exact channels that are available. Kinematically, only $\Sigma^{\pm,0}$ and $\Lambda$ can be produced in the low-energy collisions we are considering: 
\begin{align}
K^-+p&\to \Sigma^\pm \pi^\mp,\, \Sigma^0 \pi^0,\, \Lambda\pi^0\\
K^-+n&\to \Sigma^- \pi^0,\, \Sigma^0 \pi^-,\, \Lambda\pi^-\;
\end{align}
with the corresponding decay channels\footnote{Compared to previous work, we have updated some of the branching ratios with their recent PDG values \cite{ParticleDataGroup:2024cfk}.}
\begin{alignat}{2}
    & \Sigma^+ &&\to \begin{cases} p\pi^0 & 51.47\% \\ n\pi^+ & 48.43\% \end{cases} \,, \\[6pt]
    & \Sigma^- &&\to\quad n\pi^- \quad 99.85\% \,, \\[6pt]
    & \Sigma^0 &&\to\quad \Lambda\gamma \quad 100\% \,, \\[6pt]
    & \Lambda  &&\to \begin{cases} p\pi^- & 64.1\% \\ n\pi^0 & 35.9\% \end{cases} \,. \label{eq:BR_kaon}
\end{alignat}
To arrive at the effective interconversion cross section induced by charged kaons, these processes need to be summed over, weighted with the corresponding branching ratios. We show the resulting effective cross sections in figure~\ref{fig:km_comp} (top left) in appendix~\ref{app:new_xsec}. The contribution to $n/p$ is then given as
\begin{align}\label{eq:K_interconv}
    \frac{\text{d}n_{p}}{\text{d}t}\Bigg|_{K^-}= -\frac{\text{d}n_{n}}{\text{d}t}\Bigg|_{K^-}=\bigg(\langle \sigma v  \rangle^{K^-}_{n\to p}n_{K^-}  n_n - \langle \sigma v \rangle^{K^-}_{p\to n} n_{K^-} n_p\bigg)\;.
\end{align}

\noindent {\bf{Secondary pions:}}
We note that the upscattering into hyperons and their subsequent decays always produce two pions. We should therefore take this into account as a source of secondary pions, weighted according to the cross sections and branching ratios. Charge conservation implies 
\begin{align}
    & K^- n\to n \; \pi^0 \pi^- \\
    & K^- n\to p \; \pi^- \pi^- \\
    & K^- p\to p \; \pi^0 \pi^- \\
    & K^- p\to n \; \pi^+ \pi^-, n \; \pi^0 \pi^0
\end{align}
We see that only for the last process there is any ambiguity. As discussed above, the $\pi^0$ is irrelevant for the interconversion processes, so we will concentrate on the charged pions. We will follow \cite{Akita:2024ork,Reno:1988} and weight the individual hyperon channels with the respective number of pions, calculating specific pion-yield cross sections. The corresponding contribution to the pion source term can be written as
\begin{align}
    S_{\pi^+,\text{nuc}}^{K^-}&=\langle \sigma v\rangle_{p\to \pi^+}^{K^-} n_p n_{K^-} 
    \\
    S_{\pi^-,\text{nuc}}^{K^-}&=\langle \sigma v\rangle_{p\to \pi^-}^{K^-} n_p n_{K^-}+\langle \sigma v\rangle_{n\to \pi^-}^{K^-} n_n n_{K^-}\;.
\end{align}
Note that there is only one contribution to the $\pi^+$ source term as $K^-+n$ can never produce $\pi^+$. The pion production cross sections can also be found in figure~\ref{fig:km_comp} (top right).
In contrast to the nucleons, kaons can also decay, which provides an additional source of pions. 
The relevant hadronic branching ratios are given by \cite{ParticleDataGroup:2024cfk}
\begin{align}
    K^{\pm}\to \begin{cases}
        \pi^\pm\pi^0(\pi^0)\quad &, \quad 22.43\%\\
        \pi^\pm \pi^+\pi^-\quad &, \quad 5.583\%\\
    \end{cases}\;,
\end{align}
which implies that we get additional pion source terms by calculating the average number of charged pions per decay process.
\begin{align}
    S_{\pi^+,\text{dec}}^{K^\pm}&=\Gamma_K\left[(0.2243+2\times 0.05583)n_{K^+}+0.05583n_{K^-} \right]\\
    S_{\pi^-,\text{dec}}^{K^\pm}&=\Gamma_K\left[(0.2243+2\times 0.05583)n_{K^-}+0.05583n_{K^+} \right]\;
\end{align}
with $\Gamma_K=1/\tau_K$. Especially at later times, when decays typically dominate the kaonic processes, this is the dominant source of secondary pions from kaons.

Finally, we should also consider kaon annihilations, which were studied in \cite{Akita:2024ork}. While subleading, we include this term here for completeness because the dynamical equilibrium equations can still be solved analytically. Using a 2:1 splitting for $\pi^+\pi^-$ over $\pi^0 \pi^0$ final states for low-energy annihilations \cite{Akita:2024ork} we obtain
\begin{align}
    S_{\pi^+,\text{ann}}^{K^\pm}&=S_{\pi^-,\text{ann}}^{K^\pm}=\frac{2}{3}\langle \sigma v\rangle^K_{\text{ann}} n_{K^+} n_{K^-}\quad \text{with}\quad
    \langle\sigma v\rangle^K_{\text{ann}}  \simeq 17.23\,\text{mb} \;.
\end{align}
Now that we know all important processes for kaons, let us also discuss their dynamical equilibrium abundance. Starting off from the Boltzmann equation, we find
\begin{align}
    \frac{\text{d}n_{K^+}}{\text{d}t}+3Hn_{K^+}&= S_K-\frac{n_{K^+}}{\tau_K}-\langle \sigma v\rangle^K_{\text{ann}} n_{K^+}n_{K^-}\\
    \frac{\text{d}n_{K^-}}{\text{d}t}+3Hn_{K^+}&= S_K-\frac{n_{K^+}}{\tau_K}-\langle \sigma v\rangle^K_{\text{ann}} n_{K^+}n_{K^-} - \langle \sigma v\rangle^{K^-}_{p\to n} n_{p}n_{K^-}- \langle \sigma v\rangle^{K^-}_{n\to p} n_{n}n_{K^-}\\
    \Rightarrow n^{\text{de}}_{K^-}&=\frac{S_K}{\Gamma_K+\Gamma_N+\langle\sigma v\rangle^K_{\text{ann}} n^{\text{de}}_{K^+}} \quad \quad\text{and} \quad \quad n^{\text{de}}_{K^+}=\frac{\Gamma_K+\Gamma_N}{\Gamma_K}n^{\text{de}}_{K^-}\label{eq:dyn_eq_k1}\\
    \Rightarrow n^{\text{de}}_{K^-} &=\frac{2S_K}{(\Gamma_K+\Gamma_N)\left[\sqrt{1+\frac{4\langle\sigma v\rangle^K_{\text{ann}}}{(\Gamma_K+\Gamma_N)\Gamma_K}S_K}+1\right]}\;,\label{eq:dyn_eq_k2}
\end{align}
where we have defined $\Gamma_N=\langle \sigma v\rangle^{K^-}_{p\to n} n_{p}+ \langle \sigma v\rangle^{K^-}_{n\to p} n_{n}$. We have written the final result in a way which makes it apparent that it will fall back to the simple expected result in case of $\langle\sigma v\rangle^K_{\text{ann}} \to 0$.

\subsubsection*{$\mathbf{K^0_L}$}

Similar to the case of charged kaons, we have the following interconversion channels available
\begin{align}
K_L+p&\to \Sigma^+ \pi^0,\, \Sigma^0 \pi^+,\, \Lambda\pi^+\\
K_L+n&\to \Sigma^\pm \pi^\mp,\, \Sigma^0 \pi^0,\, \Lambda\pi^0\;\;,
\end{align}
where the subsequent decays are given in eq.~\eqref{eq:BR_kaon}. We then just have to combine the cross sections with their nucleon yield to determine the effective interconversion cross sections.
In full analogy to eq.~\eqref{eq:K_interconv} we find
\begin{align}
    \frac{\text{d}n_{p}}{\text{d}t}\Bigg|_{K_L}= -\frac{\text{d}n_{n}}{\text{d}t}\Bigg|_{K_L}=\bigg(\langle \sigma v  \rangle^{K_L}_{n\to p}n_{K_L}  n_n - \langle \sigma v \rangle^{K_L}_{p\to n} n_{K_L} n_p\bigg)\;.
\end{align}

\noindent {\bf{Secondary pions:}}
To estimate the production of pions, we once more calculate an effective cross section as for the charged kaons, based on the pionic final states of the processes, including the hyperon decays. We thus find
\begin{align}
    S_{\pi^+,\text{nuc}}^{K_L}&=\langle \sigma v\rangle^{K_L}_{p\to \pi^+} n_p n_{K_L}+\langle \sigma v\rangle_{n\to \pi^+}^{K_L} n_n n_{K_L}\\
    S_{\pi^-,\text{nuc}}^{K_L}&=\langle \sigma v\rangle_{p\to \pi^-}^{K_L} n_p n_{K_L}+\langle \sigma v\rangle_{n\to \pi^-}^{K_L} n_n n_{K_L}\;.
\end{align}
For more details on the interconversion and pion production cross section, we refer to figure~\ref{fig:km_comp} (centre panel).
For decays, it is completely straightforward to infer the source terms once we determine the pionic branching ratios. Focusing on the most important modes, we find
\begin{align}
    K_L\to \begin{cases}
        \pi^\pm+X\;&,\quad\quad 67.59\%\\
        \pi^+\pi^-+X\;&,\quad\quad 12.54\%
    \end{cases}\;,
\end{align}
which allows us to write
\begin{align}
    S_{\pi^+,\text{dec}}^{K_L}& =S_{\pi^-,\text{dec}}^{K_L}=\Gamma_{K_L}(0.6759/2+0.1254)n_{K_L}\;.
\end{align}
Finally, the annihilation term is the simplest if we assume the same ratio of final-state charged/neutral pions and the same cross section \cite{Akita:2024ork}
\begin{align}
    S_{\pi^+,\text{ann}}^{K_L}&=S_{\pi^-,\text{ann}}^{K_L}=\frac{2}{3}\langle \sigma v\rangle^K_{\text{ann}} n_{K_L}^2\;.
\end{align}
Regarding dynamical equilibrium, we find in complete analogy to the case of charged kaons above (now considering self-annihilations)
\begin{align}
    \frac{\text{d}n_{K_L}}{\text{d}t}+3Hn_{K_L}&= S_{K_L}-\frac{n_{K_L}}{\tau_{K_L}}-\langle \sigma v\rangle^K_{\text{ann}} n_{K_L}^2 - \langle \sigma v\rangle^{K_L}_{p\to n} n_{p}n_{K_L}- \langle \sigma v\rangle^{K_L}_{n\to p} n_{n}n_{K_L}\\
    \Rightarrow n^{\text{de}}_{K_L}&=\frac{2S_{K_L}}{(\Gamma_{K_L}+\Gamma_{N,L})\left[\sqrt{1+\frac{4\langle \sigma v\rangle^K_{\text{ann}} S_{K_L}}{(\Gamma_{K_L}+\Gamma_{N,L})^2}}+1\right]}\;,\label{eq:dyn_eq_kl}
\end{align}
defining the quantities in analogy to the charged case.

\subsubsection{Pions}

Pions provide a very important contribution to interconversions via
\begin{align}
& \pi^- p \;\to\; n  X \\
& \pi^+ n \;\to\; p  X 
\end{align}
with $X= \pi^0,\gamma$. For the corresponding interconversion cross sections, we find much smaller values compared to the other cases, as shown in figure~\ref{fig:km_comp} (bottom right) in appendix~\ref{app:new_xsec}.

Let us finally come to the pion dynamical equilibrium abundance. Note that the source term is generally asymmetric, as discussed in the previous sections. In analogy to the previous mesons, we write\footnote{For the annihilation cross section, we use the result from \cite{ovchynnikov_2025_14882777}
\begin{align}
           \langle \sigma v\rangle^\pi_{\text{ann}}= \left[1.07631+0.0944817\left(\frac{T}{\text{MeV}}\right)\right]\,\text{mb}\;.
\end{align}}
\begin{align}
    \frac{\text{d}n_{\pi^+}}{\text{d}t}+3Hn_{\pi^+}&= S_{\pi^+}-\frac{n_{\pi^+}}{\tau_\pi}-\langle \sigma v\rangle_{\text{ann}}^\pi n_{\pi^+}n_{\pi^-}-\langle \sigma v\rangle^{\pi^+}_{n\to p} n_n n_{\pi^+}\\
    \frac{\text{d}n_{\pi^-}}{\text{d}t}+3Hn_{\pi^-}&=S_{\pi^-}-\frac{n_{\pi^-}}{\tau_\pi}-\langle \sigma v\rangle^\pi_{\text{ann}}  n_{\pi^+}n_{\pi^-}-\langle \sigma v\rangle^{\pi^-}_{p\to n} n_p n_{\pi^-}\\
    \Rightarrow n^{\text{de}}_{\pi^-}&=\frac{S_{\pi^-}}{\Gamma_\pi+\langle \sigma v\rangle^\pi_{\text{ann}} n^{\text{de}}_{\pi^+}+\Gamma_p}\label{eq:dyn_eq_pi1}\\
    n^{\text{de}}_{\pi^+}&=\frac{(S_{\pi^+}-S_{\pi^-})+(\Gamma_\pi+\Gamma_p)n^{\text{de}}_{\pi^-}}{\Gamma_\pi+\Gamma_n}\;,\label{eq:dyn_eq_pi2}
\end{align}
where we have introduced the total source term
\begin{align}
    S_{\pi^\pm}=S_{\pi^\pm,\text{inj}}^\phi+\sum_{p,t}S_{\pi^\pm,t}^p\;.
\end{align}
Here, $S_{\pi^\pm,\text{inj}}^\phi$ denotes the pions sourced by the relic and $p\in [K^\pm,K_L,\bar{n},\bar{p}]$ and $t\in [\text{ann},\text{nuc},\text{dec}]$ (depending on $p$).
Solving the ensuing quadratic equation for the pion density is straightforward but leads to tedious expressions which are not very enlightening. However, a numerical implementation is straightforward.
The pions will then affect the neutron and proton abundance in the following way
\begin{align}
    \frac{\text{d}n_{p}}{\text{d}t}\Bigg|_{\pi^\pm}= -\frac{\text{d}n_{n}}{\text{d}t}\Bigg|_{\pi^\pm}=\bigg(\langle \sigma v  \rangle^{\pi^+}_{n\to p}n_{\pi^+}  n_n - \langle \sigma v \rangle^{\pi^-}_{p\to n} n_{\pi^-} n_p\bigg)\;.
\end{align}

\subsubsection{Numerical scheme for all mesons}

In this section, we want to discuss a numerical scheme that allows us to track all meson contributions in a simple dynamical equilibrium approach. This relies on the assumption that the precise value of the neutron-to-proton ratio does not impact the results too much.

The main goal is to implement this scheme into \texttt{AlterAlterBBN}, i.e.~to calculate the rate at which the neutrons and protons approach the dynamical equilibrium value. Since we have not discussed the dynamical equilibrium of neutrons and protons yet, let us quickly consider
\begin{align}
    \frac{\text{d}Y_{n}}{\text{d}t}&= -\Gamma_\text{had}^{n\to p} Y_n +\Gamma_\text{had}^{p\to n} Y_p\label{eq:ntopboltz}\\
    \frac{\text{d}Y_{p}}{\text{d}t}&=-\frac{\text{d}Y_{n}}{\text{d}t}\;\\
    \Rightarrow Y^{\text{de}}_p&=\frac{\Gamma_\text{had}^{n\to p}}{\Gamma_\text{had}^{p\to n}} Y^{\text{de}}_n\;,\label{eq:np_dyneq}
\end{align}
where we have encoded all the conversion rates in a total rate $\Gamma_\text{had}$, i.e.
\begin{align}
    \Gamma_\text{had}^{n\to p}&=\langle \sigma v\rangle^{\pi^+}_{n\to p} n^{\text{de}}_{\pi^+}+\langle \sigma v\rangle^{K_L}_{n\to p}  n^{\text{de}}_{K_L}+\langle \sigma v\rangle^{K^-}_{n\to p}  n^{\text{de}}_{K^-}+\langle \sigma v\rangle^{\bar{p}}_{n\to p} n^{\text{de}}_{\bar{p}}\\
    \Gamma_\text{had}^{p\to n}&=\langle \sigma v\rangle^{\pi^-}_{p\to n} n^{\text{de}}_{\pi^-}+\langle \sigma v\rangle^{K_L}_{p\to n}  n^{\text{de}}_{K_L}+\langle \sigma v\rangle^{K^-}_{p\to n}  n^{\text{de}}_{K^-}+\langle \sigma v\rangle^{\bar{n}}_{p\to n} n^{\text{de}}_{\bar{n}}\;.
\end{align}
Combining this with the evolution of the (comovingly conserved) baryon number density, we can calculate the dynamical-equilibrium nucleon number density.

\begin{figure}
     \centering
    \begin{subfigure}{0.48\linewidth}
      \centering
    \includegraphics[width=\linewidth]{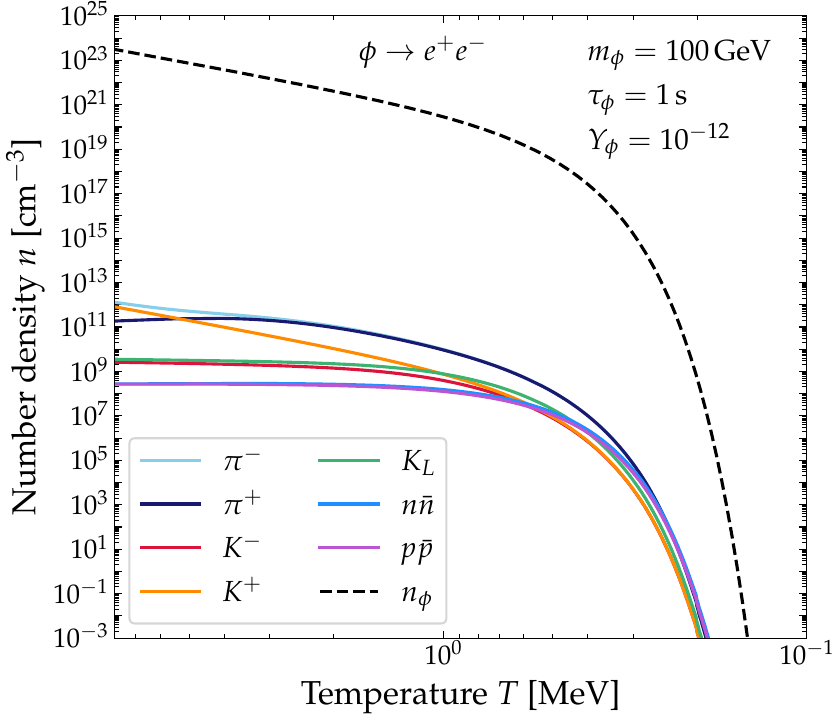}
    \end{subfigure}
    \begin{subfigure}{0.48\linewidth}
      \centering
      \includegraphics[width=\linewidth]{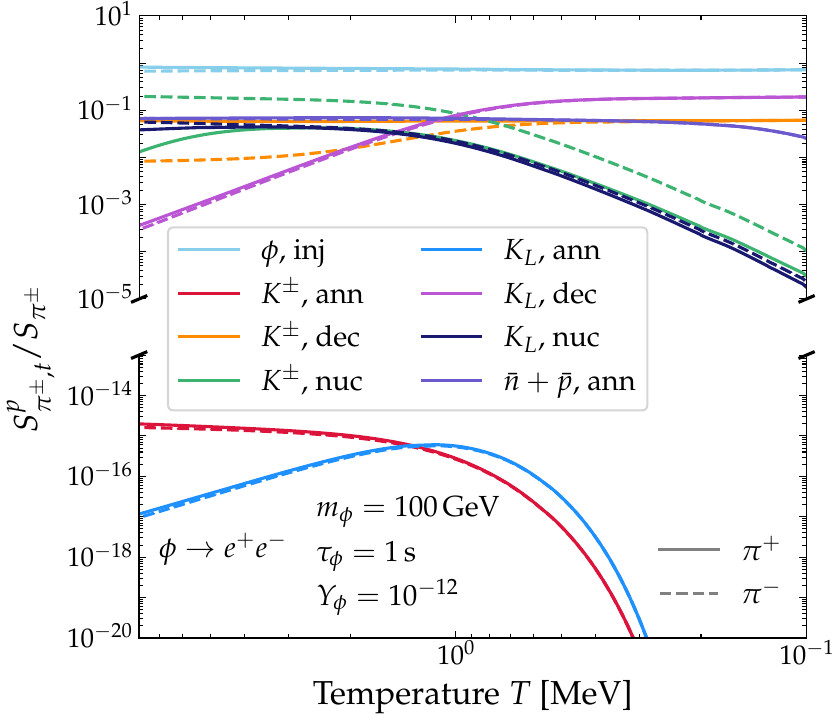}
    \end{subfigure}
    \caption{\textbf{Left:} Dynamical equilibrium number densities for a benchmark point. We also show the relic abundance as a dashed black line. \textbf{Right:} Relative contribution to the total source term for the pions.}
    \label{fig:dyneq}
\end{figure}

Let us now describe in greater detail how we explicitly implement this in the code. The challenge is to merge the direct time evolution of \texttt{AlterAlterBBN} with the dynamical equilibrium where all time derivatives vanish and the time evolution is just driven by the slow changes in source terms and the baryon number density. Therefore, we want to calculate the rate at which the nucleon abundances are driven towards equilibrium at every time step in the code. We apply the following prescription:
\begin{enumerate}
    \item we use the initial value (or the result of the previous evolution step in \texttt{AlterAlterBBN}) as our initial guess for the equilibrium neutron and proton abundance. We then use these to calculate the kaon and anti-nucleon abundances according to dynamical equilibrium (cf.\ eqs.~\eqref{eq:dyn_eq_bar}, \eqref{eq:dyn_eq_k1}, \eqref{eq:dyn_eq_k2}, and \eqref{eq:dyn_eq_kl})
    \item we calculate the source term for pions, made up of the original source term as well as the secondary pions from kaons and anti-nucleons. This allows us to also estimate the pion number densities (cf.\ eqs.~\eqref{eq:dyn_eq_pi1} and \eqref{eq:dyn_eq_pi2}),
    \item with all these number densities, we calculate the nucleon number densities required to enforce the dynamical-equilibrium condition as in eq.~\eqref{eq:np_dyneq},
    \item we take these new nucleon abundances as an input for another iteration. We repeat this process until we have converged on the consistent dynamical equilibrium result,\footnote{We have found that requiring a relative change of less than $10^{-6}$ in the neutron and proton abundances gives an excellent compromise between accuracy and runtime.}
    \item using the \emph{original} nucleon number densities and the consistent dynamical-equilibrium meson and anti-nucleon abundances, we calculate the interconversion rates towards dynamical equilibrium (cf.\ \eqref{eq:ntopboltz}) which we feed back into the main loop of \texttt{AlterAlterBBN}.
\end{enumerate}
This method allows us to incorporate the idea of dynamical equilibrium into the standard cosmological evolution of \texttt{AlterAlterBBN}. In particular, it is still functional when the weak rates are competitive with the hadronic ones. In figure~\ref{fig:dyneq} (left), we show the number densities as used in our code for a benchmark point with $m_\phi=100\,$GeV, $\tau_\phi=1\,$s, and $\mathpazoletters{Y}_\phi=10^{-12}$. We further show the relic number density to demonstrate that the hadrons just make up a tiny fraction of their source. As expected, over a wide range of temperatures, the pions are the most abundant particles due to their larger injection rate (cf.\ section~\ref{sec:pythia}). As secondary injections are small, the pion injection is very symmetric with a larger asymmetry only at the highest temperatures considered here. This difference stems from the much larger kaon asymmetry, driven by the fact that we do not consider the interactions of $K^+$ with the background nucleons.\footnote{In reality, at these high temperatures such processes are possible, but as we see that the asymmetry is quickly washed out at lower temperatures, we conclude that this approximation is justified.} As the lifetime of the relic dictates the period of interconversions, we see an exponential cutoff towards the times of BBN. In that regime, the weak interaction has already dropped out of equilibrium, and the neutrons now experience an ``hadronic freeze-out''. We should further note that for all injected hadrons, only a fraction participates in interconversions, especially for the mesons, where the small lifetime reduces the interacting fraction significantly. As a cross-check, we refer to appendix~\ref{app:dyn_eq} for a validation against a direct numerical solution of the Boltzmann equations for hadron injections.

We further show the relative contributions to the pion source term in figure~\ref{fig:dyneq} (right) for the same benchmark point as in the right figure. We explicitly use the quantity $S_{\pi^\pm,t}^p/S_{\pi^\pm}$, i.e.~the ratio of the individual source terms (now also including the relic decay: $p=\phi$, $t=\text{inj}$) divided by the sum of all source terms. Since pions are typically the most numerously injected hadron, the direct injections dominate the total budget. As expected, at late stages, kaon decays become particularly relevant as they do not depend on temperature, while the kaon annihilation and nuclear rates drop significantly. Throughout all temperatures, kaon annihilations are completely negligible (note the break in the y-axis!). However, for much larger relic abundances this contribution could grow in importance due to its quadratic scaling with relic density \cite{Akita:2024ork}. The largest asymmetry is induced by the charged kaons, where an already asymmetric abundance decays and scatters in a way that also injects pions with a bias. Of course, annihilation injections are perfectly symmetric; the marginal difference between the dashed and solid meson annihilation curves is just due to the different normalisation, i.e.~different $S_{\pi^\pm}$. The same holds true for primary injections from the relic and all $K_L$ except for nuclear scattering, where $n/p$ leads to different scattering probabilities for neutrons and protons.

\subsection{Hadrodisintegration}\label{sec:hadrodis}

Once hadrons are no longer thermalised immediately after injection, the kinetic energy can be high enough to disintegrate the previously synthesised light elements. As this process starts to become effective for $t \gtrsim 10^2$s, it generally proceeds at the same time as the standard BBN processes.
Below, we will discuss how we technically implement these reactions, where the approach is similar to the one in our previous paper \cite{Bianco:2025boy}, in which we discussed the impact of introducing hadrons following thermal BBN. The topic of hadrodisintegration has been extensively covered in \cite{Kawasaki:2005, Jedamzik:2006}, and more recently in \cite{Kawasaki:2018,Angel:2025dkw}. However, as no public code exists for accounting for the effect of injecting hadrons, we performed an independent calculation based on these publications. In the present study, we have expanded the effects of hadrodisintegration to earlier times, which required specific modifications to the \texttt{AlterBBN} code, as detailed in appendix.~\ref{app:hadrodis_abbn}. We refer the reader to \cite{Kawasaki:2005,Bianco:2025boy} for a comprehensive description of our formalism; here, we summarise only the key points.

Let us briefly introduce the main parameters that we need in order to account for the effect of hadrodisintegration. We first define two quantities $\xi_{X}^{n/p}[n_j](T, K)$ for each light element $X \in {p, n, \text{D}, \text{T}, {}^3\text{He}, {}^4\text{He}}$. This quantity encodes the net number of nuclei $X$ produced (or destroyed, if negative) by the injection of a single neutron or proton with kinetic energy $K$ at temperature $T$, and it explicitly depends on the background abundances $n_j$ with $j=p,{}^4\text{He}$.\\
The corresponding Boltzmann equation for hadrodisintegration can then be written as
\begin{align}
    \left[ \frac{\text{d} n_X}{\text{d} t} \right]_\text{hadro} = \sum_{y = n, p}\int_0^\infty \text{d}K\; \xi_{X}^y[n_j](K) \frac{\text{d}^2 n_y^\text{inj}}{\text{d} t \text{d} K}(K)\eqsp,
    \label{eq:y_hdi}
\end{align}
where $\text{d}^2 n_y^\text{inj}/(\text{d} t \text{d} K)$ denotes the injection rate of hadrons per unit time and kinetic energy.

\noindent In order to solve this equation, we further make the following two simplifications:
\begin{enumerate}
    \item An equal amount of protons and neutrons is injected.
    \item We approximate the full injection spectrum by a single, \emph{average} kinetic energy and the total injection rate.
\end{enumerate}
Under these assumptions, eq.~\eqref{eq:y_hdi} simplifies to
\begin{align}
    \left[ \frac{\text{d} n_X}{\text{d} t} \right]_\text{hadro} \simeq \sum_{y = n, p} \left( \xi_{X}^{y}[n_j](K_\text{hd}^\text{inj}) \frac{\text{d} n_\text{hd}^\text{inj}}{\text{d} t} \right) \eqsp.
    \label{eq:y_hdi_app}
\end{align}
A key difference with respect to \cite{Bianco:2025boy} is that we consider relic lifetimes $\tau_\phi \leq 10^{4}\,$s such that significant hadronic injections occur already during the early stages of BBN. While at $t\lesssim10^2\,$s all injected mesons and nucleons are efficiently stopped, this is no longer the case for neutrons already at $t>10^2\,$s. As a consequence, hadrodisintegration does not ``factorize'' with the standard BBN evolution. This requires a simultaneous treatment of the cascade and nuclear network from early times, including the computation of the hadrodisintegration rates. This is achieved by incorporating the matrix formalism developed in \cite{Kawasaki:2005,Bianco:2025boy} within \texttt{AlterAlterBBN} (see appendix \ref{app:hadrodis_abbn} for details). This way, we can consistently evolve the nuclear abundances during early times, where thermal and non-thermal effects are intertwined. The resulting abundances from \texttt{AlterAlterBBN} are then used as initial conditions for \texttt{ACROPOLIS}, which takes over the late-time evolution, and accounts for both hadrodisintegration and the onset of photodisintegration processes that become relevant at $t \gtrsim 10^4\,$s.

\textbf{A note on anti-nucleons} In our analysis, we neglect the contribution of anti-baryons, as their impact is subleading in the parameter space of interest. Since we consider decays that conserve baryon number, baryons and anti-baryons are produced in equal amounts. These injections have been studied in \cite{Kawasaki:2018}. From figure~6 of \cite{Kawasaki:2018}, one can observe that the effect of anti-neutrons (denoted by ``$\xi_{H_i,\overline{n}}$'') exceeds that of protons (``$\xi_{H_i,p}$'') only at very high energies, well beyond the regime relevant for our work. Furthermore, due to a lack of data, \cite {Kawasaki:2018} also does not truly handle direct hadronic disintegration but just the hadronic cascade initiated by elastic anti-nucleon scatterings on background protons.

\textbf{A note on meson-induced hadrodisintegration}
The literature has studied the disintegration effects of mesons (in addition to interconversions), e.g.\ in~\cite{Pospelov:2010cw}. As mentioned before, we ignore those here as we always expect to have accompanying nucleons in all decay channels, unlike the reference which treats purely mesonic injections. For all of the injections treated in this work, the number of mesons (in particular pions) is larger than the number of nucleons, and the disintegration cross sections are typically similar. Naively, this seems to imply that mesonic hadrodisintegration should also dominate in our setup. However, we further need to account for the fact that mesons will decay rapidly (cf.\ figure~\ref{fig:therm_check}). In particular, even if the mesons interact before thermalisation, where the cross sections can be enhanced by the $\Delta$-resonance, there will only be few interactions. Let us quantify this effect: for mesonic disintegration to be efficient, we would like it to happen more rapidly than the decays. From \cite{Pospelov:2010cw} we take a \emph{maximum} disintegration cross section of $\sigma^\pi_{\text{dis}}\lesssim 200\,$mb. Then we estimate the $^4$He abundance $n_{^4\text{He}}\sim \eta_b n_\gamma$. The disintegration rate is therefore roughly given as $\Gamma_{\pi,\text{dis}}\sim \sigma^\pi_{\text{dis}}\, \beta\, n_{^4\text{He}} \lesssim 0.12\,\text{s}^{-1} (T/10\,\text{keV})^3$, assuming that the pions are fully relativistic ($\beta=1$). Even this most optimistic guess is much smaller than the inverse lifetime of all mesons. Even with relativistic time dilation, e.g.~a pion would need to have $E= m_\pi/(\Gamma_{\pi,\text{dis}}\tau_\pi) \sim 4.5\cdot 10^7\,$GeV to compensate for the decay rate at rest.

\subsection{Photodisintegration}\label{sec:photodis}

The injection of high-energy electromagnetic (EM) particles -- namely electrons, positrons, and photons -- initiates an EM cascade, resulting in a spectrum of non-thermal photons (see e.g.~\cite{Kawasaki:1994sc, Kawasaki:2005, Hufnagel:2018bjp, Depta:2020zbh}). These photons can efficiently photodisintegrate light nuclei once their energies exceed the corresponding nuclear binding thresholds, $E_\gamma > E_N^\text{th}$, thereby modifying the primordial abundances. At the same time, the photon energy cannot be arbitrarily large: for sufficiently high energies, $E_\gamma \gtrsim E_{ee}^\text{th} \simeq m_e^2/(22T)$, interactions with the background photons lead to efficient $e^+e^-$ pair production, rapidly depleting the high-energy photon population. As a consequence, photodisintegration can only proceed efficiently if $E_N^\text{th} < E_{ee}^\text{th}$. This condition defines, for each element, the maximum temperature below which photodisintegration becomes effective, implying that in practice this process is only relevant for $T \lesssim O(\text{keV})$, when SBBN has already finished, so that the two processes simply factorise.

For relatively high energy, there is another useful simplification we can make: if the energy of the particles initiating this cascade is above the electron pair-creation threshold, i.e.~$ E_0 \gg E_{e e}^\text{th}$, any photon produced during this cascade with an energy above this threshold is rapidly depleted. In this case, it is well known that the cascade produces a universal spectrum of non-thermal photons, which is well approximated by \cite{Kawasaki:1994sc}\footnote{For the highly energetic injections considered in this work, we expect that the universal spectrum is always a good approximation. Therefore, we do not have to explicitly solve the equations governing the EM cascade to obtain the spectrum (cf.~e.g.~\cite{Poulin:2015woa,Poulin:2015opa})}
\begin{align}\label{eq:universal}
    \text{f}_{\gamma,\text{univ}}(T, E) \simeq \frac{S_\text{em}(T)}{\Gamma_\gamma(T, E)} \times \begin{cases}
        K_0 (E/E_X)^{-3/2}\quad &,\quad E<E_X\eqsp,\\
        K_0 (E/E_X)^{-2}\quad &,\quad E_X<E< E^\text{th}_{ee}\eqsp,\\
        0\quad &,\quad E>E^\text{th}_{ee}\eqsp.
    \end{cases}
\end{align}
Here, $K_0=E_0 E_X^{-2}\left[2+\ln(E^\text{th}_{ee}/E_X)\right]^{-1}$, $E_X=m_e^2/(80T)$, $\Gamma_\gamma$ is the total scattering rate of non-thermal photons with the background, and $S_\text{em}$ is the source-term describing the total amount of injected EM material.
Additionally, the distribution function $\text{f}_\gamma$ is defined to be differential in energy instead of momentum $f_\gamma$, with the two being related via
\begin{align}
    \text{f}_\gamma(T,E)=g_\gamma\frac{Ep}{2\pi^2}f_\gamma(T,p)\eqsp.
\end{align}
Here, $g_\gamma = 2$ is the number of photon degrees of freedom.
To evaluate eq.~\eqref{eq:universal} for our scenario, we have to compute the source term $S_\text{em}$ originating from the cascade.
It can be written as
\begin{align} 
S_\text{em}(T) = \zeta_{\rm em}(m_\phi) \frac{2\Delta n_\phi}{\Delta t}\eqsp ,
\label{eq:def_Sem}
\end{align}
where $\Delta n_\phi$ is the number density of $\phi$ particles decaying during the interval $\Delta t$, and $\zeta_\text{em}$ parameterises the fraction of the injected energy transferred to the electromagnetic cascade (cf.\ section~\ref{sec:pythia}). Note that we include the entire output of the shower process inside this term; in particular, it also contains the tree-level contributions in electron-positron or photon-photon injections.

After the EM cascade, the non-thermal photons originating from the EM cascade, i.e.~the ones encoded in  $\text{f}_{\gamma, \text{univ}}$, will initiate photodisintegration reactions, e.g.~$\gamma \text{D} \rightarrow n p$ among others. The Boltzmann equations governing these reactions can be written as~\cite{Depta:2020mhj} (dropping the $t/T$ dependence for convenience)
\begin{align}
\left[\frac{\text{d} n_X}{\text{d} t}\right]_\text{photo} = \sum_{j} n_{j} N_{j\gamma\rightarrow X} \int_{0}^{\infty}\text{d} E\;\text{f}_\gamma(E)\sigma_{j\gamma \rightarrow X}(E) - n_X \sum_{j'} \int_{0}^{\infty} \text{d} E\; \text{f}_\gamma(E)\sigma_{X\gamma \rightarrow j'}(E)\eqsp.
\label{eq:y_pdi}
\end{align}

Here, $\text{f}_\gamma = \text{f}_{\gamma, \text{univ}}$, $X$ denotes any of the light nuclei under consideration, and $N_{j\gamma\rightarrow X}$ is the number of $X$ nuclei produced in the reaction $j\gamma\rightarrow X$. The quantities $\sigma_{j\gamma\rightarrow X}$ ($\sigma_{X\gamma\rightarrow j'}$) represent the cross sections for the reactions that create (destroy) $X$, and can be found in~\cite{Depta:2020mhj}.

Given an expression for $S_\text{em}(T)$, one could in principle use either the most recent stable version \texttt{v1.3.2} or the updated development version \texttt{v2.0.0-dev} of \texttt{ACROPOLIS}\footnote{The difference between the two versions lies in the implementation of hadrodisintegration in version \texttt{v2.0.0-dev}.} to compute $\text{f}_{\gamma, \text{univ}}$ and solve eq.~\eqref{eq:y_pdi}, thereby obtaining the final light-element abundances after photodisintegration. However, since hadrodisintegration is also effective at early times, we must first run \texttt{AlterAlterBBN}. Only after obtaining the resulting abundances can \texttt{ACROPOLIS} be applied. In other words, the initial abundances used in \texttt{ACROPOLIS} are the final products of \texttt{AlterAlterBBN}.

We can then combine eq.~\eqref{eq:y_pdi} with eq.~\eqref{eq:y_hdi}, to get the total Boltzmann equation governing non-thermal BBN (NBBN):
\begin{align}
    \left[ \frac{\text{d} n_X}{\text{d} t} \right]_\text{NBBN} = \left[ \frac{\text{d} n_X}{\text{d} t} \right]_\text{photo} + \left[ \frac{\text{d} n_X}{\text{d} t} \right]_\text{hadro}\eqsp.
    \label{eq:y_nth}
\end{align}

\subsection{Effects on the background cosmology}\label{sec:bgcosmo}

The injection of additional energy into the SM thermal bath modifies the effective number of relativistic species at recombination, $N_\text{eff}$, which is constrained via CMB observations~\cite{Planck:2018vyg}. This can be expressed as: 
\begin{align}
    N_\text{eff} = \frac87  \left(\frac{11}{4}\right)^{4/3} \left( \frac{\rho_\text{rad} - \rho_\gamma }{\rho_\gamma} \right) \equiv \big[ 3 + \Delta N_\text{eff}(t_\text{rec})\big] \left(\frac{11}{4}\right)^{4/3}\left[\frac{T_\nu(t_\text{rec})}{T(t_\text{rec})}\right]^4,
\end{align}
where $\rho_\text{rad}$ is the total radiation energy density and $\rho_\gamma$ is the photon energy density. If the relic $\phi$ decays into EM particles after neutrino decoupling, as is the case for $\phi\to e^+ e^-$, it injects entropy into the SM plasma without heating the neutrino sector, as long as injections happen after neutrino decoupling. This alters the ratio $T_\nu/T$, reducing the neutrino temperature relative to the photon temperature. As a consequence, the value of $N_\text{eff}$ at recombination is suppressed in this case, yielding $N_\text{eff} < 3.04$ (for a recent and more detailed discussion, see \cite{Escudero:2026mgw}).
However, for many SM final states, the injection will be into both the photon and the neutrino sectors. The exact determination of the $N_\text{eff}$ bounds requires a more detailed treatment of the injections. In particular, it requires accounting for the injection of neutrinos into the plasma (increasing $N_\text{eff}$) while at the same time tracking the EM energy loss of charged particles before their decay (which will decrease $N_\text{eff}$).
As neutrinos typically do \emph{not} interact with the background plasma \cite{Boyarsky:2021yoh,Bianco:2025boy}, this part is relatively straightforward. However, for general injections, the decomposition into EM and neutrino energy depositions is non-trivial to extract. Let us discuss muons as a prime example. Muons, when injected, will eventually decay into an electron and two neutrinos. On average, we can assume that the energy is split equally among the three effectively massless particles. However, if the muon is stopped in the plasma (cf.\ figure~\ref{fig:therm_check}), it will give most of its kinetic energy into the EM plasma before decaying so that the neutrinos resulting from the decay will only carry 2/3 of the rest mass, potentially much below the initial kinetic energy. Thus, the decomposition of the injection is highly dependent on the temperature at injection, the injected particle, and the relic mass as well. This consideration is analogous for many more particles, in particular those with hadronic final states. 
In the following, we will therefore only show the $N_\text{eff}$ limits for $\phi \to e^+ e^-$, with the understanding that for other decay channels the limits will be somewhat weaker due to partial cancellations.
However, as our direct BBN bounds are so much stronger than the $N_\text{eff}$ constraints, this simplified treatment will not impact our final results.

The injection of additional particles around BBN can also affect the neutrino decoupling temperature \cite{Escudero:2018mvt, EscuderoAbenza:2020cmq, Froustey:2019owm,Bennett:2019ewm}. In general, neutrinos remain in thermal equilibrium with the electromagnetic plasma through weak interactions with electrons and positrons. The interaction rate scales as $\Gamma_\nu \sim G_F^2\,T^5$, while the Hubble expansion rate during radiation domination scales as $H \sim T^2$. Neutrino decoupling occurs when $\Gamma_\nu \simeq H$ which fixes the decoupling temperature $T_{\nu,\text{dec}}$.
We work under the assumption of instantaneous decoupling, a common approach which is known to be a very good approximation \cite{deSalas:2016ztq}. 
Within this approximation, we take a decoupling temperature of 
\begin{equation}
T_{\nu\text{-dec, SM}} \simeq 1.3\,\mathrm{MeV},
\end{equation}
where the numerical value is chosen to reproduce $N_\text{eff} \simeq 3.04$ within our code.

In the presence of an additional relic, the expansion rate is modified, $H \to H'$. The decoupling condition $\Gamma_\nu(T_{\nu\text{-dec}}) \sim H'(T_{\nu\text{-dec}})$ then implies
\begin{equation}
G_F^2\,T_{\nu\text{-dec}}^5 \simeq H'(T_{\nu\text{-dec}}).
\end{equation}
Taking the ratio with the Standard Model case, and using the same condition $G_F^2\,T_{\nu\text{-dec, SM}}^5 \simeq H_{\text{SM}}(T_{\nu\text{-dec, SM}})$, we obtain
\begin{align}
\left(\frac{T_{\nu\text{-dec}}}{T_{\nu\text{-dec, SM}}}\right)^5 \simeq \frac{H'(T_{\nu\text{-dec}})}{H_{\text{SM}}(T_{\nu\text{-dec, SM}})}.
\end{align}
from which:
\begin{align}\label{eq:T_dec_scaling}
\frac{T_{\nu\text{-dec}}}{T_{\nu\text{-dec, SM}}} \simeq \left(\frac{H'(T_{\nu\text{-dec}})}{H_{\text{SM}}(T_{\nu\text{-dec, SM}})}\right)^{1/5}.
\end{align}
This expression follows directly from the temperature scaling of the interaction rates, $\Gamma_\nu~\propto~T^5$, and provides a consistent estimate within the instantaneous-decoupling approximation.

The neutrino decoupling temperature has a direct effect on the light-element abundances through its impact on the weak interaction rates that govern neutron–proton interconversion. A shift in $T_{\nu\text{-dec}}$ modifies the temperature at which the neutron-to-proton ratio departs from equilibrium, which in turn modifies the neutron-to-proton ratio, finally affecting the $^4$He abundance.

\subsection{Treatment of the errors} 

Let us also quickly comment on the handling of the theoretical errors entering the determination of the light-element abundances. We consider (i) the errors in the nuclear reaction rates used in \texttt{AlterAlterBBN}, (ii) the errors in our calculation of $\xi_X^y$, which amount to $\epsilon_\xi \approx 20\%$ \cite{Kawasaki:2005,Kawasaki:2018,Bianco:2025boy}, and (iii) the errors in the interconversion rates which we also estimate as a $\pm20\%$ relative change for all rates.\footnote{The literature \cite{Jung:2025dyo} does not provide us with error bars, and so we estimate them to be similar to the hadrodisintegration rate uncertainty.} A detailed treatment of the error propagation on the final abundances would require a Monte-Carlo approach, running the code many times for a single data point and varying the parameters within their errors.

This approach is prohibitively time-consuming for the large scans that we perform, and thus, we use a simpler, well-established approach to estimate the uncertainties.
In order to do so, we run \texttt{AlterAlterBBN} with high, mean, and low values for all relevant rates, specifically the interconversion, reaction, and disintegration rates.

This yields three sets of 
abundances $Y_X^{\Gamma,\text{high},0}$, $Y_X^{\Gamma,\text{mean},0}$, and $Y_X^{\Gamma,\text{low},0}$ which we use as initial values for \texttt{ACROPOLIS}. Evolving these through eq.~\eqref{eq:y_nth} in \texttt{ACROPOLIS}, we obtain the corresponding final abundances $Y_X^{\Gamma,\text{high}}$, $Y_X^{\Gamma,\text{mean}}$, and $Y_X^{\Gamma,\text{low}}$.\footnote{For small relic lifetimes this last step is not necessary.}

Additionally, since hadrodisintegration is also part of \texttt{ACROPOLIS}, we solve eq.~\eqref{eq:y_nth} with the central result $Y_X^{\Gamma\;\text{mean},0}$ as initial conditions, varying once more the disintegration parameters $\xi_X^y \{ (1+\epsilon_\xi), (1-\epsilon_\xi) \}$ to obtain two more sets of final abundances  $Y_X^{\xi,\text{high}}$ and $Y_X^{\xi,\text{low}}$, which represents the uncertainty arising in hadrodisintegration from $\texttt{ACROPOLIS}$. We then calculate the theoretical error on the different abundances $Y_X$ as the minimal difference between the central result and the high/low result for the sets $[Y_X^{\Gamma,\text{high}},Y_X^{\Gamma,\text{mean}},Y_X^{\Gamma,\text{low}}]$ and
$[Y_X^{\xi,\text{high}},Y_X^{\xi,\text{mean}},Y_X^{\xi,\text{low}}]$.\footnote{Note that in this notation $Y_X^{\Gamma,\text{mean}}=Y_X^{\xi,\text{mean}}$. } To arrive at the final result, we add up both contributions in quadrature. Finally, this theoretical uncertainty is combined with the observational uncertainties on the abundances presented in section~\ref{sec:abundances} to derive the overall exclusion limits. In light of the mild tension in the theory predictions for D/$^1$H \cite{EscuderoAbenza:2025tsi}, let us quickly quote the results that our code provides in SBBN:
\begin{align}
    \text{D}/^1\text{H}&=\left(25.23\pm0.53\right)\times 10^{-5}\;,\\
    \mathcal{Y}_p&=\left(0.2472\pm 0.0008\right),\\
    ^3\text{He}/\text{D}&=\left(0.410\pm0.003\right)\;.
\end{align}
This result makes immediately clear that the deuterium error budget is dominated by theory uncertainty even in the SBBN scenario. The comparatively large uncertainties on the non-thermal rates will further enhance the theoretical error. As long as the uncertainties from non-thermal sources dominate the total error budget we can be certain that the tension in the SBBN prediction of D/$^1$H does not impact our results.

\section{From decays to final-state particles utilising \texttt{PYTHIA}}\label{sec:pythia}

Having discussed the leading effects of generic injections, we now focus on the case of a relic decaying into a pair of SM particles. To this end, we use \texttt{PYTHIA} to shower an initial two-particle state whose centre-of-mass energy is set equal to the relic mass.
This procedure captures two distinct physical effects. First, it hadronises coloured initial states — such as quarks or gluons — yielding a final state composed of electromagnetically interacting particles, mesons, baryons, and neutrinos.\footnote{We retain only those particles that are (meta-)stable on the timescales of interest, allowing all others to decay; in particular, we include all other hadrons not treated explicitly above.} Second, \texttt{PYTHIA} automatically incorporates the leading final-state radiation (FSR) corrections to the primary spectrum, so that even purely electromagnetic decay channels such as $e^+e^-$
can give rise to hadronic and neutrino final states.

A priori, the relic particle could be of scalar or vector nature, in general with different coupling structures to left- and right-handed SM particles.
When initialising the final state in \texttt{PYTHIA} which is to be showered, one has to make a corresponding choice.
For definiteness, we have chosen a scalar relic, but we have checked that when assuming equal couplings to left- and right-handed states, the results for scalar and vector relics are identical within uncertainties, so our results apply to both cases.\footnote{Of course, a vector particle could not decay into massless vectors such as photons or gluons directly.}
A major difference between the two cases comes from the degrees of freedom (1 for a real scalar compared to 3 for a massive vector). In the limit plots we show below, this is already absorbed into the yield $\mathpazoletters{Y}_\phi$, so that both limits are identical.

As we have seen in the previous section, we need the following quantities to cover all effects relevant to the light-element abundances:
\begin{itemize}
    \item $\zeta_\text{EM}$, i.e.~the fraction of EM energy in the final state,\footnote{Since photodisintegration only becomes relevant at late times, we also let all mesons decay since we do not expect them to lose their energy into the EM sector by scattering down (c.f.\ section~\ref{sec:bgcosmo}). Thus, we count all electrons, positrons, and photons after mesonic decays into $\zeta_\text{EM}$.
    }
    \item $N_{n/p}$ and $K_{n/p,\text{avg}}$, i.e.~the typical number of baryons per decay and their average kinetic energy, which we use to approximate the injection spectrum of neutrons and protons
    \item $N_m$ with $m=\pi^\pm,K^\pm, K^0_L$, i.e.~the number of mesons per decay\,.
\end{itemize}
Let us now discuss on a more technical level how we extract these numbers. For a given pair of SM particles that we want to analyse, we utilise the \texttt{VINCIA} module of \texttt{PYTHIA8.3} to obtain the relevant particles after final-state radiation and showering. We initialise an \texttt{.lhe} file with events of the tree-level decay as input for \texttt{PYTHIA} and then read out all quantities of interest from the output and average them over all events. We adjust the number $N$ of considered events to the case of interest: a low mass relic 
with mass below the weak scale decaying into $e^+e^-$ has only very suppressed final-state radiation and thus requires a large number of events (of the order of $10^7$) 
to get reasonable statistics for all relevant quantities, such as the number of pions, $N_\pi$. For heavier particles, we can lower the number of events as FSR gets more likely, which improves the statistics. 
In fact, if coloured particles appear at tree-level, it is helpful to further reduce the number of generated events, in particular for heavy relics: the large energies available for the shower lead to an enormous multiplicity, significantly slowing down the code. 
Note also that for showers close to the mass threshold of coloured particles, \texttt{PYTHIA} ceases to give reliable results due to a number of complicating effects, which is why we only show constraints where the relic particle is sufficiently heavier than the considered final state. 

\begin{figure}
    \centering
    \includegraphics{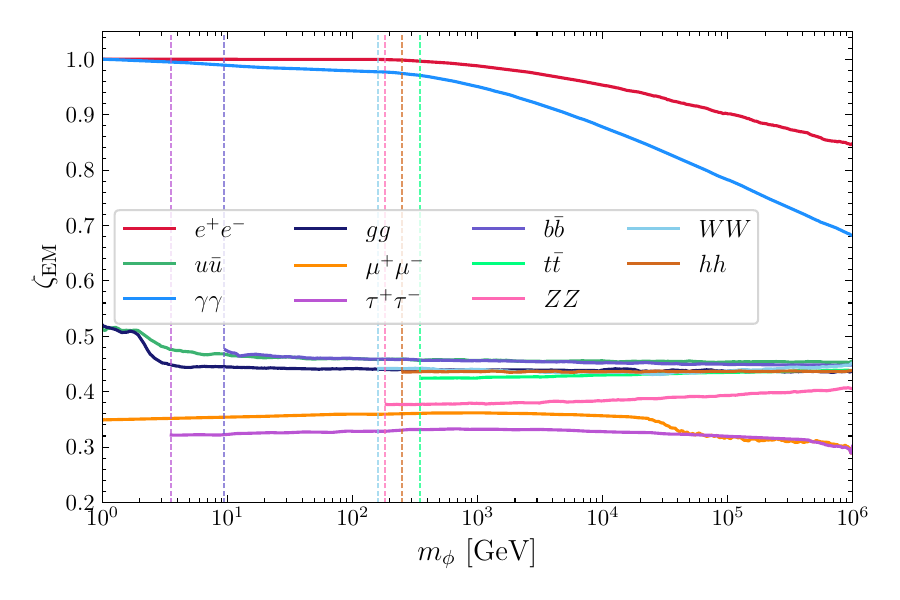}
    \caption{Average fraction of injected EM energy for different SM final states at late times, i.e.\ after decays of intermediate mesons.}
    \label{fig:zeta_em}
\end{figure}

To demonstrate our results, we show the average electromagnetic component after FSR and hadronisation for all injections we probe in figure~\ref{fig:zeta_em}, including the benchmark cases $e^+e^-$, $u\bar{u}$, and $\gamma\gamma$. As expected, starting with an electromagnetic final state (before FSR) will also typically lead to a mostly EM state after showering. For a hadronic setup, we see that typically a relatively constant (w.r.t.~the relic mass) value is reached soon after the kinematic threshold is crossed. We find that, excluding threshold effects, the relic mass dependence of the electromagnetic fraction is relatively mild for all particles in the range of $[10,10^4]\,$GeV which we consider for the constraints. Since the unstable charged leptons tend to produce neutrinos in their decay chain, they have the smallest EM fraction. At this point, it should be emphasised once more that for the heavier quarks the results too close to the mass threshold are not reliable. Therefore, we do not consider these cases in our analysis of the constraints.

\begin{figure}
    \centering
    \includegraphics{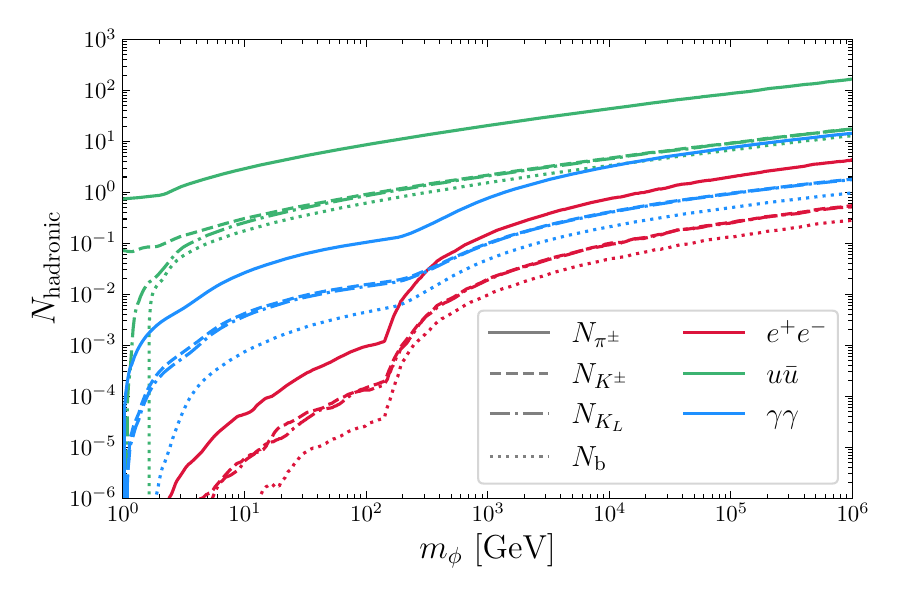}
    \caption{Hadronic final states for $e^+e^-$, $u\bar{u}$, and $\gamma\gamma$ injections. We focus on the (meta-) stable particles relevant to our analysis and let all other hadrons decay immediately.}
    \label{fig:N_had}
\end{figure}

Let us now turn to the hadronic components, which we present in figure~\ref{fig:N_had}. We focus on the three benchmark cases $e^+e^-$, $u\bar{u}$, and $\gamma\gamma$, indicating the various final-state hadrons with different linestyles. In our convention, the lines denote individual particles' species, so $N_{\pi^\pm} = N_{\pi^+} = N_{\pi^-}$ is the number of charged pions, $N_{K^\pm} = N_{K^+} = N_{K^-}$ the number of charged kaons, and $N_\text{b}= N_p = N_{\bar p}= N_n = N_{\bar{n}}$ the number of (anti-) protons and neutrons individually.
We see clearly that pions are consistently the most numerous hadrons in any final state. We further see that the number of kaons is largely identical between $K^+$, $K^-$, and $K_L$ and that nucleons are the least abundant species. We refer to appendix~\ref{app:pythia} for the other injection channels.

Finally, let us briefly discuss the mass dependence of the kinetic energy of the different injections.
As a rule of thumb, we find that (sufficiently far away from all mass thresholds), there is an approximately linear trend with $K_\text{b}\sim 0.01m_\phi$ for all final states.

\section{Results \& Discussion}\label{sec:results}

In this section, we present the resulting limits for particles decaying into different final states over all lifetimes relevant for BBN, ranging from $\tau_\phi = 10^{-2}\,\mathrm{s}$ to $\tau_\phi=10^{11}\,\mathrm{s}$, where the lower end is motivated by an exponential loss in sensitivity for lifetimes below neutrino decoupling, while the upper end is motivated by the expected takeover of CMB bounds \cite{Poulin:2016anj}. 
We examine all two-body decays into SM final states except the case of decays into neutrinos, which we leave for a future study.\footnote{We have explored decays into neutrinos with longer lifetimes in \cite{Bianco:2025boy}. However, the case of early decays requires us to consider additional effects that are beyond the scope of the present work.}

\subsection{Decay into $e^+e^-$}\label{sec:ee_inj}

In this section, we will discuss the constraints on $e^+e^-$ injections in detail. This discussion will not only clarify the shape and strength of these particular bounds but also help in understanding the constraints on other decay channels better.

\begin{figure}[!t]
    \centering
    \includegraphics{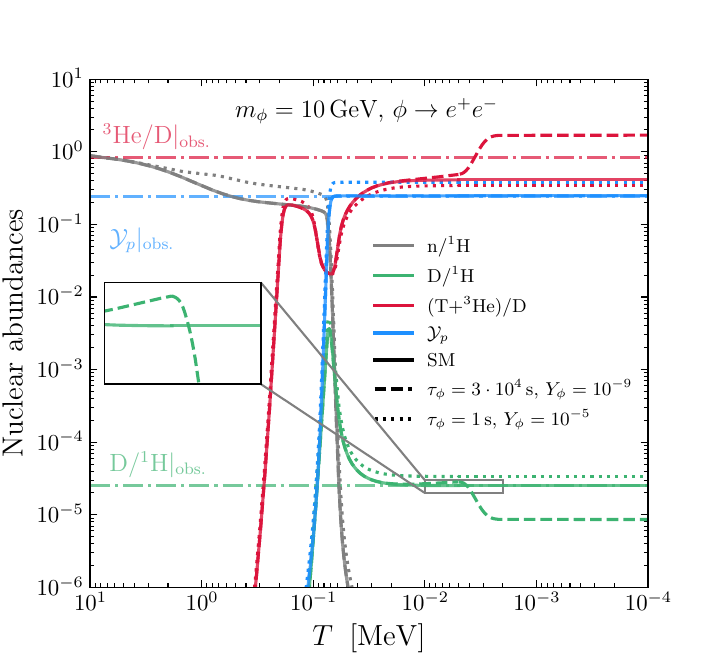}
    \caption{Evolution of the light-element abundances for $m_\phi = 10$GeV for two benchmark points, with one demonstrating the impact of interconversions (dotted) while the other captures both hadro- and photodisintegration effects (dashed).}
    \label{fig:evo_dis_2}
\end{figure}

However, before we discuss the bounds in the full parameter space, let us discuss the full evolution of the light elements in figure~\ref{fig:evo_dis_2}.
Let us begin with one of the two benchmark points (dotted line, $m_\phi=10\,$GeV, $\tau_\phi=1\,$s, $\mathpazoletters{Y}_\phi=10^{-5}$ for injections of $e^+e^-$). For these early injections, disintegration is  negligible, and the dominant effect is due to the change in the neutron-to-proton ratio $n/p$ driven by interconversion processes. This becomes clear from the deviation of the neutron abundance from the SM prediction. The neutron abundance is then mostly transferred into the $^4$He abundance, which also strongly deviates from the SM prediction now. We further note that the other abundances feel the effect of the change in $n/p$ as well, but due to the high precision of the $^4$He abundance measurement, it remains the leading bound for this benchmark point.

For the second parameter point we have chosen $m_\phi=10\,$GeV, $\tau_\phi=3\cdot 10^4\,$s, and $\mathpazoletters{Y}_\phi=10^{-9}$. We have selected this point as it decays sufficiently late so that the strongest bounds are dominated by disintegration (and not interconversions) while still decaying early enough to have significant hadrodisintegration happening in both \texttt{ACROPOLIS} \emph{and} \texttt{AlterAlterBBN}. Furthermore, this point lies close to the transition between D/$^1$H over- and underproduction due to the advent of photodisintegration around these lifetimes. For better visibility, we have zoomed into the transition region below $T\approx 100\,$keV. Early on, deuterium is produced by $^4$He hadrodisintegration, but later on, the trend is inverted, and deuterium is mostly photodisintegrated. We further note that $^4$He is barely affected in this scenario as its overall abundance is much larger than the rest of the light elements (except for protons), so that converting a tiny fraction of $^4$He already leads to a massive change in D, T, and $^3$He.

\begin{figure}[!t]
 \centering
    \begin{subfigure}{0.48\linewidth}
      \centering
      \includegraphics[width=0.72\linewidth, trim= 3cm 0 2cm 0cm]{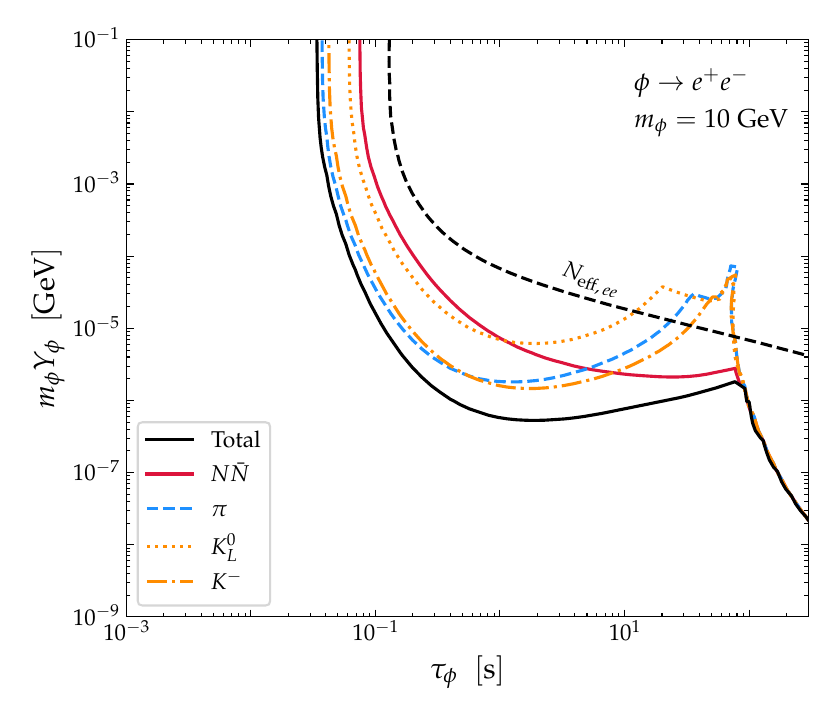}
    \end{subfigure}
    \begin{subfigure}{0.48\linewidth}
      \centering
      \includegraphics[width=0.72\linewidth, trim= 2cm 0 3cm 0cm]{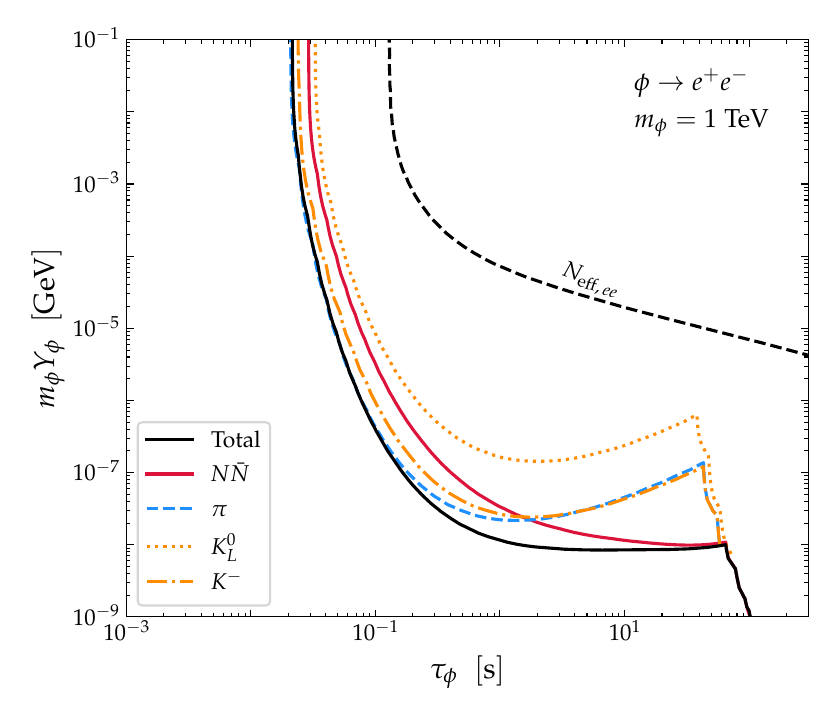}
    \end{subfigure}
    \caption{\textbf{Left}: Decomposition of the individual interconversion bounds for $\phi\to e^+e^-$ and $m_\phi=10\,$GeV. \textbf{Right}: Same but for $m_\phi=1\,$TeV.}
    \label{fig:mphi_comp_inter}
\end{figure}

Let us now focus on the regime where interconversions dominate the bounds. Unlike hadro- and photodisintegration, where only few different particles drive the constraints (neutrons/protons and photons, respectively), many different hadrons contribute to the bounds by interconverting the nucleons before BBN. To provide a better understanding on the pion, kaon, and nucleon contribution, we show them individually in figure~\ref{fig:mphi_comp_inter} for two benchmark points ($\phi\to e^+e^-$, $m_\phi=[10,\,1000]\,$GeV). In both cases, we see that pions are leading the bounds at the lowest lifetimes because they are the most abundant particle in the injection shower (cf.\ figure~\ref{fig:N_had}). However, as their lifetime is similar to the kaon lifetimes, but their interconversion cross sections are lower, they will become increasingly less relevant for larger lifetimes, where temperatures at the injection are lower and thus decays are more relevant. In this case, the charged kaons take over as they profit from their larger interaction rate. Neutral kaons, however, are always subleading which further justifies our treatment lined out in section~\ref{sec:interconversion}. Finally, all three meson contributions become weaker than the $N_\text{eff}$ bound for $m_\phi=10\,$GeV and hit the irreducible background cosmology bounds. The case for the antinucleon-nucleon pairs is the exact opposite. Due to their low injection rate, they are consistently weaker than the (charged) meson contributions for short lifetimes. However, as they are stable on the timescales considered here, the constraints do not suffer a suppression with increasing relic lifetime, unlike the mesons where their short lifetime prevents efficient interconversion processes. This leads to the nucleons dominating the overall limit up to lifetimes of $\tau_\phi\sim 10^2\,$s where the hadrodisintegration bounds take over. Due to the larger hadronic branching ratio, all individual $m_\phi=1\,$TeV bounds are overall stronger and in particular, they are always stronger than the background cosmology bounds. We expect these results to qualitatively hold for all injections with similar ratios of pion, kaon, and nucleon injections. We will discuss the special case of $\tau^+\tau^-$ in more detail later. Finally, let us emphasise again the confirmation of our assumption that the interconversions, independently of the type of hadron, tend to \emph{increase} $n/p$ due to the larger proton abundance in the SM, cf.\ our discussion at the beginning of section~\ref{sec:interconversion}.

\begin{figure}[!t]
 \centering
    \begin{subfigure}{0.48\linewidth}
      \centering
      \includegraphics[width=0.72\linewidth, trim= 3cm 0 2cm 0cm]{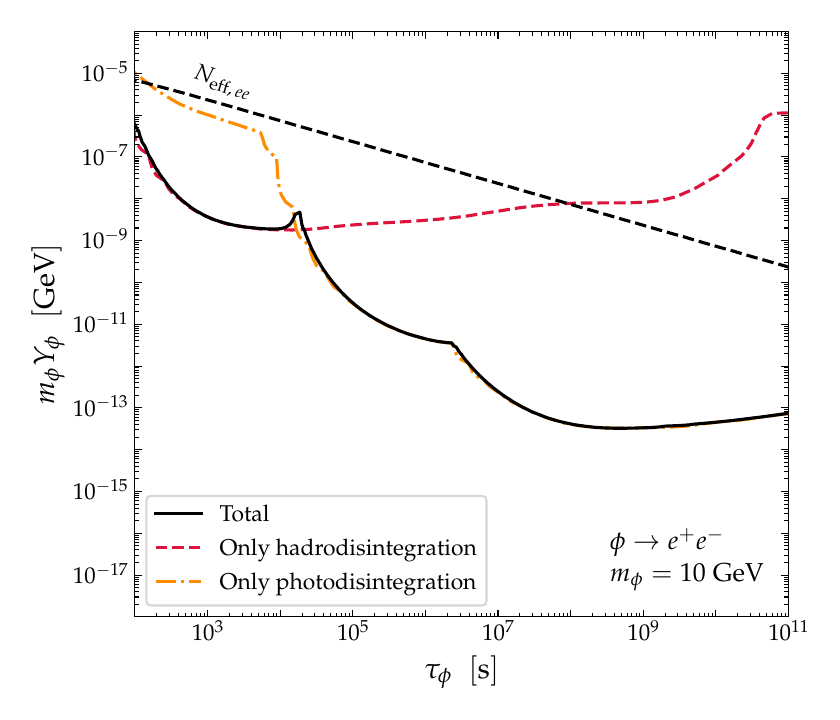}
    \end{subfigure}
    \begin{subfigure}{0.48\linewidth}
      \centering
      \includegraphics[width=0.72\linewidth, trim= 2cm 0 3cm 0cm]{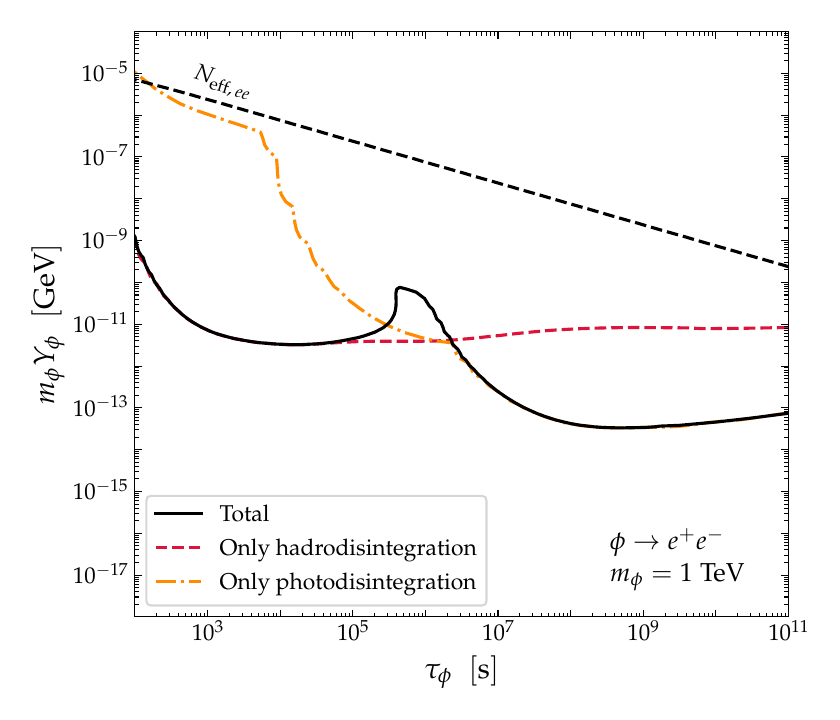}
    \end{subfigure}
    \caption{\textbf{Left}: Decomposition of the photodisintegration and hadrodisintegration bounds for $\phi\to e^+e^-$ and $m_\phi=10\,$GeV. \textbf{Right}: Same but for $m_\phi=1\,$TeV.}
    \label{fig:mphi_comp_pdihdi}
\end{figure}

In figure~\ref{fig:mphi_comp_pdihdi}, we show the effect of accounting separately for hadrodisintegration and photodisintegration, again for the benchmark case of $\phi\to e^+e^-$ and masses of $m_\phi=[10,\,1000]\,$GeV. In the former case, hadrodisintegration constitutes the leading limit up until $\tau_\phi \sim 10^4\,$s, after which photodisintegration dominates. We note the cancellation between the regime of deuterium over- and underproduction.
For 1\,TeV instead, hadrodisintegration bounds are stronger until approximately $10^6\,$s where once again photodisintegration takes over. As both bounds independently become relatively flat in that regime, the cancellation is much more pronounced for this relic mass. Incidentally, the points where photo- and hadrodisintegration cancel each other is close to the point where $^3$He/D becomes the dominant bound, driven by the photodisintegration of $^4$He.

\begin{figure}[!t]
 \centering
    \begin{subfigure}{0.48\linewidth}
      \centering
      \includegraphics[width=0.72\linewidth, trim= 3cm 0 2cm 0cm]{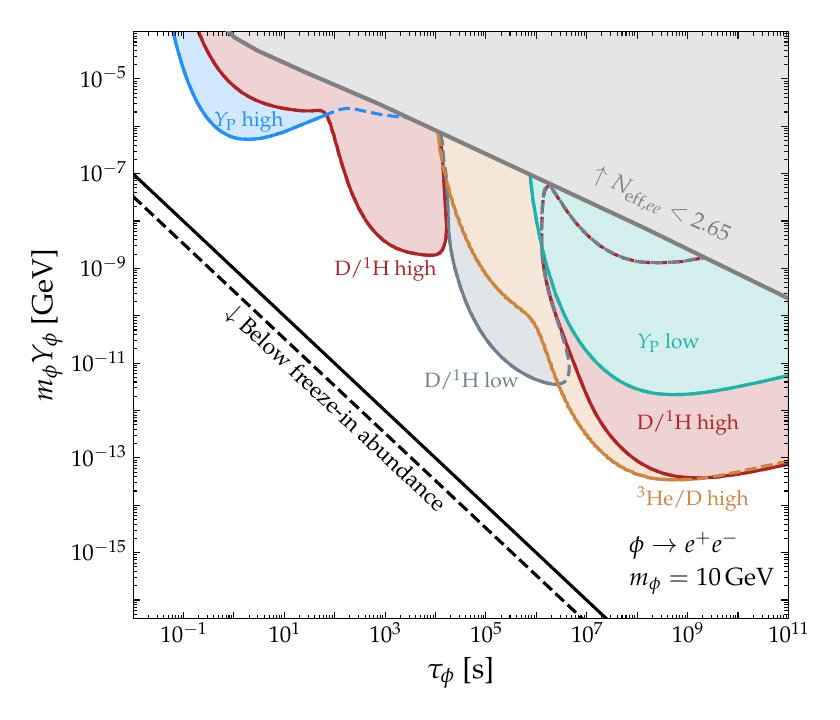}
    \end{subfigure}
    \begin{subfigure}{0.48\linewidth}
      \centering
      \includegraphics[width=0.72\linewidth, trim= 2cm 0 3cm 0cm]{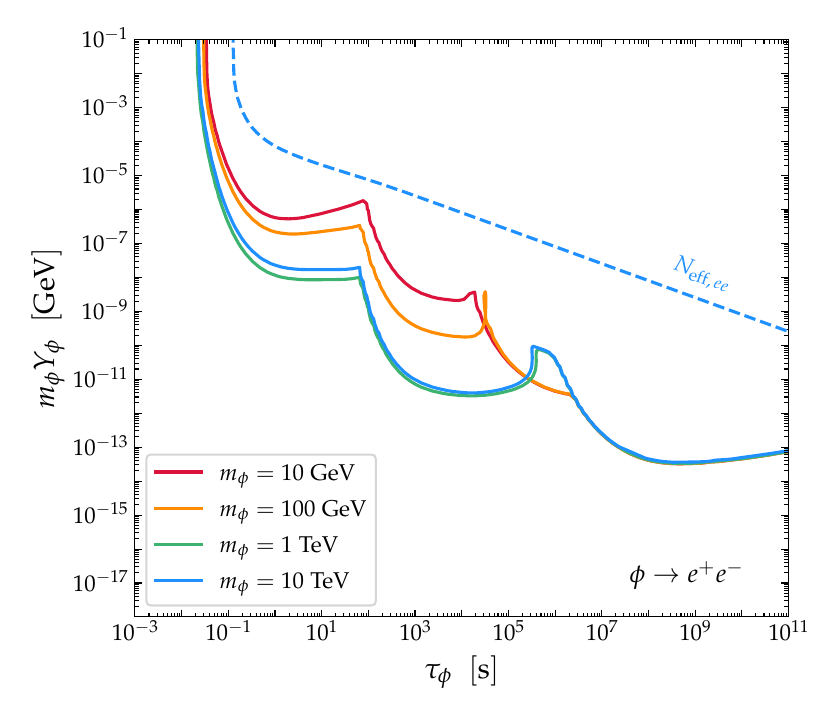}
    \end{subfigure}
    \caption{\textbf{Left:} BBN constraints from the different elements from the decay $\phi\rightarrow e^+e^-$ for $m_\phi$ = 10 GeV. Coloured regions indicate parameter space excluded by bounds on individual elements, as labelled. The corresponding $\Neff$ constraint is shown in grey. We also indicate the region where the irreducible freeze-in abundance is above the assumed value of $m_\phi \mathpazoletters{Y}_\phi$ and can therefore not be consistently reached in standard cosmology. We show it for both a vector relic (solid) and a scalar relic (dashed). \textbf{Right:} Overall BBN constraints for different masses $m_\phi$ (solid, different colours). In addition, we also show the corresponding $\Neff$ constraint for $m_\phi = 10\,\mathrm{TeV}$ (blue dashed).}
    \label{fig:mphi_ee}
\end{figure}

With this detailed understanding of the individual contributions, we show the resulting constraints in the $m_\phi \mathpazoletters{Y}_\phi - \tau_\phi$ plane in figure\,\ref{fig:mphi_ee}. 
To emphasise the findings of the last two paragraphs, we highlight the contributions from the different elements that drive the constraints in the case $m_\phi = 10\,\mathrm{GeV}$ (left figure). For shorter lifetimes, approximately $\tau_\phi \lesssim 10^2$\,s, the limits come from the efficient neutron-proton interconversions driven by the injected hadrons, as discussed in \ref{sec:interconversion}, and are dominated by a too high value of $Y_\mathrm{P}$ which is ``inherited'' from the increased neutron abundance compared to the SM scenario. For $\tau_\phi \gtrsim 10^2$\,s, injected neutrons are no longer immediately stopped in the plasma and can start to disintegrate $^4$He efficiently, leading to an overproduction of D/$^1$H. For lifetimes $\tau_\phi \gtrsim 10^4\,\mathrm{s}$, photodisintegration processes become important, as the universal spectrum contains photons with high enough energy to deplete deuterium. At later times, even higher energies are possible, allowing photons to effectively destroy the more abundant $^4$He and to drive the overproduction of D and $^3$He. 

Not all regions in the $(m_\phi \mathpazoletters{Y}_\phi,\tau_\phi)$ plane are physically realisable. 
Even if the primordial abundance of $\phi$ is assumed to vanish, inverse decays 
and related freeze-in processes from the thermal bath generically produce a minimal 
irreducible abundance. For decays into electrons, $\phi \to e^+e^-$, 
the inverse decay process $e^+e^- \to \phi$ yields approximately~\cite{Hall:2009bx}

\begin{equation}
\mathpazoletters{Y}_\phi^{\rm FI}
\simeq
\frac{135\, g_\phi}{8\pi^3\,1.66\,g_{*s}\sqrt{g_*}}
\frac{M_{\rm Pl}}{m_\phi^2 \tau_\phi}\, ,
\end{equation}

up to $\mathcal{O}(1)$ corrections from quantum statistics and the temperature 
dependence of $g_*$. In fact, for a fixed lifetime $\tau_\phi$, 
the freeze-in abundance is largely independent 
of the specific decay channel. Although inverse decays into e.g.\ coloured particles 
receive additional colour multiplicities, these factors are already encoded in 
the total decay width $\Gamma_\phi = \tau_\phi^{-1}$. Consequently, the minimal 
freeze-in abundance is approximately determined only by $m_\phi$ and $\tau_\phi$, 
up to small corrections which will also apply to other cases discussed below. The production is dominated at temperatures 
$T \sim m_\phi$, and therefore $g_*$ and $g_{*s}$ should be evaluated around 
that epoch. We indicate the corresponding freeze-in floor in the 
$(m_\phi \mathpazoletters{Y}_\phi,\tau_\phi)$ plane; regions below this line are not populated in standard cosmology.

In the right panel of figure\,\ref{fig:mphi_ee}, we show the resulting constraints for masses between $10\,$GeV and $10\,$TeV.\footnote{This range is motivated by the breakdown of \texttt{PYTHIA} at too low and too high masses. Furthermore, we expect the bounds to rapidly vanish with the mass approaching the threshold for hadron production.} 
We see that for heavier relics, above the electroweak scale, the structure of the bumps is mildly modified. The behaviour is the same as the one we found in \cite{Bianco:2025boy}: once the $Z$-boson production is no longer kinematically suppressed, hadron production from electrons and positrons increases significantly (cf.\ figure~\ref{fig:N_had}). Since $^4$He is much more abundant than D, its destruction and the ensuing deuterium production are able to outperform the direct photodisintegration of deuterium more and more, shifting the transition towards larger lifetimes. Ultimately, the $^3$He constraints become dominant, before any region of parameter space can be excluded solely by deuterium underproduction.

A general feature of EM injections is that the bounds tend to become independent of the mass of the relic towards large lifetimes (assuming the same energy density). This is due to the universal spectrum, which is only sensitive to the total injected energy and not the injection spectrum. This explains the convergence of the bounds towards a universal constraint for large lifetimes. Since we expect the amount of EM energy and the number of hadrons not to change too much for very high energies, the other limits will also become very similar. In the hadronically dominated regime, we make the interesting observation that the bounds tend to decrease again for $m_\phi\gtrsim 1\,$TeV. Especially in the interconversion region, this can be explained using figure~\ref{fig:N_had}: at low masses, the growth of the number of hadrons is typically larger than linear, especially when crossing the EW scale. However, eventually this growth decreases and becomes less than linear. In that case, for fixed $m_\phi \mathpazoletters{Y}_\phi$, we find a \emph{smaller} number of hadrons injected, thus decreasing the limits (mildly). For the region where hadrodisintegration is most relevant, there is a more complicated interplay of effects since both the number of baryons and their energy influence the bounds.
Overall, we find a net reduction in the constraints also in this regime.

\begin{figure}[!t]
 \centering
     \begin{subfigure}{0.48\linewidth}
      \centering
      \includegraphics[width=0.72\linewidth, trim= 3cm 0 2cm 0cm]{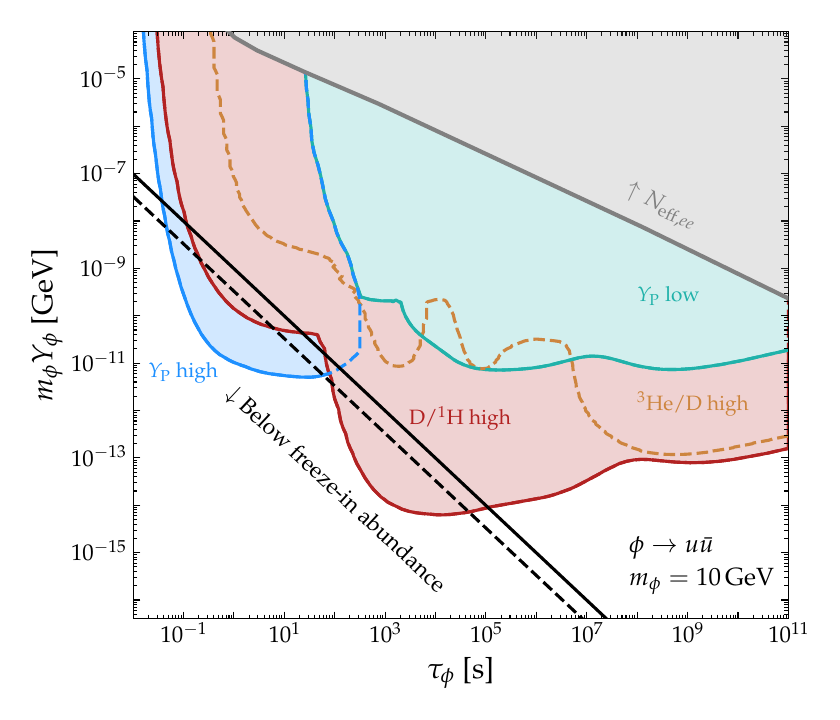}
    \end{subfigure}
    \begin{subfigure}{0.48\linewidth}
      \centering
      \includegraphics[width=0.72\linewidth, trim= 2cm 0 3cm 0cm]{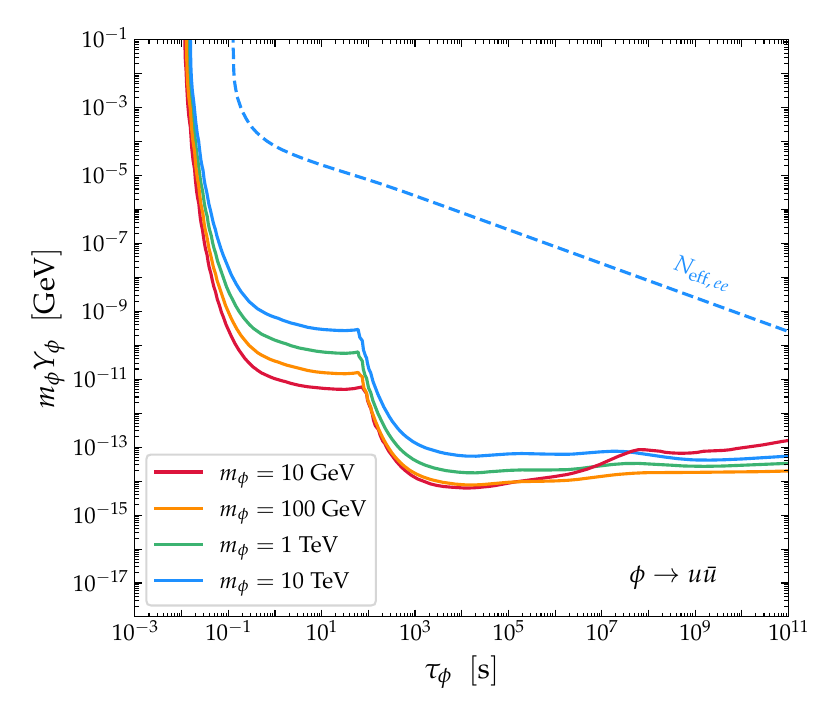}
    \end{subfigure}
    \caption{Same as figure\,\ref{fig:mphi_ee} but for $\phi\rightarrow u\bar{u}$. Note that the $N_\text{eff}$ limit is only indicative as described in the text.}
    \label{fig:mphi_uu}
\end{figure}

\subsection{Decay into  $u\bar{u}$ }

Let us now discuss the case of hadronic injections. We focus on $u\bar{u}$ injections, which we present in figure~\ref{fig:mphi_uu}. In this case, the bump structure among different masses is mostly identical, and generally, the bounds become weaker for larger masses.
Only for the $10\,$GeV case, we observe a somewhat different shape of the limit. Compared to the case of $e^+ e^-$, the bounds are significantly stronger for lifetimes where interconversions and hadrodisintegration set the leading limits due to the larger injected hadronic energy for $m_\phi$ below the electroweak scale. In fact, the limits are so strong that even the irreducible freeze-in abundance is fully excluded for lifetimes 0.1s $\lesssim \tau_\phi \lesssim 10^5 $s for $m_\phi = 10\,$GeV.\footnote{This conclusion could be avoided with very low reheat temperatures.}

For low lifetimes, the constraints are again driven by an overabundance of $^4$He, due to the efficient interconversion of protons into neutrons. As expected these bounds are significantly stronger than for the $e^+e^-$ decay channel, as the hadronic component is unsuppressed (cf.~figure~\ref{fig:N_had}). Also in this case, for $\tau_\phi > 10^2$\,s, the injected nucleons disintegrate $^4$He, leading to an overproduction of D/$^1$H, as can be seen in figure\,\ref{fig:mphi_uu}.

\subsection{Decay into $\gamma\gamma$}\label{sec:gammagamma}
\begin{figure}
    \centering
    \includegraphics[width=0.7\linewidth]{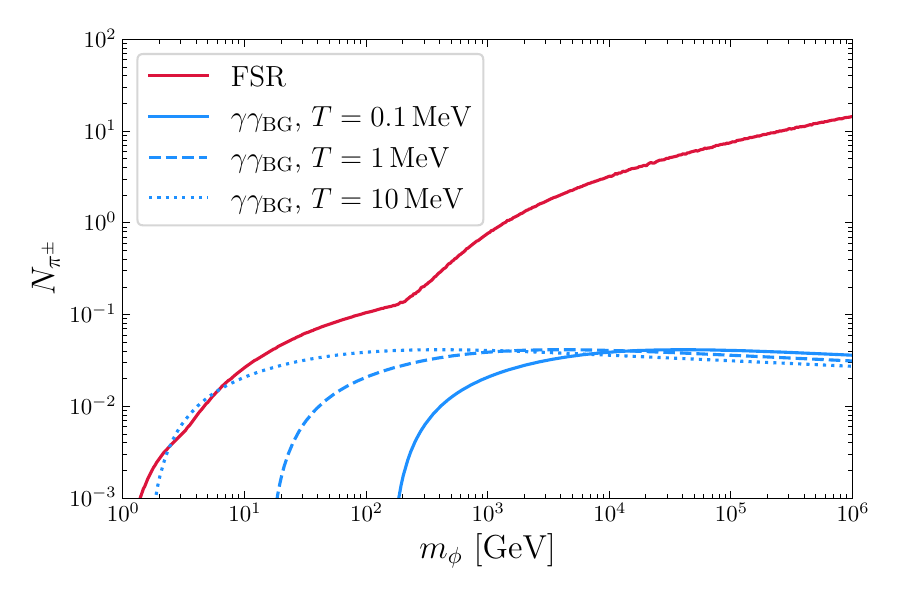}
    \caption{Comparison of the average number of pions per double photon injection from FSR (red) and from scattering with the background photons (blue) for different injection temperatures.}
    \label{fig:ggpi}
\end{figure}

Before we start to discuss the other SM injections in a briefer format, let us investigate the injection of photons with more scrutiny, as the scattering of high-energy photons on the background photons can lead to $\gamma+\gamma_\text{BG}\to \pi^+\pi^-$, which sources pion-driven interconversions as studied in  ref.~\cite{Kawasaki:2005}. This process could appear relevant as the photon is the only background particle (except neutrinos) that still has a large thermal population at temperatures $\sim 100\,$keV, where interconversions dominate the resulting constraints.
To evaluate whether this process could ever be relevant, let us 
investigate the corresponding pion-production cross section. Following~\cite{Kawasaki:2005}, we only consider the primary scattering of an injected pair of photons, which implies that the average number of pions per injected photon pair is two times the ratio of the thermally averaged pion-production cross section and the total cross section
\begin{align}
    n_\pi^{\gamma\gamma}\simeq 2 \frac{\langle \sigma v\rangle_{\gamma\gamma\to\pi^+\pi^-}}{\langle\sigma v\rangle_{\gamma\gamma,\text{tot}}}\simeq2 \frac{\langle \sigma v\rangle_{\gamma\gamma\to\pi^+\pi^-}}{\langle \sigma v\rangle_{\gamma\gamma\to e^+ e^-}+\langle\sigma v\rangle_{\gamma\gamma\to\mu^+\mu^-}+\langle\sigma v\rangle_{\gamma\gamma\to\pi^+\pi^-}}\;.
\end{align}
We can evaluate this ratio numerically using results from e.g.~\cite{Brodsky:1971ud}. We find that the pion cross section is significantly smaller than the leptonic ones. Overall, the number of pions in these processes always remains below 0.05. When comparing this to the injection due to FSR (see figure~\ref{fig:ggpi}), we see that it is clearly subdominant in the interesting temperature range, i.e.~briefly before the fusion processes start. 

We observe that there is an exponential (kinematic) cutoff which drops with temperature as expected. 
At the temperatures of relevance, we can thus conclude that pions from this source would never be relevant to our phenomenology, and we can neglect this process.
Having clarified the case of photon injections, we show the constraints derived in our framework in figure~\ref{fig:gammagamma+gg} (left). The bounds are found somewhere between the hadronic and EM benchmark, as would also be expected from figure~\ref{fig:N_had}. 
We see the typical three-bump structure we have encountered before, where interconversion dominance transitions into hadrodisintegration of $^4$He, which is eventually overcome by photodisintegration of $^4$He.

\subsection{Decay into other channels}

In this section, we will briefly summarise the remaining SM decay channels, pointing out the difference to the two benchmark cases when necessary.

\begin{figure}[!t]
 \centering
    \begin{subfigure}{0.48\linewidth}
      \centering
      \includegraphics[width=0.72\linewidth, trim= 3cm 0 2cm 0cm]{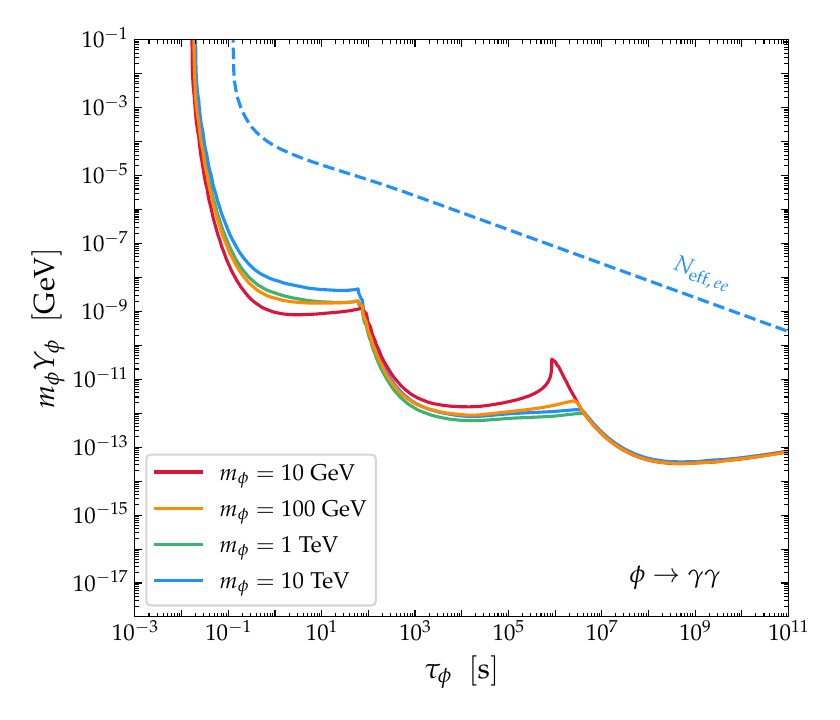}
    \end{subfigure}
    \begin{subfigure}{0.48\linewidth}
      \centering
      \includegraphics[width=0.72\linewidth, trim= 2cm 0 3cm 0cm]{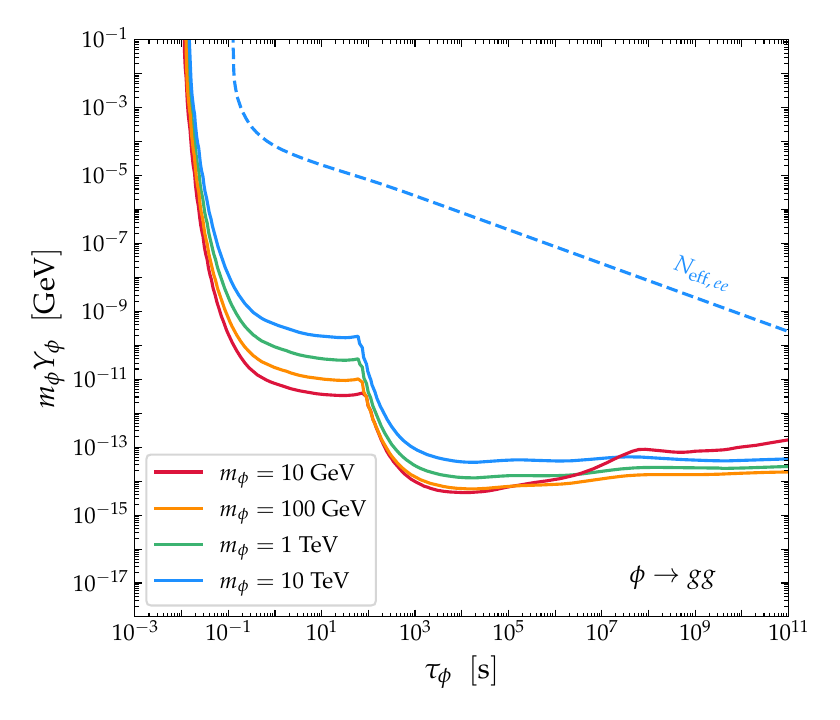}
    \end{subfigure}
    \caption{\textbf{Left:} Same as figure~\ref{fig:mphi_ee} but for  $\phi\to \gamma\gamma$. \textbf{Right}: Same but for $\phi\to gg$.}
    \label{fig:gammagamma+gg}
\end{figure}

Let us start off with an injection of gluons as shown in figure~\ref{fig:gammagamma+gg} (right). The bounds are almost identical to the ones for $u\bar{u}$ injections. In fact, we find that all channels with significant branching ratios into quarks or gluons will lead to very similar bounds.

\begin{figure}[!t]
 \centering
    \begin{subfigure}{0.48\linewidth}
      \centering
      \includegraphics[width=0.72\linewidth, trim= 3cm 0 2cm 0cm]{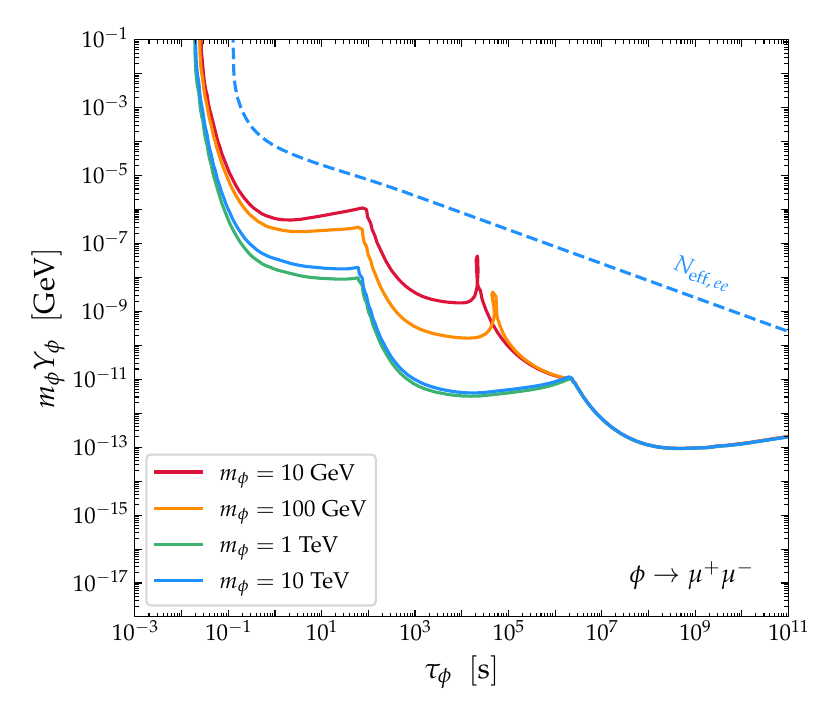}
    \end{subfigure}
    \begin{subfigure}{0.48\linewidth}
      \centering
      \includegraphics[width=0.72\linewidth, trim= 2cm 0 3cm 0cm]{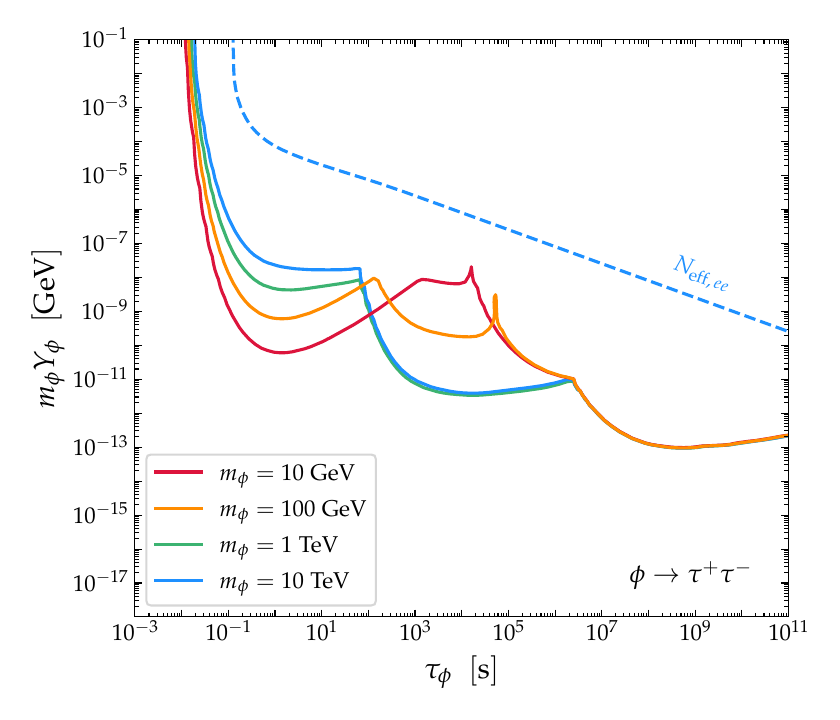}
    \end{subfigure}
    \caption{\textbf{Left:} Same as figure\,\ref{fig:mphi_ee} but for $\phi\rightarrow \mu^+\mu^-$. \textbf{Right}: Same but for $\phi\to \tau^+\tau^-$.}
    \label{fig:mumu+tautau}
\end{figure}

For the other charged leptons shown in figure~\ref{fig:mumu+tautau}, 
it is not directly clear if all relevant effects are captured in our approach. In particular, for small relic masses where FSR is subleading, the additional neutrinos associated with the leptonic decay chains might have an effect that is unaccounted for in this work. 
Ref.~\cite{Chang:2024mvg} investigated the neutrino-driven interconversions of protons and neutrons via $\bar{\nu}_e/\nu_e+p/n\to n/p+e^\pm$. 
Comparing our limit plots to 
their figure~2 we observe that their bounds could only potentially be relevant for muon decays. However, at the times relevant for interconversions, muons quickly thermalise and will decay at rest, so that
the neutrinos in the final state will have energies well below 100 MeV. Comparing the figures then makes it evident that our limits are about an order-of-magnitude stronger, and neutrino-driven interconversions are not relevant for the regions of parameter space we consider.

At large lifetimes, the $\mu^+\mu^-$ bounds are mildly weaker than the $e^+e^-$ ones, since these limits are dominated by photodisintegration and the EM injections are suppressed by the increase of neutrinos in the final state (cf.\ figure~\ref{fig:N_had}). For a $\tau^+\tau^-$ injection at relic masses beyond $1\,$TeV, the bounds driven by hadrons become universal, like for the other leptons. However, at lower relic masses, the large pionic branching ratio of the $\tau$ decay gives the bounds a ``head start'' compared to the other leptons. With the increasing relevance of FSR, we then approach the universal bounds. Since the overall EM fraction is the smallest out of all the charged leptons, so are the bounds for the largest lifetimes.

One particularity of the $\tau$ injection is the gap between interconversions and photodisintegration for $m_\phi=10\,$GeV. The enormously large pion injection leads to very strong interconversion bounds for low lifetime, whereas the photodisintegration only picks up at $\tau_\phi\sim10^4\,$s. Only in a small region of the parameter space, between approximately $\tau_\phi\sim10^3\,$s and $\tau_\phi\sim10^4\,$s, hadrodisintegration governs the limits with deuterium overproduction, eventually leading to a small cancellation in the region where photodisintegration becomes active and starts to destroy deuterium. In fact, the baryon injection is identical for all charged leptons, but the large interconversion effects can barely be overcome by the hadrodisintegration bounds, unlike in the case of $e^\pm$ and $\mu^\pm$.\footnote{With such a large discrepancy between pion and nucleon injection, pionic hadrodisintegration could become dominant. However, as the disintegration of $^4$He is cumulative from all sources, we conservatively neglect it here.}

\begin{figure}[!t]
 \centering
    \begin{subfigure}{0.48\linewidth}
      \centering
      \includegraphics[width=0.72\linewidth, trim= 3cm 0 2cm 0cm]{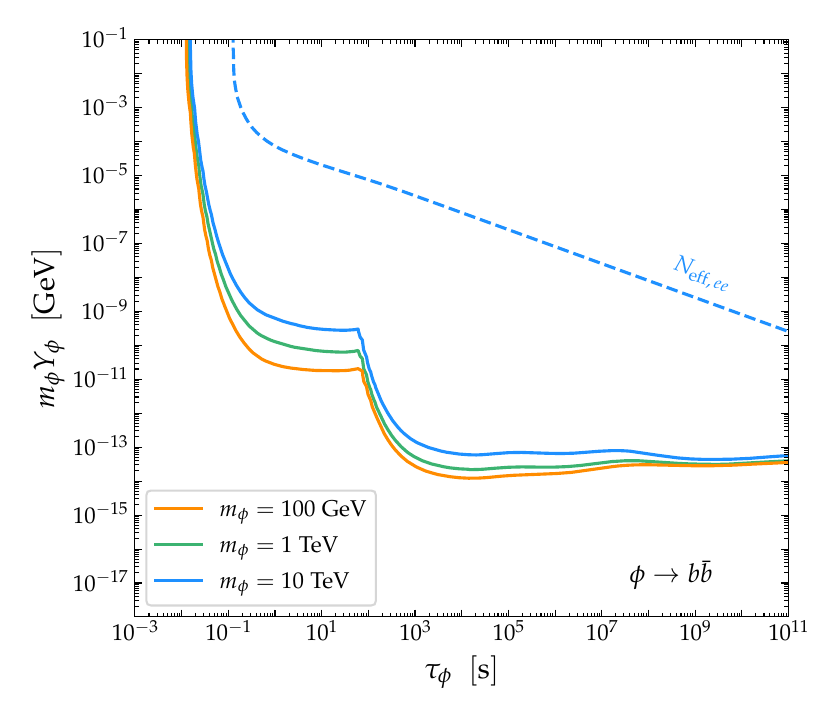}
    \end{subfigure}
    \begin{subfigure}{0.48\linewidth}
      \centering
      \includegraphics[width=0.72\linewidth, trim= 2cm 0 3cm 0cm]{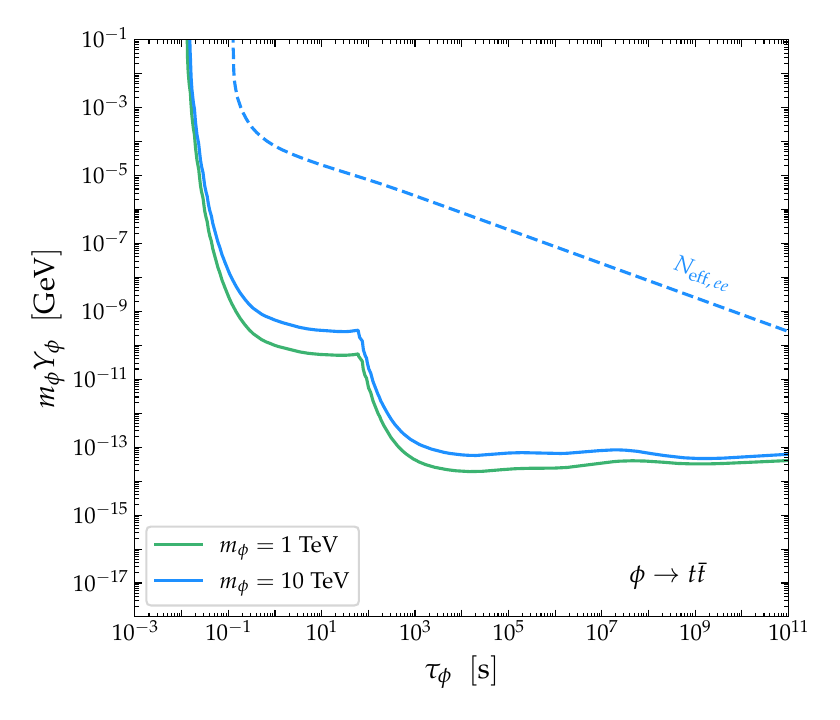}
    \end{subfigure}
    \caption{\textbf{Left:} Same as figure~\ref{fig:mphi_ee} but for  $\phi\to b\bar{b}$. \textbf{Right}: Same but for $\phi\to t\bar{t}$.}
    \label{fig:bb+tt}
\end{figure}

We have explicitly checked that the exact type of quark barely plays a role in our setup. The only major difference is the mass threshold around which the bounds deviate from universality. In particular, the shower in \texttt{PYTHIA} is unreliable too close to the mass threshold. As one example, we show $b\bar{b}$ in figure~\ref{fig:bb+tt} (left). The differences to $u\bar{u}$ are hardly visible as we refrain from showing the $10\,$GeV bound where we cannot trust the results. However, we find deviations also for higher masses, albeit at a much smaller level.
We make the same conclusion for $t$ quarks, where we only show the bounds starting from $1\,$TeV (figure~\ref{fig:bb+tt}, right).

\begin{figure}[!t]
 \centering
    \begin{subfigure}{0.48\linewidth}
      \centering
      \includegraphics[width=0.72\linewidth, trim= 3cm 0 2cm 0cm]{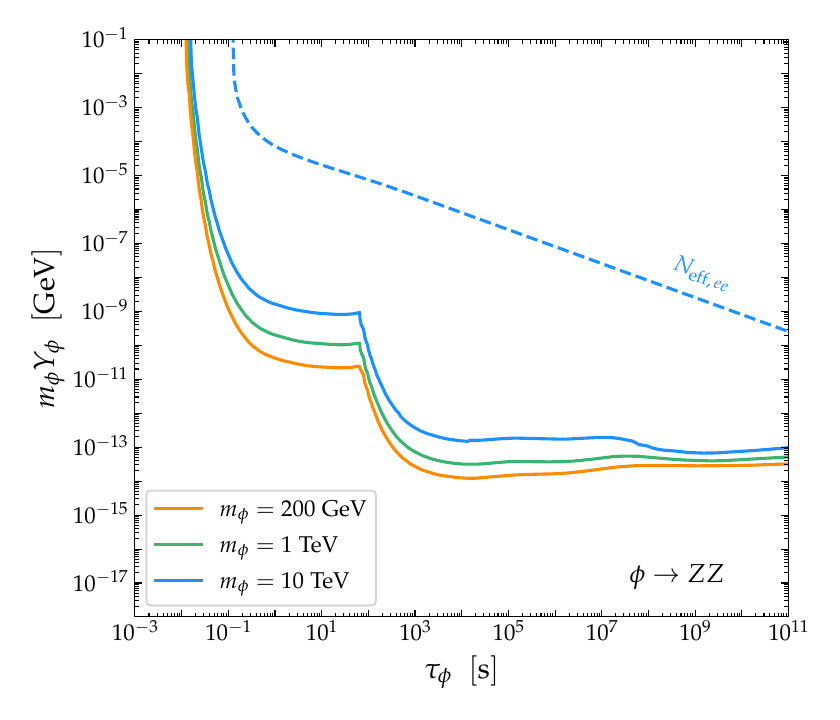}
    \end{subfigure}
    \begin{subfigure}{0.48\linewidth}
      \centering
      \includegraphics[width=0.72\linewidth, trim= 2cm 0 3cm 0cm]{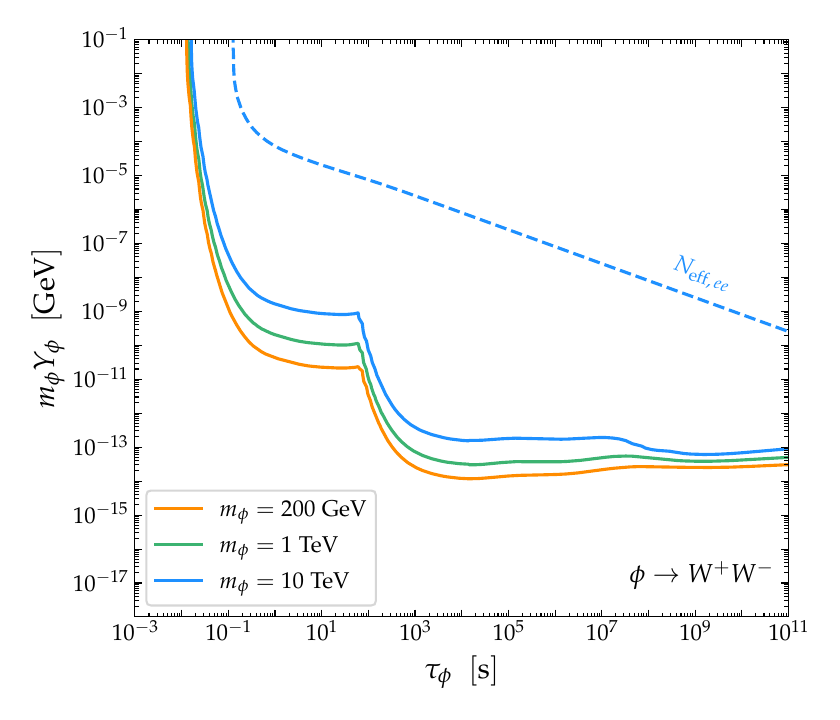}
    \end{subfigure}
    \caption{\textbf{Left:} Same as figure~\ref{fig:mphi_ee} but for  $\phi\to ZZ$. \textbf{Right}: Same but for $\phi\to W^+ W^-$.}
    \label{fig:ZZ+WW}
\end{figure}

\begin{figure}[!t]
    \centering
    \includegraphics[width=83mm,scale=0.5]{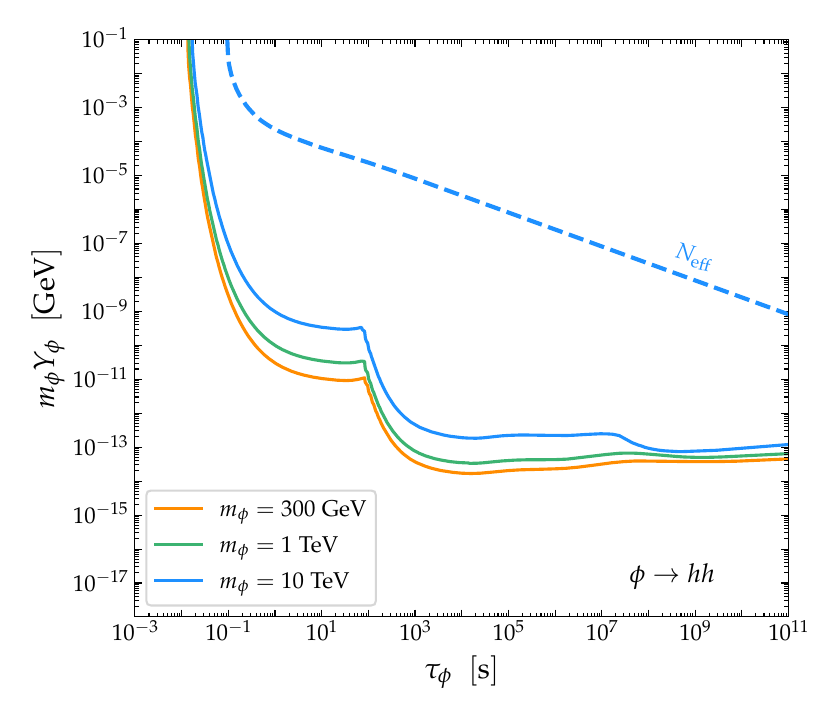}
    \caption{Same as figure\,\ref{fig:mphi_ee} but for $\phi\rightarrow hh$.}
    \label{fig:mphi_hh}
\end{figure}

For the massive vector bosons, we find bounds that are typically dominated by the hadronic injections, as is expected from their decay channels. We show the results for $ZZ$ and $W^+W^-$ injections in figure~\ref{fig:ZZ+WW} (left and right, respectively). As \texttt{PYTHIA} is expected to handle the decays of the massive vectors without problems, we also show the results here closer to the mass threshold.

Finally, let us also briefly discuss the injection of a pair of Higgs bosons as shown in figure~\ref{fig:mphi_hh}. Expectedly, also here the bounds are mostly hadronic due to the large branching ratio of the Higgs into the heavy quarks.

\subsection{Comparison with previous literature results}

\begin{figure}[!t]
    \centering
    \includegraphics[width=83mm,scale=0.5]{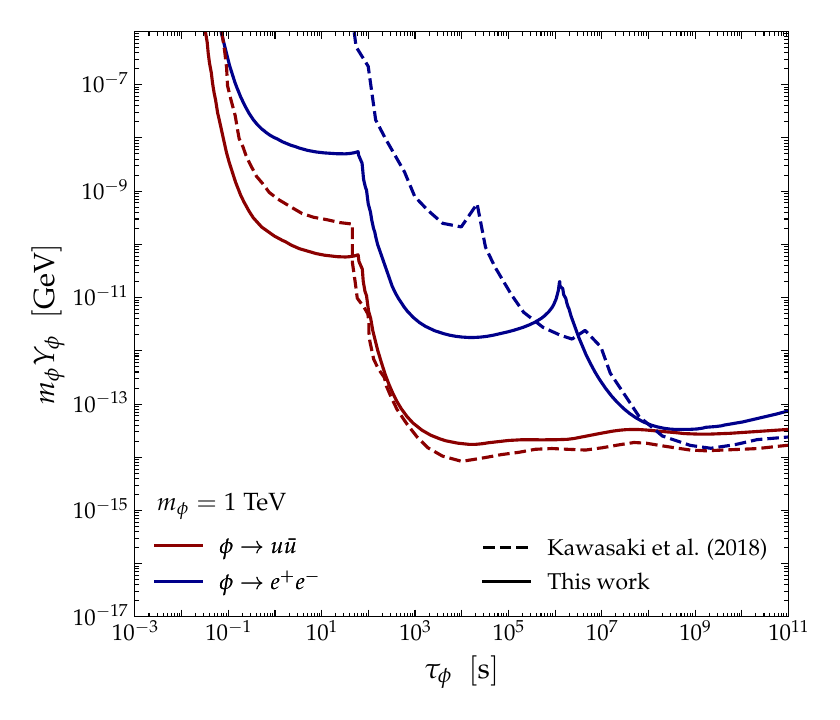}
    \caption{Comparison of our findings (solid line) to previous literature \cite{Kawasaki:2018} (dashed line). In dark red and dark blue, we show their two benchmark injections of $u\bar{u}$ and $e^+e^-$, respectively.}
    \label{fig:mphi_comp}
\end{figure}

Before we conclude our work, we want to discuss the differences we find with the previous literature for the two benchmark injections $e^+e^-$ and $u\bar{u}$. We show these differences in figure~\ref{fig:mphi_comp} as solid (this work) and dashed (ref.~\cite{Kawasaki:2018}).\footnote{Ref.~\cite{Kawasaki:2018} cut their data at $m_\phi \mathpazoletters{Y}_\phi=10^{-6}\,$GeV which is why we only show this range.} For long lifetimes, we find a qualitative agreement, but still, our results diverge  from the literature for both benchmarks. For the electrons, this is most likely due to the different modelling of photodisintegration in the different works, which makes it hard to pinpoint the exact difference. Also, the hadrodisintegration yields different results since our formalism is different from the original proposal (for a discussion we refer to appendix~\ref{app:hadrodis_abbn} and ref.~\cite{Bianco:2025boy}). We should emphasise that there are even some changes here beyond the treatment of the hadrodisintegration: since the publication of \cite{Kawasaki:2018}, new measurements for the light-element abundances and fusion rates became available. Furthermore, we have checked that the new $^4$He measurement improves the interconversion bounds significantly.

We observe that at low lifetimes, our bounds are much stronger than the previous results. We assume that this is due to the new interconversion rates (especially for $u\bar{u}$) and the use of $\texttt{PYTHIA}$ for the FSR (especially for $e^+e^-$, where interconversions do not seem to have been considered at all in \cite{Kawasaki:2018}). Especially the latter finding leads to clear deviations wherever hadrodisintegration is relevant for $e^+e^-$. 
For a detailed discussion, we refer to section~\ref{sec:ee_inj} and in particular figure~\ref{fig:mphi_comp_inter}.

\section{Conclusions}
\label{sec:conclusion}

In this work, we have presented updated and improved BBN constraints on heavy decaying relics, covering a comprehensive set of SM two-body decay channels. Our analysis builds on and substantially extends the framework of earlier studies through a series of important refinements, which we summarise here.

We have implemented new cross sections for $p \leftrightarrow n$ interconversion processes \cite{Jung:2025dyo}, incorporating kaon-induced channels that were previously neglected in parts of the literature. In conjunction with this, we have implemented a detailed treatment of hadronic abundances within the dynamical-equilibrium approximation, carefully tracking the full composition of the hadronic injection products. This allows for a more faithful description of the early-time interconversion epoch and its interplay with standard BBN.

A central methodological achievement of this work is the unification of hadronic and electromagnetic injection effects into a single, self-consistent pipeline. Hadronic interconversions and hadrodisintegration during the BBN epoch are handled within \texttt{AlterAlterBBN}, while late-time hadrodisintegration and photodisintegration are computed with \texttt{ACROPOLIS}. The two stages are carefully interfaced, ensuring that the handoff between regimes is physically consistent and free of double-counting disintegration processes. This integrated approach improves upon our earlier analysis \cite{Bianco:2025boy} (focusing on neutrinos), where we had to restrict the lifetime range to only cover hadrodisintegration after BBN has concluded. This innovation enables a uniform treatment across the full range of relic lifetimes, masses, and abundances considered here.

On the observational side, we have incorporated up-to-date primordial-abundance measurements, most notably the revised $^4$He abundance from ref.~\cite{Yeh:2026pil}, alongside updated nuclear reaction rates for standard BBN. The tightened helium constraint in particular has a significant impact on the resulting bounds, especially at short lifetimes where $^4$He overproduction  via interconversions is the dominant signal. 
Furthermore, we have performed a systematic study of final-state radiation and hadronisation using \textsc{Pythia\,8}, treating these effects consistently across all decay channels. This provides us with all relevant parameters to describe interconversion and disintegration processes, in particular the fraction of EM energy, the number of injected hadrons, and the kinetic energy of the baryonic injections, relevant for hadrodisintegration.

Finally, we have corrected an error in the electromagnetic energy-loss curve that we had already noted in~\cite{Bianco:2025boy}, leading to artificially enhanced
hadrodisintegration rates by energetic protons. This correction reduces the predicted disintegration yields at intermediate and long lifetimes and softens constraints in the corresponding parameter regions. We have also refined the treatment of elastic nucleon scattering in the hadronic cascade, following ~\cite{Cugnon:1996kh,Falter:2004uc,Larionov:2025wce}, which even further reduces the disintegration power.

Taken together, these improvements constitute a state-of-the-art treatment of BBN constraints on decaying relics, and yield updated exclusion regions across a wide range of masses $m_\phi$ and lifetimes $\tau_\phi$ for each considered decay channel. The results are presented in section~\ref{sec:results}, where we also provide comparisons with previous literature. We expect our analysis to serve as a useful reference for constraining specific BSM models, whose particle content includes heavy, long-lived states decaying to SM final states. In particular, we find that constraints on colored particles like $u\bar{u}$ can be strong enough to even rule out abundances as small as the irreducible freeze-in abundance. Extensions of this framework via the inclusion of neutrino injection channels and a more careful treatment of lighter, MeV-scale relics are left to future work.

\acknowledgments
We thank the authors of \cite{Jung:2025dyo} for helping us out with questions on the new interconversion cross sections and in particular for sharing their updated results with us.
SB and JF thank Nicolas Grimbaum Yamamoto for many discussions throughout different stages of this project.
JF thanks CERN TH for their hospitality and in particular Maksym Ovchynnikov and Miguel Escudero for useful discussions.
This work is funded by the Deutsche Forschungsgemeinschaft (DFG) through Germany's Excellence Strategy --- EXC 2121 ``Quantum Universe'' --- 390833306.
JF is supported by an ERC StG grant (“AstroDarkLS”, Grant No. 101117510).
The work of MH is supported by the Belgian IISN convention 4.4503.15 as well as by the Brussels laboratory of the Universe - BLU-ULB.

\clearpage

\appendix

\section{Updated interconversion cross sections}
\label{app:new_xsec}

\begin{figure}[!t]
 \centering
    \begin{subfigure}{0.49\linewidth}
    \centering
    \includegraphics[width=\linewidth]{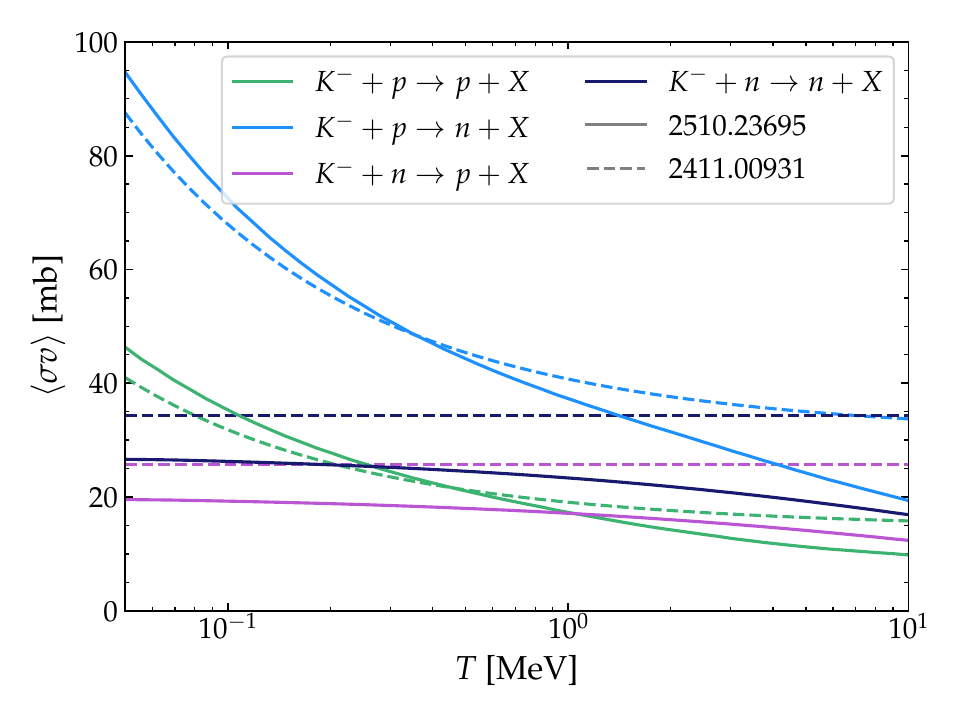}
    \end{subfigure}
    \begin{subfigure}{0.49\linewidth}
      \centering
      \includegraphics[width=\linewidth]{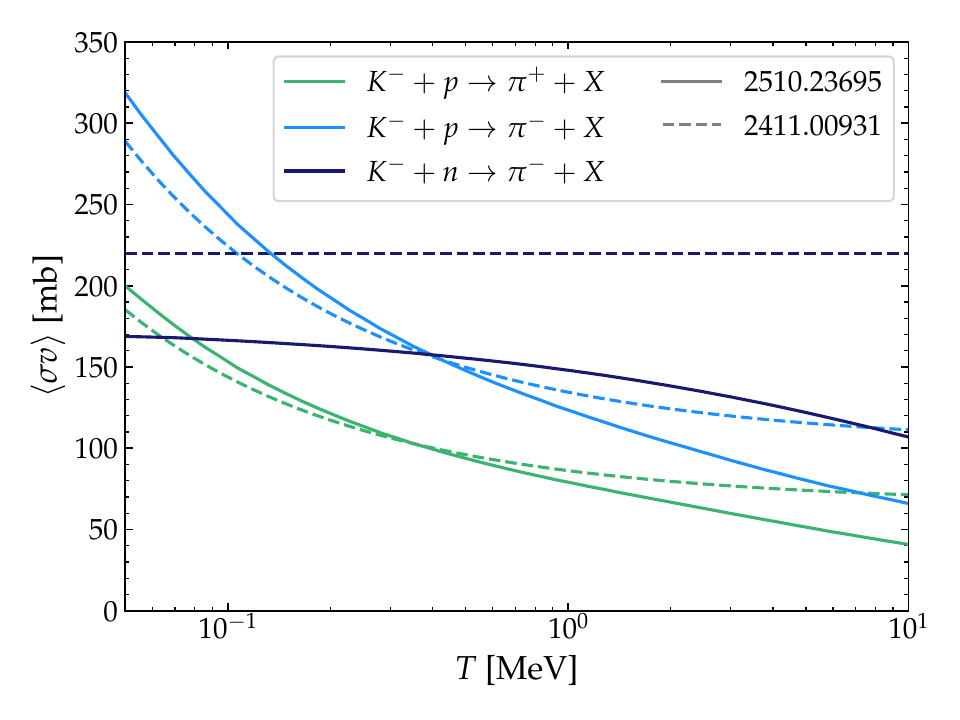}
    \end{subfigure}
        \begin{subfigure}{0.49\linewidth}
      \centering
      \includegraphics[width=\linewidth]{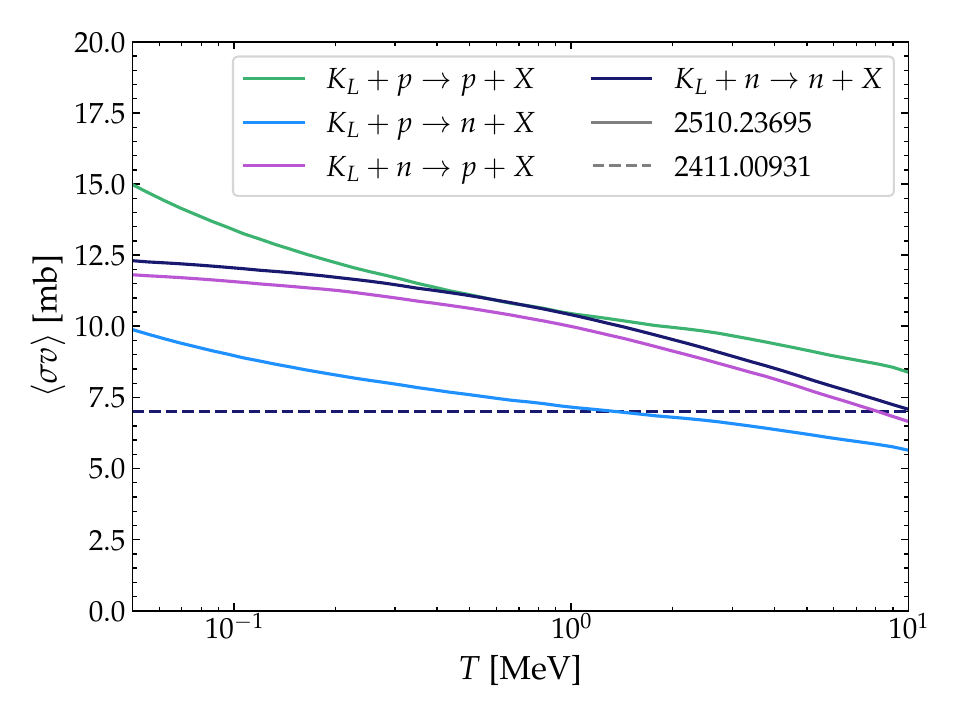}
    \end{subfigure}
    \begin{subfigure}{0.49\linewidth}
      \centering
      \includegraphics[width=\linewidth]{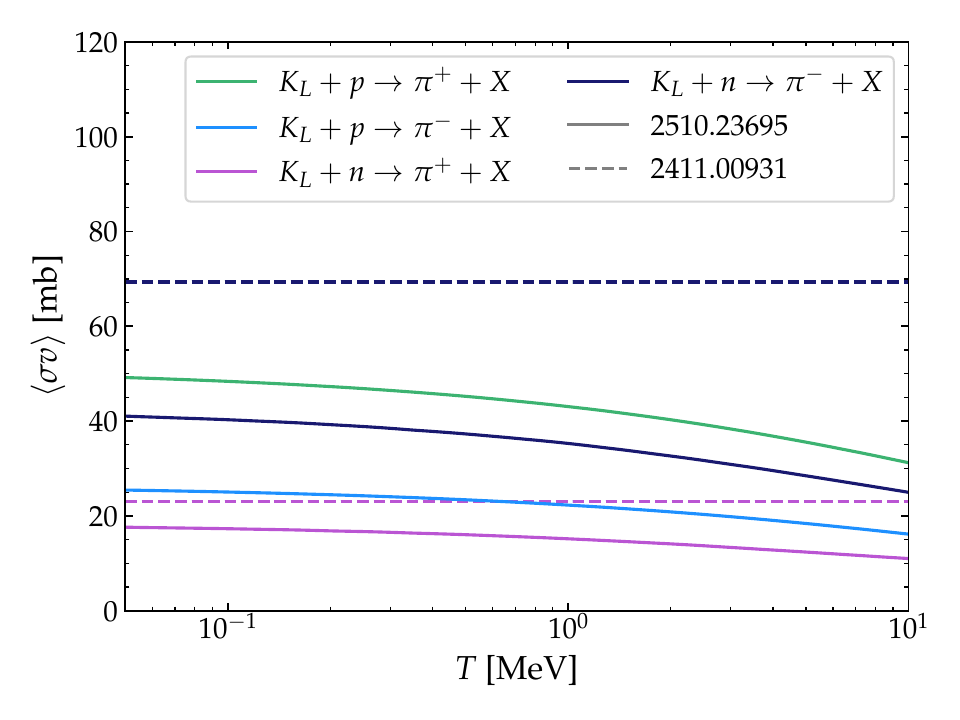}
    \end{subfigure}   \begin{subfigure}{0.49\linewidth}
      \centering
      \includegraphics[width=\linewidth]{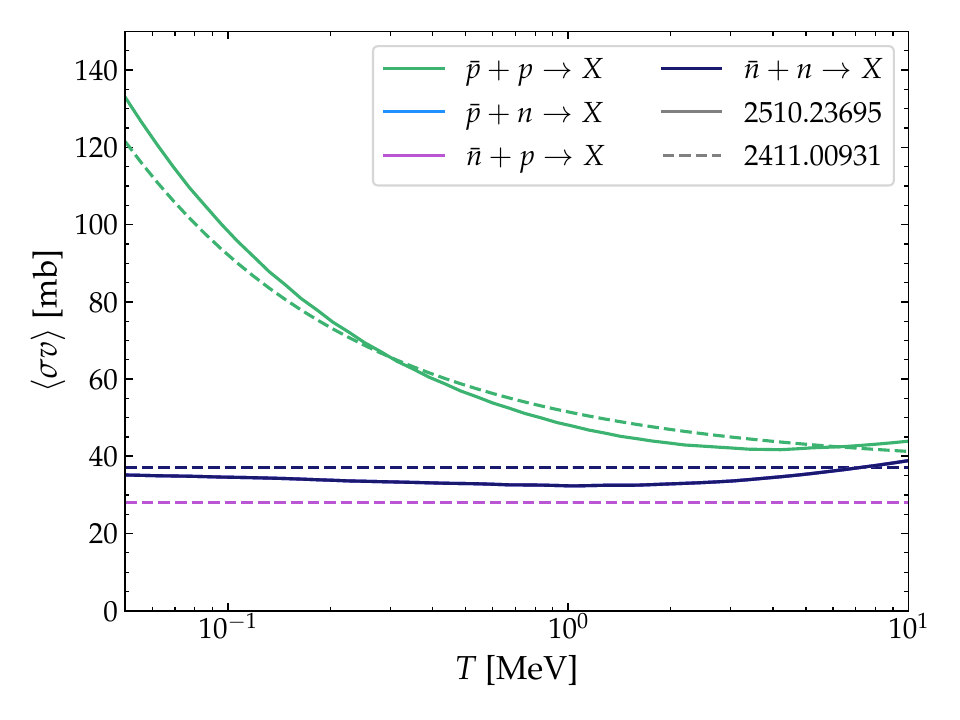}
    \end{subfigure}
    \begin{subfigure}{0.49\linewidth}
      \centering
      \includegraphics[width=\linewidth]{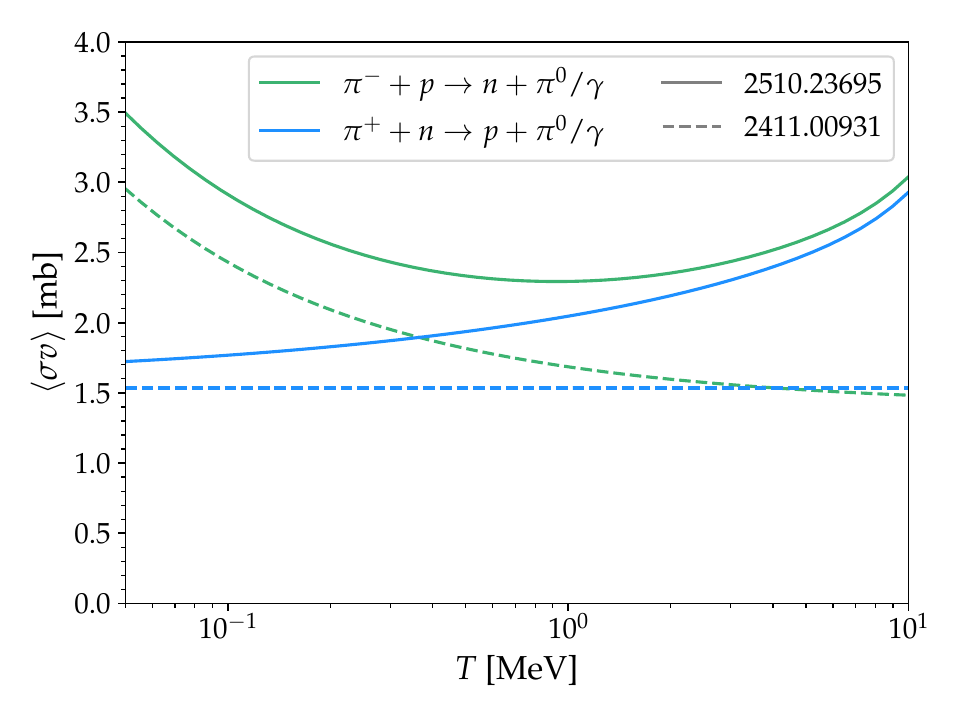}
      \end{subfigure}
\caption{Comparison of the previous literature cross sections (dashed lines) with the updated results from \cite{Jung:2025dyo} (full lines). Note the different scales on the $y$-axes.
\textbf{Top:} Updated effective interconversion cross sections for charged kaons (left) and effective pion production cross section for charged kaons (right). The respective interconversion processes are indicated in the figures. 
\textbf{Centre}: Same but for $K_L$.
\textbf{Right}: Updated effective interconversion cross sections for injected nucleon-antinucleon pairs (left) and pions (right). Note that the pion cross sections have been updated w.r.t. the initial publication in \cite{Jung:2025dyo}.
}
\label{fig:km_comp}
\end{figure}

In this section, we will present the relevant thermally-averaged interconversion cross sections \cite{Jung:2025dyo} compared to the previous literature \cite{Reno:1988, Pospelov:2010cw,Akita:2024ork}. We have implemented both in order to allow for a cross-check of the dynamical-equilibrium description (cf.\ appendix~\ref{app:dyn_eq}). 

In the top row of figure~\ref{fig:km_comp}, we present the effective interconversion cross sections (left) as well as the effective pion-production cross section (right) for charged kaons. These can be inferred from a weighted average taking into account the various hyperon cross sections and branching ratios from results presented in~\cite{Jung:2025dyo}. Note that the $K^-+n$ initial state cannot produce any $\pi^+$, so that the effective cross section vanishes.

In the middle row of figure~\ref{fig:km_comp},
we show the same effective cross sections for the case of neutral long-lived kaons. Note the degeneracy in the previous literature, where it was assumed that all interconversion processes are identical. The same assumption leads to the two-fold degeneracy in the right plot, where the production of $\pi^+$ from protons and $\pi^-$ from neutrons is the same, as well as the two remaining processes, respectively.

For the (anti-)nucleonic processes, we show the cross sections in the bottom row of figure~\ref{fig:km_comp}
(left). We note the degeneracies in the cross sections: for the new results, all but the $\bar{p}+p$ are (approximately) the same, while for the older results the identical and non-identical particle initial states ($\bar{p}p/\bar{n}n$ and $\bar{p}n/\bar{n}p$, respectively) are identical modulo the Coulomb enhancement.

Finally, let us discuss the interconversion cross sections of the pions.\footnote{We are deeply thankful to the authors of \cite{Jung:2025dyo} who have provided us with corrected pion cross sections in private communication. Please consult the article once updated for more information.} We find them to qualitatively agree with previous literature at low temperatures as presented in the bottom right panel of figure~\ref{fig:km_comp}.

\section{Checking dynamical equilibrium against full-time evolution}\label{app:dyn_eq}

In this section, we will check the dynamical-equilibrium approximation against the full evolution code of refs.~\cite{Akita:2024ork, ovchynnikov_2025_14882777}. This code was originally written for lighter relics and focuses more on the impact of the additional neutrinos from mesonic processes on neutrino decoupling. Thus, we consider a $m_\phi=5\,$GeV relic with a low initial abundance of $\mathpazoletters{Y}_\phi=10^{-8}$, so that the background cosmology is barely impacted by the presence and decay of the relic. Due to the different focus of our code and \cite{ovchynnikov_2025_14882777}, we further need to deactivate all weak rates and fusion processes in \texttt{AlterAlterBBN}. Finally, instead of the updated cross sections from \cite{Jung:2025dyo}, we need to implement matching cross sections from the previous literature \cite{Reno:1988, Pospelov:2010cw} to ensure consistency between the results. In agreement with the literature, we assume that for the old rates, the thermally-averaged cross sections are identical to the threshold cross sections $\sigma v$ (without thermal average) for $s\to (m_1+m_2)^2$ with $m_{1/2}$ being the masses of the incoming particles.
In this regime, we expect the typical nuclear cross section to behave like $\sigma \propto 1/v$ for decreasing velocity. Therefore, the thermal average $\langle \sigma v\rangle$ is trivial for the non-relativistic velocities considered in this context. Our main observables are the neutron and proton number densities at around the $100\,$keV range when BBN begins. 

We choose to investigate the pre-implemented ``Toy model kaons'' in which for every decaying particle, a $K^+K^-$ pair is injected. This scenario will capture all the important aspects of the dynamical equilibrium. In particular, it comes with a direct source term for the kaons and a ``clean'' secondary sourcing of pions. To reproduce this model, we need to manipulate the injection (cf.\ section~\ref{sec:pythia}): We choose the injection of electron-positron pairs. The electromagnetic component will be irrelevant at these high temperatures and low abundances, and so is the number of injected baryons at this low mass. We then rescale the charged-kaon injection term to one pair per decay and set all other mesonic injection rates to 0.

\begin{figure}
    \centering
    \includegraphics{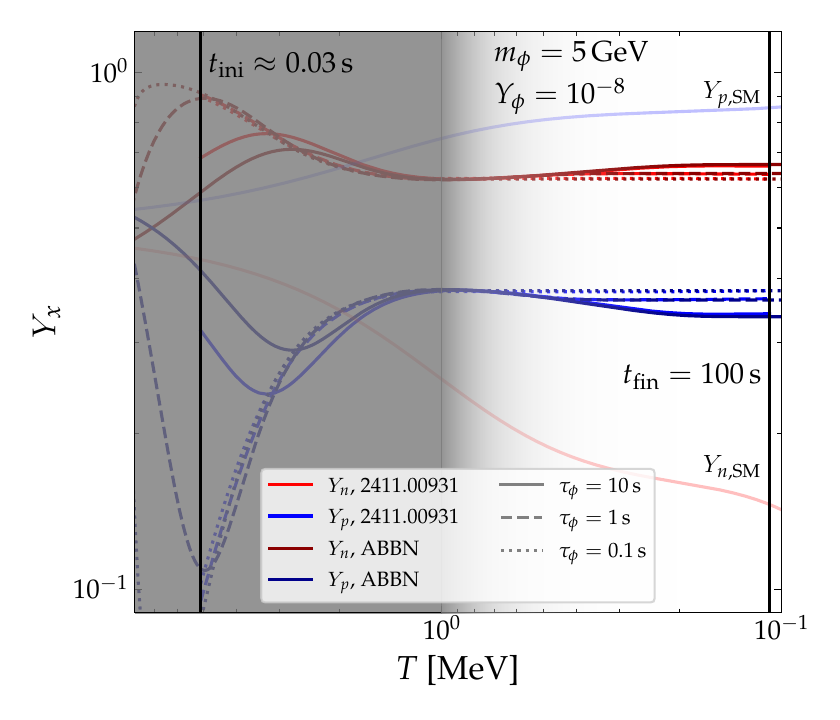}
    \caption{Temperature evolution of the neutron/proton (red/blue) abundances in \cite{ovchynnikov_2025_14882777} compared to the dynamical equilibrium approach, which we implemented in \texttt{AlterAlterBBN} (darkred/darkblue). We see good agreement for a set of different lifetimes: $\tau_\phi=10\,$s (solid), $1\,$s (dashed), and $0.1\,$s (dotted) in the relevant area beyond the grey shading. }
    \label{fig:dyneq_conf}
\end{figure}

In figure~\ref{fig:dyneq_conf}, we show the results of our implementation contrasted against the results from \cite{ovchynnikov_2025_14882777}. We have modified the code of \cite{ovchynnikov_2025_14882777} minimally to let the time evolution run for longer. To be precise, we have set $t_\text{fin}=100\,$s (instead of $10\tau_\phi$) for better comparability between the different benchmark points. We show the neutron and proton abundances in red and blue for the literature result and in darker shades for our approach. The different lifetimes are shown as different linestyles.

We see that our implementation agrees well with the literature result in the relevant temperature range.\footnote{The reason is probably that at later times, the meson abundances are dominated by decays, which makes their and the nucleon abundances less inter-dependent and thus more robust (cf.\ figure~\ref{fig:dyneq}).}
For the longer lifetime, we find the largest discrepancy with an asymptotic relative deviation of a little more than $1\%$ for the neutron abundance. Overall, we expect the uncertainty in the rates to have a much bigger effect than the systematic uncertainty introduced in the dynamical-equilibrium approximation. To guide the eye to the relevant temperature range, we have coloured the regions that have a negligible impact on the fusion of the light elements in grey.

We should keep in mind that this comparison is strictly meant as a consistency check of the \textit{implementation} of the hadronic interconversions and does not correspond to a physical point. In reality, the electroweak reactions as well as fusion processes will also enter the evolution. The important takeaway is that the relevant temperature region around the deuterium bottleneck is accurately modelled in the dynamical-equilibrium approach. We also show the SM abundances as semi-transparent lines for comparison.

\section{Implementing hadrodisintegration into \texttt{AlterAlterBBN}} \label{app:hadrodis_abbn}

In this appendix, we will discuss a few of the technicalities when integrating hadrodisintegration into our previous version of \texttt{AlterAlterBBN}. We further describe the matching procedure with \texttt{ACROPOLIS}. We use the framework presented in \cite{Kawasaki:2005,Kawasaki:2018} and implemented (with minor modifications) in \cite{Bianco:2025boy}. To conform with the base code of \texttt{AlterBBN} we port this code to \texttt{C}. For details on the framework, the interested reader can consult the aforementioned references. Thus, we will not elaborate on the details of the formalism here but rather summarise the specific changes required to merge it into the main code.

As discussed in section~\ref{sec:hadrodis}, to determine the rate of hadrodisintegration, we need two key ingredients: the first one being the injection spectrum of hadronic material, which we approximate by using the average kinetic energy, and the total injection rate (cf.\ also section~\ref{sec:pythia}). However, this needs to be convoluted with the ``efficiency'' of the disintegration, which we measure using the $\xi$-parameters, which determine the number of light elements produced/destroyed \emph{per} injected nucleon, whose calculation requires the actual hadrodisintegration code as shown in eq.~\eqref{eq:y_hdi}.

Let us summarise how we previously tackled the hadrodisintegration in \texttt{ACROPOLIS}:
all photodisintegration rates are perfectly linear in the number density (or the yield $Y_x=n_x/n_b$). We can therefore write a linear system
\begin{align}
    \frac{\text{d}\mathbf{Y}}{\text{d}T}=\mathfrak{R}(T)\mathbf{Y}\;,
\end{align}
where $\mathbf{Y}$ is the vector of abundances and $\mathfrak{R}(T)$ is a matrix with the corresponding rates. We immediately see that the solution to this equation can be written as 
\begin{align}
    \mathbf{Y}(T)=\exp\left(\int_{T}^{T_\text{min}}\text{d}T^\prime\,\mathfrak{R}(T^\prime)\right)\mathbf{Y}_0\label{eq:lin_sys}
\end{align}
with $\mathbf{Y}_0$ being the initial conditions inferred from running a BBN code. This elegant solution is fully based on the fact that $\mathfrak{R}$ does \emph{not} depend on  $\mathbf{Y}$. In the case of the $\xi$-parameters, this is not the case anymore. However, we realised that we can find a good linear approximation of eq.~\eqref{eq:y_hdi}
\begin{align}
    \left[ \frac{\text{d} n_X}{\text{d} t} \right]_\text{hadro} &= \sum_{y = n, p}\int_0^\infty \text{d}K\; \xi_{X}^y[n_j](K) \frac{\text{d}^2 n_y^\text{inj}}{\text{d} t \text{d} K}(K)\\
     &\simeq \sum_{y = n, p} \left(\xi_{X}^{y}[n_j](K_\text{hd}^\text{inj}) \frac{\text{d} n_\text{hd}^\text{inj}}{\text{d} t}\right)\\
    \text{with}\quad&\xi_X^y[n_j](T, K) \simeq \frac{\xi_X^y[n_j^0](T, K)}{n_{A(X)}^0} \times n_{A(X)}(T)\;,
\end{align}
see ref.~\cite{Bianco:2025boy} for more details.
In the first step, we have again made use of the approximation of mono-energetic hadron injections. Furthermore, we have expanded the $\xi$-parameter linearly around the SM abundance for the background protons and $^4$He which we have checked to be a good approximation. In these expressions, we used $A(X) = p$ if $X \in \{p, n\}$ and $A(X) = {}^4\text{He}$ otherwise. This is the key step for the linearisation because it transforms hadrodisintegration into a ``linear'' process which allows us to remain within the \texttt{ACROPOLIS} framework. 

In the case of \texttt{AlterAlterBBN}, we cannot use the same method: BBN naturally has many non-linear reactions. In fusion processes, there is often a higher multiplicity of nuclei in the initial and/or final state. Nevertheless, we can still use linear algebra methods to solve the system,\footnote{Note that the typical fusion reactions are power laws in the abundances, so that linearisation is straightforward $Y_x^n=Y_x^{n-1}\times Y_x$.} but we need to do this on smaller time/temperature intervals because, simply speaking, the corresponding $\mathfrak{R}$ for the fusion rates will now depend strongly on the light-element abundances, and thus we cannot just integrate up the system as in eq.~\eqref{eq:lin_sys}. 

To linearise the resulting expressions we can write $\mathcal{R}_\text{lin}(T)=[\mathcal{R}_x(T,Y_x)/Y_x(T)]\times Y_x(T)$ for every time step. We have tested this prescription and found it to run efficiently and produce robust results. For the technical details on how to solve such a system efficiently and a deeper discussion of the linearisation technique, we refer to \cite{Arbey:2011nf,Arbey:2018zfh}.

As a final step, we need to match the two codes together consistently. To do so, we need to find a sweet spot in temperature where photodisintegration is not relevant yet while all the fusion rates have already become subdominant. The reason is that thermal BBN is exclusively handled in \texttt{AlterAlterBBN}, which, however, cannot handle photodisintegration, while photodisintegration is handled in \texttt{ACROPOLIS}, which cannot handle nuclear fusion; yet, both codes are able to perform hadrodisintegration. We have found $T=5\,$keV to be a good compromise for the transition from one code to the other as fusion rates are already low and the universal spectrum starts to contain photons sufficiently energetic to disintegrate deuterium at $T=5.34\,$keV. In particular, we have checked that the final result varies only weakly under a change of the transition temperature and the abundances connect smoothly.

\begin{figure}[!t]
    \centering
    \includegraphics{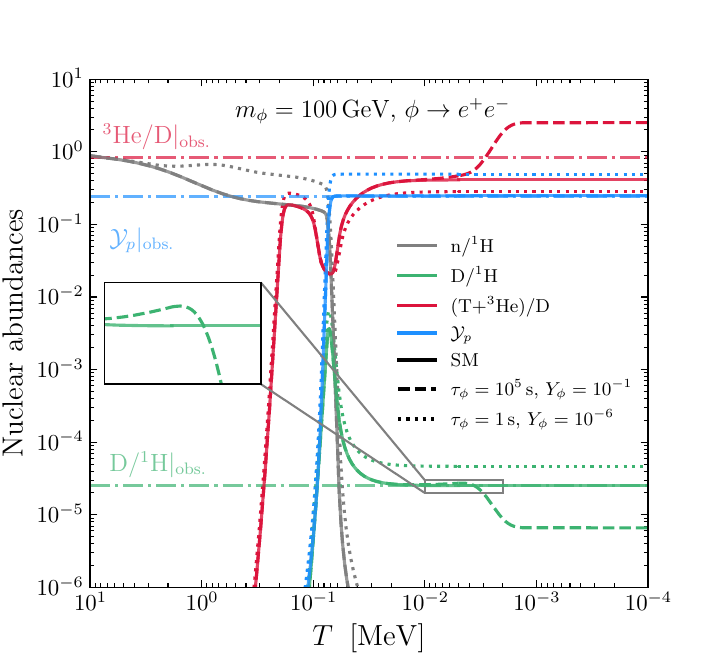}
    \caption{Evolution of the light-element abundances with matching between \texttt{AlterAlterBBN} and \texttt{ACROPOLIS} for two benchmark points.}
    \label{fig:evo_dis}
\end{figure}

To demonstrate that this procedure connects the two evolution periods sufficiently smoothly, we can revisit the full evolution of the light elements in figure~\ref{fig:evo_dis_2}, which makes it evident that the matching does not introduce any artefacts. We emphasise this point by showing two other benchmark points in figure~\ref{fig:evo_dis} ([$m_\phi=100\,$GeV, $\tau_\phi=10^5\,$s, $\mathpazoletters{Y}_\phi=10^{-11}$] and [$m_\phi=100\,$GeV, $\tau_\phi=1\,$s, $\mathpazoletters{Y}_\phi=10^{-6}$], respectively).

\begin{figure}[!t]
 \centering
    \begin{subfigure}{0.47\linewidth}
      \centering
      \includegraphics[width=0.72\linewidth, trim= 3cm 0 2cm 2cm]{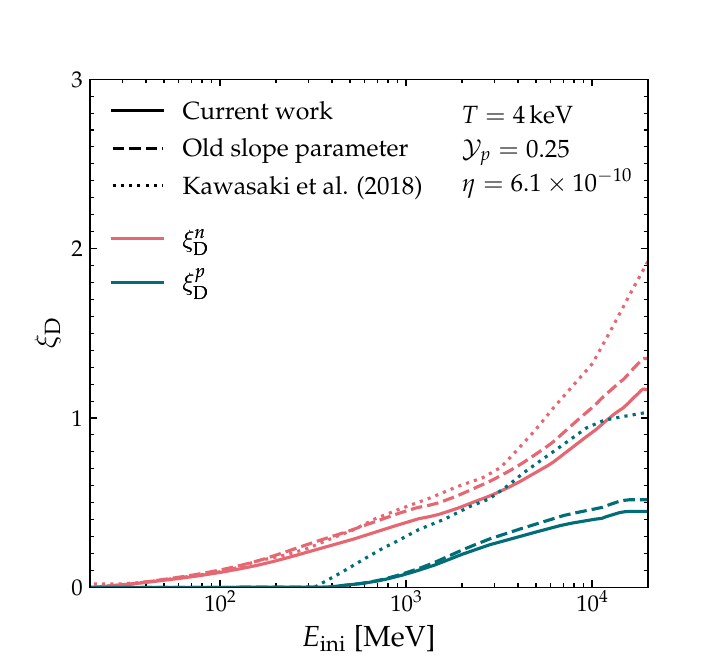}
    \end{subfigure}
    \begin{subfigure}{0.47\linewidth}
      \centering
      \includegraphics[width=0.72\linewidth, trim= 2cm 0 3cm 2cm]{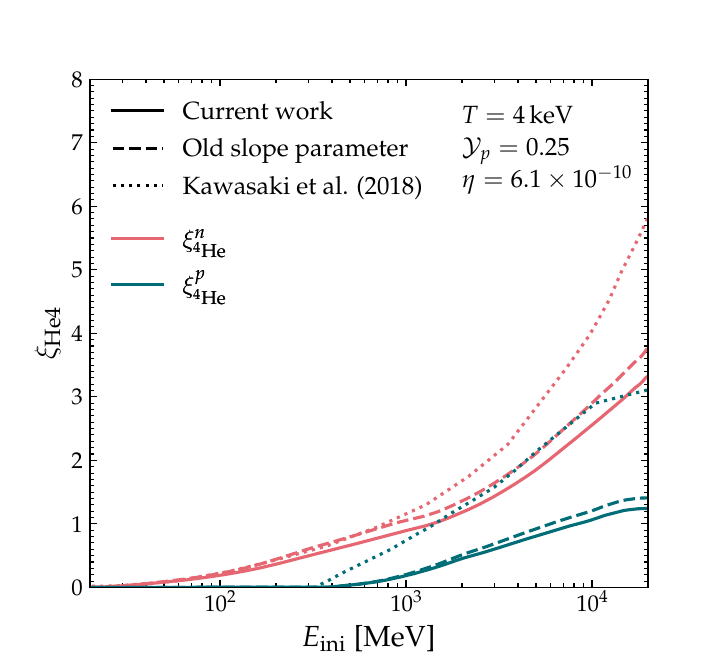}
    \end{subfigure}
    \caption{Comparison of our code with (solid) and without (dashed) the new slope parameter against the literature result (dotted).
    \textbf{Left:} Deuterium yield per injected neutron (red) and proton (blue-green) for fixed temperature compared to the literature results. \textbf{Right:} \textit{Destruction} of  $^4$He. Note that in this panel, we have flipped the sign of $\xi^y_{{}^4\text{He}}$ in order for it to be positive.}
    \label{fig:xis}
\end{figure}

As a final note of this section, let us briefly mention one additional update that we have implemented in \texttt{AlterAlterBBN} and \texttt{ACROPOLIS}. To improve upon the elastic scattering that drives the hadronic cascade \cite{Kawasaki:2005,Kawasaki:2018,Bianco:2025boy}, we have updated the so-called slope parameter with results from \cite{Cugnon:1996kh,Falter:2004uc,Larionov:2025wce}. This parameter determines the outgoing energies in scattering between injected nucleons and the background protons. We show the effect in figure~\ref{fig:xis} where we see that the changes in elastic scattering make the hadronic cascade less efficient in terms of disintegration for both protons and neutrons.

\section{An improved numerical scheme}\label{app:numerical}

In this section, we will discuss our numerical approach for the time/temperature evolution of the background cosmology in detail. This procedure is improving upon the time evolution explained in our previous work \cite{Bianco:2025boy}, leveraging the fact that we do not have to track the neutrino cascade, which provides us with additional options and significantly shortened run times. Tracking the background cosmology is highly relevant to correctly determining the injections, which depend on the evolution of the relic abundance.

To simplify the discussion, we will first discuss our upgrades in the case of standard $\Lambda$CDM. However, it is instructive to discuss the time-temperature relation in the general case. From energy conservation across all sectors, we can derive
\begin{align}
    \frac{\text{d}T}{\text{d}t}&=-\frac{\dot{\rho}_\phi+3H(\rho_\phi+\rho_\text{SM}+P_\text{SM})}{\frac{\text{d}\rho_\text{SM}}{\text{d}T}}=\frac{\dot{q}_\phi-3H(\rho_\text{SM}+P_\text{SM})}{\frac{\text{d}\rho_\text{SM}}{\text{d}T}}\qquad T\geq T_{\nu,\text{dec}}\;, \label{eq:tT_SM}
\end{align}
where we denote the energy-injection rate of the relic as $\dot{q}_\phi=\rho_\phi/\tau_\phi$. We note that this is valid until neutrino decoupling, which we model as instantaneous. Below neutrino decoupling, the $\nu$ sector is completely independent of the SM one.\footnote{We ignore possible reheating effects due to large relic abundances as well as the neutrinos following from the FSR.} We thus find
\begin{align}
\begin{aligned}
    \frac{\text{d}T}{\text{d}t}&=\frac{\dot{q}_\phi-3H(\rho_\text{EM}+P_\text{EM})}{\frac{\text{d}\rho_\text{EM}}{\text{d}T}}\\
    \frac{\text{d}T_\nu}{\text{d}t}&=-H T_\nu
    \end{aligned}\qquad T<T_{\nu,\text{dec}}\;,
\end{align}
where only the remaining electromagnetic particles contribute to the time-temperature relation of the electromagnetic bath. The neutrinos just redshift down from the time of decoupling with the initial condition $T_\nu(T_{\nu,\text{dec}})=T_{\nu,\text{dec}}$.
We have discussed how to determine the temperature of neutrino decoupling in a modified background cosmology, cf.\ eq.~\eqref{eq:T_dec_scaling}.

We will now introduce a numerical scheme based on the midpoint method to solve the time-temperature evolution. To ensure high precision and stability, we use the implicit version of this method. We determine the next temperature point for a fixed time step as
\begin{align}
    T_{n+1}=T_{n}+\frac{\text{d}T}{\text{d}t}\left(\frac{t_n+t_{n+1}}{2},\frac{T_n+T_{n+1}}{2} \right)\;.\label{eq:midpoint}
\end{align}
We note that in the pure SM case of $\dot{q}_\phi=0$, the term on the right-hand side only depends on the temperature. To get accurate results, we iterate over different $T_{n+1}^i$, starting with $T_{n+1}^0=T_n$ until $(T_{n+1}^{i+1}-T_{n+1}^i)/T_{n+1}^i<\epsilon_T\ll 1$ where we can choose $\epsilon_T$ to find a good compromise between run time and precision.\footnote{For early times when we do not have to track injections, we use $\epsilon_T=10^{-8}$ while later on we reduce the precision to $10^{-5}$ to improve the run time.} We will generalise this later on to also include non-SM cases. At this point, we should emphasise that this implicit method generally provides better results than the explicit counterpart; however, it requires additional (iterative) evaluations of the derivative in eq.~\eqref{eq:midpoint}.

We have confirmed the accuracy of this method against the evolution of the SM. We were able to reproduce the exact solution to within $\mathcal{O}(10^{-4})$ for time steps as large as $\delta t =10^{-2}/H$.

We further want to track the scale factor $R$ in our code, in particular in the case of a non-vanishing relic density of $\phi$. Nevertheless, it is instructive to once again discuss the SM scenario first. For the time evolution, we are then confronted with two possible avenues. Let us begin with the more restrictive, but simple, method of entropy conservation
\begin{align}
    \dot{s}=-3Hs \Rightarrow s\propto R^{-3}\Rightarrow R_{n+1}=R_n\left[\frac{s(T_n)}{s(T_{n+1})}\right]^{1/3}\;.\label{eq:ent_cons}
\end{align}
This approach is the preferred choice in the case of strict entropy conservation in the SM because it is an exact solution of the evolution equations. If we have solved the time-temperature evolution with sufficient precision, this method will always provide an excellent approximation of the function $R(t)$. Alternatively, we can also use the first Friedmann equation to find
\begin{align}
    \frac{\dot{R}}{R}=H(t) \Rightarrow R(t)=R_0 e^{\int_{t_0}^t \text{d}t^\prime H(t^\prime)}\;.
\end{align}
We see that this approach requires us to solve an integral to determine the time evolution. To find a suitable numerical scheme, we write
\begin{align}
    R_{n+1}&=R_n e^{\int_{t_n}^{t_{n+1}} \text{d}t^\prime H(t^\prime)}\\
    \int_{t_n}^{t_{n+1}} \text{d}t^\prime H(t^\prime)&\simeq \int_{t_n}^{t_{n+1}} \text{d}t^\prime \left(H(t_n)+\frac{\text{d}H}{\text{d}t}\left(t_n\right)\left(t^\prime-t_0\right)\right)\\
    &=H(t_n)(t_{n+1}-t_n)+\frac{\text{d}H}{\text{d}t}\left(t_n\right)\frac{\left(t_{n+1}-t_n\right)^2}{2}\;.\label{eq:sec_ord}
\end{align}
Under the integral, we have expanded the Hubble rate to first order to allow for higher precision. In particular, the second Friedmann equation provides the quadratic term (in the time step) for free 
\begin{align}
    \frac{\Ddot{R}}{R}&=-\frac{4\pi G}{3}(\rho+3P)\\
    \Rightarrow \frac{\text{d}H}{\text{d}t}&=\left(\frac{\Ddot{R}}{R}-\frac{\dot{R}^2}{R^2}\right)=-4\pi G (\rho+P)\;,
\end{align}
where we have introduced $\rho$ and $P$ as the total energy density and pressure of the Universe, respectively. We can evaluate this expression at $T(t_n)$ and same goes, of course, for $H(T(t_n))$. The complete evaluation scheme therefore is
\begin{align}
    R_{n+1}&\simeq R_n e^{H(t_n)(t_{n+1}-t_n)-2\pi G [\rho(t_n)+P(t_n)]\frac{\left(t_{n+1}-t_n\right)^2}{2}}\;.
\end{align}
We find these two methods to be efficient and fast in reproducing the correct SM results. However, note that we have not made any reference to entropy conservation for the ``Hubble method''. It is therefore also valid in the case of non-zero entropy injection. We should clarify that the appearance of the density and pressure is to be understood as the total density and pressure, including the relic.\footnote{In our scenario of a cold relic only the total \emph{energy density} is impacted.}

Before we continue with the implementation of non-SM effects, let us quickly analyse the accuracy of the method.
Since in the case of vanishing injection, the entropy conservation method is exact, we have compared the Hubble method to the simple relation in eq.~\eqref{eq:ent_cons}. 
In particular, we have found that it is important to take into account the expansion of the Hubble rate to the quadratic term in eq.~\eqref{eq:sec_ord}. We have ensured that the method is stable and provides sub-percent precision for time steps with $\delta t=10^{-2}/H$.

Now, let us try to make the implementation ``BSM-friendly''. For that, it is clear that the scale factor is easier to describe. In the Hubble method explained above, we just need to add the density of the relic in the current time step (since we do not rely on an implicit method there). Updating the relic abundance is then straightforward
\begin{align}
    \rho_{\phi,n+1}\simeq\rho_{\phi,n}\left(\frac{R_n}{R_{n+1}}\right)^3e^{-\frac{t_{n+1}-t_n}{\tau_\phi}}\;.
\end{align}
For the time-temperature relation, however, we require the SM heating rate, ideally evaluated at the midpoint. The heating rate must be proportional to $\rho_\phi$ itself. We can therefore evaluate the injection term at the midpoint via
\begin{align}
    \dot{q}_\phi\left(\frac{t_n+t_{n+1}}{2}\right)&\simeq\dot{q}_\phi(t_n)\left[\frac{R(t_n)}{R((t_{n+1}+t_n)/2)}\right]^4e^{-\frac{t_{n+1}-t_n}{2\tau_\phi}}\\
    R(\,(t_{n+1}+t_n)/2\,)&\simeq R(t_n) e^{H(t_n)(t_{n+1}-t_n)/2-2\pi G [\rho(t_n)+P(t_n)]\frac{\left(t_{n+1}-t_n\right)^2}{4}}\;.
\end{align}
We can use this expression, as well as the density evaluated via this method, to implicitly determine the change in temperature. In particular, a relic with extremely high initial density may reheat the SM bath, which is captured in this formalism.

We should disclaim that the main goal of our code is to accurately describe the background cosmology in the case where hadronic and EM injections have the leading impact on BBN and its products, so that even approximating the SM entropy to be exactly conserved is an excellent approach for large parts of the parameter space. Nevertheless, we have two secondary goals which require the code to also be reliable elsewhere. First of all, we would like to make sure that there are no cancellations between background cosmology and ``direct'' disintegration/interconversion effects. To this end, we also need to consistently reproduce background-cosmology bounds, which are typically weaker than the latter and thus require the correct treatment of larger injections. Secondly, towards very short lifetimes, we expect that all bounds eventually have to disappear. In our approximation of instantaneous neutrino decoupling, the effect of a particle decaying is parametrically suppressed by $\exp(-t_{\nu,\text{dec}}/\tau_\phi)$. Thus, even for large hadronic branching ratios, we require a large initial relic abundance to overcome the exponential suppression.

\subsection{A note on the effective degrees of freedom around neutrino decoupling}\label{app:dofs}

In our initial temperature evolution, we were confronted with local $\mathcal{O}(10\%)$ errors in the time temperature evolution, which we tracked back to our tabulated values for the degrees of freedom. Therefore, we needed to find more precise values for the effective degrees of freedom below $\sim 5\,$MeV. Fortunately, at these low temperatures, we are just left with photons, electrons/positrons, and neutrinos. Further, we can assume that all chemical potentials are 0. Photons have the simplest behaviour as they neither decouple nor have a mass. For electrons, we can write for the effective degrees of freedom \cite{Drees:2015exa}
\begin{align}
    g_{\ast,\rho,e}=\frac{\frac{m^4}{2\pi^2}\int_1^\infty\text{d}y\,\frac{y^2\sqrt{y^2-1}}{e^{ym/T}+1}}{\frac{\pi^2}{30}T^4}=\frac{15}{\pi^4}\left(\frac{m}{T}\right)^4\int_1^\infty\text{d}y\,\frac{y^2\sqrt{y^2-1}}{e^{ym/T}+1}\;,
\end{align}
and analogous for the pressure and entropy DoFs. The great advantage of this approach is that we can easily evaluate the derivative of the DoFs on any arbitrary grid without the problem of inducing additional errors from taking derivatives on discretised data.

Now, we should also discuss the neutrinos. In a pure SM scenario, we can always link the electromagnetic and neutrino degrees of freedom via entropy conservation after neutrino decoupling. In particular, we find that
\begin{align}
    \frac{s_\nu(T)}{s_\text{EM}(T)}&=\text{const.}=\frac{s_\nu(T_{\nu,\text{dec}})}{s_\text{EM}(T_{\nu,\text{dec}})}\\
    \Rightarrow \frac{g_\nu T_\nu^3}{g_\nu T_{\nu,\text{dec}}^3}&= \frac{[g_\gamma+g_{\ast,s,e}(T)] T^3}{[g_\gamma+g_{\ast,s,e}(T_{\nu,\text{dec}})] T_{\nu,\text{dec}}^3} \Leftrightarrow T_\nu=\left[\frac{(g_\gamma+g_{\ast,s,e}(T)) }{(g_\gamma+g_{\ast,s,e}(T_{\nu,\text{dec}}))}\right]^{1/3}T\;,
\end{align}
where we have used the fact that there are no additional injections of entropy for the two different sectors. We can therefore write
\begin{align}
    g_{\ast,\rho,\nu}&=6\cdot \frac{7}{8}\frac{\frac{\pi^2}{30}T_\nu^4}{\frac{\pi^2}{30}T^4}=6\cdot \frac{7}{8}\left[\frac{(g_\gamma+g_{\ast,s,e}(T)) }{(g_\gamma+g_{\ast,s,e}(T_{\nu,\text{dec}}))}\right]^{4/3}\\
    g_{\ast,P,\nu}&=6\cdot \frac{7}{8}\frac{\frac{\pi^2}{90}T_\nu^4}{\frac{\pi^2}{90}T^4}=6\cdot \frac{7}{8}\left[\frac{(g_\gamma+g_{\ast,s,e}(T)) }{(g_\gamma+g_{\ast,s,e}(T_{\nu,\text{dec}}))}\right]^{4/3}\\
    g_{\ast,s,\nu}&=6\cdot \frac{7}{8}\frac{\frac{2\pi^2}{45}T_\nu^3}{\frac{2\pi^2}{45}T^3}=6\cdot \frac{7}{8}\frac{(g_\gamma+g_{\ast,s,e}(T)) }{(g_\gamma+g_{\ast,s,e}(T_{\nu,\text{dec}}))}\;,
\end{align}
where the 6 stands for the number of left-handed (active) neutrinos (3 flavours, particles and anti-particles) and the $7/8$ stems from the fermionic nature of the neutrinos.
In the presence of non-negligible entropy injections, this straightforward rescaling, which only depends on the temperature, cannot be applied anymore. Nevertheless, since we are accurately tracking the neutrino temperature throughout our code, we do not have to work much harder in the general case.\footnote{Note, however, that using pre-tabulated values cannot describe the neutrino degrees of freedom anymore in this scenario.} As the ratio $T_\nu/T$ is always available, it is straightforward to determine the effective neutrino contribution to the total SM DoFs. Finally, let us note that these expressions are only valid after neutrino decoupling, whereas before, all the neutrino DoFs are constant ($6\cdot 7/8=21/4$).

\section{More detailed results from \texttt{PYTHIA}}\label{app:pythia}

\begin{figure}[t]
 \centering
    \begin{subfigure}{0.48\linewidth}
      \centering
      \includegraphics[width=0.72\linewidth, trim= 3cm 0 2cm 0cm]{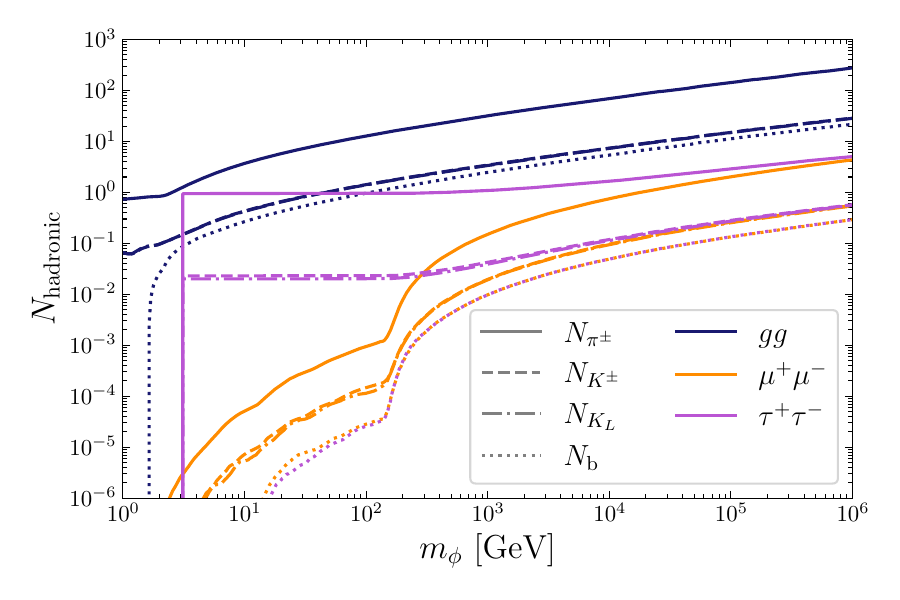}
    \end{subfigure}
    \begin{subfigure}{0.48\linewidth}
      \centering
      \includegraphics[width=0.72\linewidth, trim= 2cm 0 3cm 0cm]{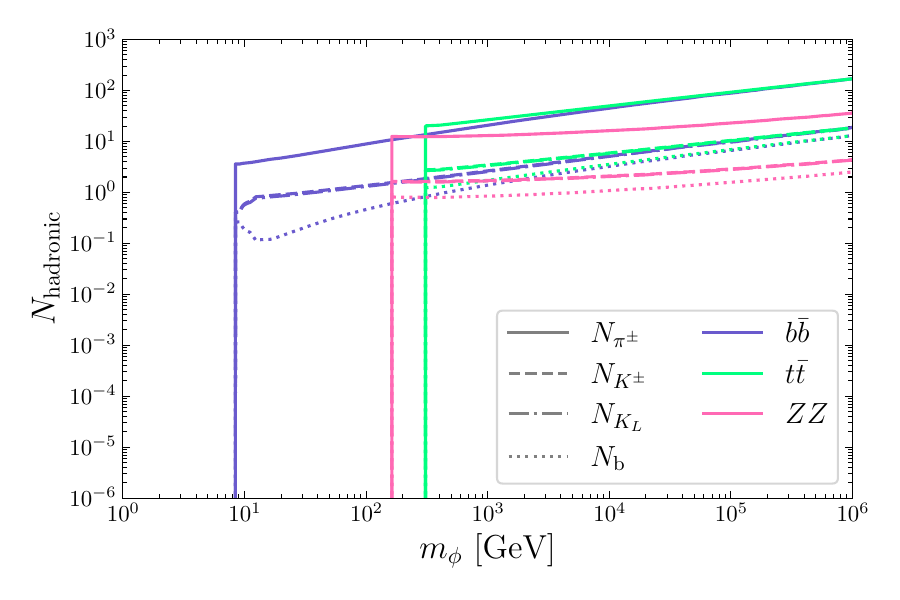}
    \end{subfigure}
    \caption{\textbf{Left:} Same as figure~\ref{fig:N_had} but for $gg$, $\mu^+\mu^-$, and $\tau^+\tau^-$. \textbf{Right}: Same but for $b\bar{b}$, $t\bar{t}$, and $ZZ$.}
    \label{fig:N_had2}
\end{figure}

In this brief section, we quickly discuss the hadronic injection of the other SM channels, beginning with $gg$, $\mu^+\mu^-$, and $\tau^+\tau^-$ (figure~\ref{fig:N_had2}, left). We find the gluon injection to be very similar to the quark injection, as we have already seen from the constraints. The charged leptons, however, show a universal behaviour: the muon results are identical to the electron ones within the statistical uncertainty in this approach. For the $\tau$, this is only true for the baryons as there are direct kaonic and pionic decay channels which populate the hadronic sector \emph{beyond} FSR.

\begin{figure}[!t]
 \centering
    \begin{subfigure}{0.48\linewidth}
      \centering
      \includegraphics[width=0.72\linewidth, trim= 3cm 0 2cm 0cm]{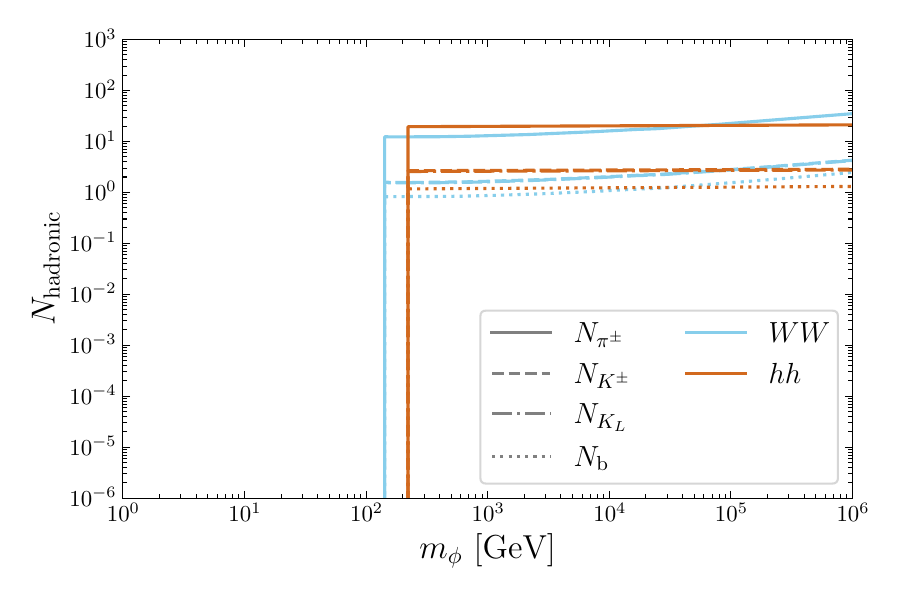}
    \end{subfigure}
    \begin{subfigure}{0.48\linewidth}
      \centering
      \includegraphics[width=0.72\linewidth, trim= 2cm 0 3cm 0cm]{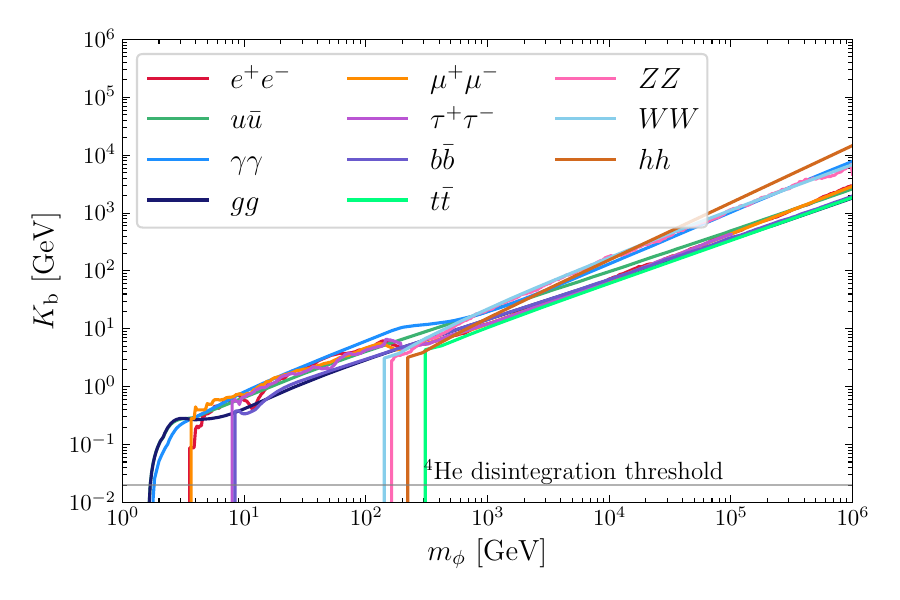}
    \end{subfigure}
    \caption{\textbf{Left:} Same as figure~\ref{fig:N_had} but for $W^+W^-$ and $hh$. \textbf{Right}: Kinetic energy of the injected baryons as a function of relic mass.}
    \label{fig:N_had_K_had}
\end{figure}

In the right panel of figure~\ref{fig:N_had2}, we can observe the injected hadrons for $b\bar{b}$, $t\bar{t}$, and $ZZ$. For the colored particles, we once more find a universal behaviour (away from the mass thresholds). However, the $ Z$ boson also shows a very similar behaviour with only mildly lower hadronic injections due to the large hadronic branching ratios. It is useful to remember that the results close to the mass threshold are not reliable.

We close the discussion on the hadronic injections with figure~\ref{fig:N_had_K_had} (left). The remaining particles are $W^+W^-$ and $hh$, which, once again, are similar to ``pure QCD'' injections because of their significant branching ratios into quarks.

As a last point, let us quickly discuss figure~\ref{fig:N_had_K_had} (right), which shows the average kinetic energy of the injected nucleons. We see that the function is approximately linear in $m_\phi$ with a factor $\sim 0.01$ between it and the kinetic energy.

\clearpage
\bibliographystyle{JHEP}
\bibliography{biblio.bib}
\end{document}